

\chardef\tempcat=\the\catcode`\@ \catcode`\@=11
\def\cyracc{\def\u##1{\if \i##1\accent"24 i%
    \else \accent"24 ##1\fi }}
\newfam\cyrfam
\font\tencyr=wncyr10
\newfam\cyrfam
\font\eightcyr=wncyr8
\def\cyr{\fam\cyrfam\tencyr\cyracc}
\def\smallcyr{\fam\cyrfam\eightcyr\cyracc}


\input epsf

\def\figin{\epsfcheck\figin}\def\figins{\epsfcheck\figins}
\def\epsfcheck{\ifx\epsfbox\UnDeFiNeD
\message{(NO epsf.tex, FIGURES WILL BE IGNORED)}
\gdef\figin##1{\vskip2in}\gdef\figins##1{\hskip.5in}
\else\message{(FIGURES WILL BE INCLUDED)}%
\gdef\figin##1{##1}\gdef\figins##1{##1}\fi}
\def\DefWarn#1{}
\def\figinsert{\goodbreak\topinsert}
\def\ifig#1#2#3#4{\DefWarn#1\xdef#1{fig.~\the\figno}
\writedef{#1\leftbracket fig.\noexpand~\the\figno}%
\figinsert\figin{\centerline{\epsfxsize=#3mm \epsfbox{#2}}}
\bigskip\medskip\centerline{\vbox{\baselineskip12pt
\advance\hsize by -1truein\noindent\footnotefont{\sl Fig.~\the\figno:}\sl\ #4}}
\bigskip\endinsert\noindent\global\advance\figno by1}

\def\Figx#1#2#3{
\bigskip
\vbox{\centerline{\epsfxsize=#1 cm \epsfbox{NOpic#2.eps}}
\centerline{{\bf Fig.\the\figno} #3}}\bigskip\global\advance\figno by1}

\def\Figy#1#2#3{
\bigskip
\vbox{\centerline{\epsfysize=#1 cm \epsfbox{NOpic#2.eps}}
\centerline{{\bf Fig.\the\figno} #3}}\bigskip\global\advance\figno by1}
\newcount\figno
 \figno=1
 \def\fig#1#2#3{
 \par\begingroup\parindent=0pt\leftskip=1cm\rightskip=1cm\parindent=0pt
 \baselineskip=11pt
 \global\advance\figno by 1
 \midinsert
 \epsfxsize=#3
 \centerline{\epsfbox{#2}}
 \vskip 12pt
 {\bf Fig.\ \the\figno: } #1\par
 \endinsert\endgroup\par
 }
 \def\figlabel#1{\xdef#1{\the\figno}}
 \def\encadremath#1{\vbox{\hrule\hbox{\vrule\kern8pt\vbox{\kern8pt
 \hbox{$\displaystyle #1$}\kern8pt}
 \kern8pt\vrule}\hrule}}


\def\unlockat{\catcode`\@=11}

\unlockat

\global\newcount\secno \global\secno=0
\global\newcount\prono \global\prono=0
\def\newsec#1{\vfill\eject\global\advance\secno by1\message{(\the\secno. #1)}
\global\subsecno=0\global\subsubsecno=0
\global\deno=0\global\teno=0
\eqnres@t\noindent
{\titlefont\the\secno. #1}
\writetoca{{\bf\secsym} {\rm #1}}\par\nobreak\medskip\nobreak}
\global\newcount\subsecno \global\subsecno=0
\def\subsec#1{\global\advance\subsecno
by1\message{(\secsym\the\subsecno. #1)}
\ifnum\lastpenalty>9000\else\bigbreak\fi
\global\subsubsecno=0
\global\deno=0
\global\teno=0
\noindent{\bf\secsym\the\subsecno. #1}
\writetoca{\bf \string\quad {\secsym\the\subsecno.} {\it  #1}}
\par\nobreak\medskip\nobreak}
\global\newcount\subsubsecno \global\subsubsecno=0
\def\subsubsec#1{\global\advance\subsubsecno by1
\message{(\secsym\the\subsecno.\the\subsubsecno. #1)}
\ifnum\lastpenalty>9000\else\bigbreak\fi
\noindent\quad{\bf \secsym\the\subsecno.\the\subsubsecno.}{\ \sl \ #1}
\writetoca{\string\qquad\bf { \secsym\the\subsecno.\the\subsubsecno.}{\sl  \ #1}}
\par\nobreak\medskip\nobreak}

\global\newcount\deno \global\deno=0
\def\de#1{\global\advance\deno by1
\message{(\bf Definition\quad\secsym\the\subsecno.\the\deno #1)}
\ifnum\lastpenalty>9000\else\bigbreak\fi
\noindent{\bf Definition\quad\secsym\the\subsecno.\the\deno}{#1}
\writetoca{\string\qquad{\secsym\the\subsecno.\the\deno}{#1}}}

\global\newcount\prono \global\prono=0
\def\pro#1{\global\advance\prono by1
\message{(\bf Proposition\quad\secsym\the\subsecno.\the\prono 
)}
\ifnum\lastpenalty>9000\else\bigbreak\fi
\noindent{\bf Proposition\quad
\the\prono\quad}{\ninepoint #1}
}

\global\newcount\teno \global\prono=0
\def\te#1{\global\advance\teno by1
\message{(\bf Theorem\quad\secsym\the\subsecno.\the\teno #1)}
\ifnum\lastpenalty>9000\else\bigbreak\fi
\noindent{\bf Theorem\quad\secsym\the\subsecno.\the\teno}{#1}
\writetoca{\string\qquad{\secsym\the\subsecno.\the\teno}{#1}}}
\def\subsubseclab#1{\DefWarn#1\xdef #1{\noexpand\hyperref{}{subsubsection}%
{\secsym\the\subsecno.\the\subsubsecno}%
{\secsym\the\subsecno.\the\subsubsecno}}%
\writedef{#1\leftbracket#1}\wrlabeL{#1=#1}}

\def\unredoffs{} \def\redoffs{\voffset=-.40truein\hoffset=-.40truein}
\def\speclscape{}

\newbox\leftpage \newdimen\fullhsize \newdimen\hstitle \newdimen\hsbody
\tolerance=1000\hfuzz=2pt

\catcode`\@=11
\def\bigans{b }
\def\answ{b }

\ifx\answ\bigans\message{(This will come out unreduced.}
\magnification=1200\unredoffs\baselineskip=16pt plus 2pt minus 1pt
\hsbody=\hsize \hstitle=\hsize

\else\message{(This will be reduced.} \let\l@r=L
\magnification=1200\baselineskip=16pt plus 2pt minus 1pt \vsize=7truein
\redoffs \hstitle=8truein\hsbody=4.75truein\fullhsize=10truein\hsize=\hsbody
\output={\ifnum\pageno=0

   \shipout\vbox{{\hsize\fullhsize\makeheadline}
     \hbox to \fullhsize{\hfill\pagebody\hfill}}\advancepageno
   \else
   \almostshipout{\leftline{\vbox{\pagebody\makefootline}}}\advancepageno
   \fi}
\def\almostshipout#1{\if L\l@r \count1=1 \message{[\the\count0.\the\count1]}
       \global\setbox\leftpage=#1 \global\let\l@r=R
  \else \count1=2
   \shipout\vbox{\speclscape{\hsize\fullhsize\makeheadline}
       \hbox to\fullhsize{\box\leftpage\hfil#1}}  \global\let\l@r=L\fi}
\fi

\newcount\yearltd\yearltd=\year\advance\yearltd by -2000

\def\Title#1#2{\nopagenumbers
\abstractfont\hsize=\hstitle\rightline{#1}%
\vskip 5pt\centerline{\titlefont #2}\abstractfont\vskip .5in\pageno=0}
\def\Date#1{\vfill\centerline{#1}\tenpoint\supereject\global\hsize=\hsbody%
\footline={\hss\tenrm\folio\hss}}


\def\draftmode{\message{ DRAFTMODE }\def\draftdate{{\rm preliminary draft:
\number\month/\number\day/\number\yearltd\ \ \hourmin}}%

\writelabels\baselineskip=20pt plus 2pt minus 2pt
  {\count255=\time\divide\count255 by 60 \xdef\hourmin{\number\count255}
   \multiply\count255 by-60\advance\count255 by\time
   \xdef\hourmin{\hourmin:\ifnum\count255<10 0\fi\the\count255}}}

\def\nolabels{\def\wrlabeL##1{}\def\eqlabeL##1{}\def\reflabeL##1{}}
\def\writelabels{\def\wrlabeL##1{\leavevmode\vadjust{\rlap{\smash%
{\line{{\escapechar=` \hfill\rlap{\sevenrm\hskip.03in\string##1}}}}}}}%
\def\eqlabeL##1{{\escapechar-1\rlap{\sevenrm\hskip.05in\string##1}}}%
\def\reflabeL##1{\noexpand\llap{\noexpand\sevenrm\string\string\string##1}}}
\nolabels
%


\global\newcount\secno \global\secno=0
\global\newcount\meqno
\global\meqno=1
\def\eqnres@t{\xdef\secsym{\the\secno.}\global\meqno=1
\bigbreak\bigskip}
\def\sequentialequations{\def\eqnres@t{\bigbreak}}
\def\appendix#1#2{\vfill\eject\global\meqno=1\global\subsecno=0\xdef\secsym{\hbox{#1.}}
\bigbreak\bigskip\noindent{\bf Appendix #1. #2}\message{(#1. #2)}
\writetoca{Appendix {#1.} {#2}}\par\nobreak\medskip\nobreak}

\def\eqnn#1{\xdef #1{(\secsym\the\meqno)}\writedef{#1\leftbracket#1}%
\global\advance\meqno by1\wrlabeL#1}
\def\eqna#1{\xdef #1##1{\hbox{$(\secsym\the\meqno##1)$}}
\writedef{#1\numbersign1\leftbracket#1{\numbersign1}}%
\global\advance\meqno by1\wrlabeL{#1$\{\}$}}
\def\eqn#1#2{\xdef #1{(\secsym\the\meqno)}\writedef{#1\leftbracket#1}%
\global\advance\meqno by1$$#2\eqno#1\eqlabeL#1$$}

\newskip\footskip\footskip14pt plus 1pt minus 1pt

\def\footnotefont{\ninepoint}\def\f@t#1{\footnotefont #1\@foot}
\def\f@@t{\baselineskip\footskip\bgroup\footnotefont\aftergroup\@foot\let\next}
\setbox\strutbox=\hbox{\vrule height9.5pt depth4.5pt width0pt}
\global\newcount\ftno \global\ftno=0
\def\foot{\global\advance\ftno by1\footnote{$^{\the\ftno}$}}

\newwrite\ftfile
\def\footend{\def\foot{\global\advance\ftno by1\chardef\wfile=\ftfile
$^{\the\ftno}$\ifnum\ftno=1\immediate\openout\ftfile=foots.tmp\fi%
\immediate\write\ftfile{\noexpand\smallskip%
\noexpand\item{f\the\ftno:\ }\pctsign}\findarg}%
\def\footatend{\vfill\eject\immediate\closeout\ftfile{\parindent=20pt
\centerline{\bf Footnotes}\nobreak\bigskip\input foots.tmp }}}
\def\footatend{}

\global\newcount\refno \global\refno=1
\newwrite\rfile
\def\ref{[\the\refno]\nref}
\def\nref#1{\xdef#1{[\the\refno]}\writedef{#1\leftbracket#1}%
\ifnum\refno=1\immediate\openout\rfile=refs.tmp\fi \global\advance\refno
by1\chardef\wfile=\rfile\immediate \write\rfile{\noexpand\item{#1\
}\reflabeL{#1\hskip.31in}\pctsign}\findarg}

\def\findarg#1#{\begingroup\obeylines\newlinechar=`\^^M\pass@rg}
{\obeylines\gdef\pass@rg#1{\writ@line\relax #1^^M\hbox{}^^M}%
\gdef\writ@line#1^^M{\expandafter\toks0\expandafter{\striprel@x #1}%
\edef\next{\the\toks0}\ifx\next\em@rk\let\next=\endgroup\else\ifx\next\empty%
\else\immediate\write\wfile{\the\toks0}\fi\let\next=\writ@line\fi\next\relax}}
\def\striprel@x#1{} \def\em@rk{\hbox{}}
\def\lref{\begingroup\obeylines\lr@f}
\def\lr@f#1#2{\gdef#1{\ref#1{#2}}\endgroup\unskip}
\def\semi{;\hfil\break}
\def\addref#1{\immediate\write\rfile{\noexpand\item{}#1}}

\def\startrefs#1{\immediate\openout\rfile=refs.tmp\refno=#1}
\def\xref{\expandafter\xr@f}\def\xr@f[#1]{#1}
\def\refs#1{\count255=1[\r@fs #1{\hbox{}}]}
\def\r@fs#1{\ifx\und@fined#1\message{reflabel \string#1 is undefined.}%
\nref#1{need to supply reference \string#1.}\fi%
\vphantom{\hphantom{#1}}\edef\next{#1}\ifx\next\em@rk\def\next{}%
\else\ifx\next#1\ifodd\count255\relax\xref#1\count255=0\fi%
\else#1\count255=1\fi\let\next=\r@fs\fi\next}


\def\writetoc{\immediate\openout\tfile=betagamma.tmp
    \def\writetoca##1{{\edef\next{\write\tfile{\noindent  ##1
    \string\leaderfill {\noexpand\number\pageno} \par}}\next}}}


%
\catcode`\@=12 
%
\edef\tfontsize{\ifx\answ\bigans scaled\magstep3\else scaled\magstep4\fi}
\font\titlerm=cmr10 \tfontsize \font\titlerms=cmr7 \tfontsize
\font\titlermss=cmr5 \tfontsize \font\titlei=cmmi10 \tfontsize
\font\titleis=cmmi7 \tfontsize \font\titleiss=cmmi5 \tfontsize
\font\titlesy=cmsy10 \tfontsize \font\titlesys=cmsy7 \tfontsize
\font\titlesyss=cmsy5 \tfontsize \font\titleit=cmti10 \tfontsize
\skewchar\titlei='177 \skewchar\titleis='177 \skewchar\titleiss='177
\skewchar\titlesy='60 \skewchar\titlesys='60 \skewchar\titlesyss='60
\def\titlefont{\def\rm{\fam0\titlerm}
\textfont0=\titlerm \scriptfont0=\titlerms \scriptscriptfont0=\titlermss
\textfont1=\titlei \scriptfont1=\titleis \scriptscriptfont1=\titleiss
\textfont2=\titlesy \scriptfont2=\titlesys \scriptscriptfont2=\titlesyss
\textfont\itfam=\titleit
\def\it{\fam\itfam\titleit}\rm}
\font\authorfont=cmcsc10 \ifx\answ\bigans\else scaled\magstep1\fi
\ifx\answ\bigans\def\abstractfont{\tenpoint}\else \font\abssl=cmsl10 scaled
\magstep1 \font\absrm=cmr10 scaled\magstep1 \font\absrms=cmr7
scaled\magstep1 \font\absrmss=cmr5 scaled\magstep1 \font\absi=cmmi10
scaled\magstep1 \font\absis=cmmi7 scaled\magstep1 \font\absiss=cmmi5
scaled\magstep1 \font\abssy=cmsy10 scaled\magstep1 \font\abssys=cmsy7
scaled\magstep1 \font\abssyss=cmsy5 scaled\magstep1 \font\absbf=cmbx10
scaled\magstep1 \skewchar\absi='177 \skewchar\absis='177
\skewchar\absiss='177 \skewchar\abssy='60 \skewchar\abssys='60
\skewchar\abssyss='60
\def\abstractfont{\def\rm{\fam0\absrm}
\textfont0=\absrm \scriptfont0=\absrms \scriptscriptfont0=\absrmss
\textfont1=\absi \scriptfont1=\absis \scriptscriptfont1=\absiss
\textfont2=\abssy \scriptfont2=\abssys \scriptscriptfont2=\abssyss
\textfont\itfam=\bigit \def\it{\fam\itfam\bigit}\def\footnotefont{\tenpoint}%
\textfont\slfam=\abssl \def\sl{\fam\slfam\abssl}%
\textfont\bffam=\absbf \def\bf{\fam\bffam\absbf}\rm}\fi
\def\tenpoint{\def\rm{\fam0\tenrm}
\textfont0=\tenrm \scriptfont0=\sevenrm \scriptscriptfont0=\fiverm
\textfont1=\teni  \scriptfont1=\seveni  \scriptscriptfont1=\fivei
\textfont2=\tensy \scriptfont2=\sevensy \scriptscriptfont2=\fivesy
\textfont\itfam=\tenit \def\it{\fam\itfam\tenit}\def\footnotefont{\ninepoint}%
\textfont\bffam=\tenbf
\def\bf{\fam\bffam\tenbf}\def\sl{\fam\slfam\tensl}\rm}
\font\ninerm=cmr9 \font\sixrm=cmr6 \font\ninei=cmmi9 \font\sixi=cmmi6
\font\ninesy=cmsy9 \font\sixsy=cmsy6 \font\ninebf=cmbx9 \font\nineit=cmti9
\font\ninesl=cmsl9 \skewchar\ninei='177 \skewchar\sixi='177
\skewchar\ninesy='60 \skewchar\sixsy='60
\def\ninepoint{\def\rm{\fam0\ninerm}
\textfont0=\ninerm \scriptfont0=\sixrm \scriptscriptfont0=\fiverm
\textfont1=\ninei \scriptfont1=\sixi \scriptscriptfont1=\fivei
\textfont2=\ninesy \scriptfont2=\sixsy \scriptscriptfont2=\fivesy
\textfont\itfam=\ninei \def\it{\fam\itfam\nineit}\def\sl{\fam\slfam\ninesl}%
\textfont\bffam=\ninebf \def\bf{\fam\bffam\ninebf}\rm}
%
%

\hyphenation{anom-aly anom-alies coun-ter-term coun-ter-terms}
\def\inv{^{\raise.15ex\hbox{${\scriptscriptstyle -}$}\kern-.05em 1}}

\def\Dsl{\,\raise.15ex\hbox{/}\mkern-13.5mu D} 
\def\dsl{\raise.15ex\hbox{/}\kern-.57em\partial}

\def\tr{{\rm tr}} 
\font\bigit=cmti10 scaled \magstep1
\def\lspace{\ifx\answ\bigans{}\else\qquad\fi}
\def\lbspace{\ifx\answ\bigans{}\else\hskip-.2in\fi} 
\def\boxeqn#1{\vcenter{\vbox{\hrule\hbox{\vrule\kern3pt\vbox{\kern3pt
     \hbox{${\displaystyle #1}$}\kern3pt}\kern3pt\vrule}
    }}}
\def\mbox#1#2{\vcenter{\hrule \hbox{\vrule height#2in
         \kern#1in \vrule} \hrule}}  


\newwrite\ffile\global\newcount\figno \global\figno=1
\def\nfig#1{\xdef#1{fig.~\the\figno}%
\writedef{#1\leftbracket fig.\noexpand~\the\figno}%
\ifnum\figno=1\immediate\openout\ffile=figs.tmp\fi\chardef\wfile=\ffile%
\immediate\write\ffile{\noexpand\medskip\noexpand\item{Fig.\ \the\figno. }
\reflabeL{#1\hskip.55in}\pctsign}\global\advance\figno by1\findarg}
\def\xfig{\expandafter\xf@g}
\def\xf@g fig.\penalty\@M\ {}
\def\figs#1{figs.~\f@gs #1{\hbox{}}}
\def\f@gs#1{\edef\next{#1}\ifx\next\em@rk\def\next{}\else
\ifx\next#1\xfig #1\else#1\fi\let\next=\f@gs\fi\next}
\newwrite\lfile
{\escapechar-1\xdef\pctsign{\string\%}\xdef\leftbracket{\string\{}
\xdef\rightbracket{\string\}}\xdef\numbersign{\string\#}}

\def\writestop{\def\writestoppt{\immediate\write\lfile{\string\pageno%
\the\pageno\string\startrefs\leftbracket\the\refno\rightbracket%
\string\def\string\secsym\leftbracket\secsym\rightbracket%
\string\secno\the\secno\string\meqno\the\meqno}\immediate\closeout\lfile}}
\def\writestoppt{}\def\writedef#1{}
\def\seclab#1{\xdef #1{\the\secno}\writedef{#1\leftbracket#1}\wrlabeL{#1=#1}}
\def\subseclab#1{\xdef #1{\secsym\the\subsecno}%
\writedef{#1\leftbracket#1}\wrlabeL{#1=#1}}
\newwrite\tfile \def\writetoca#1{}
\def\leaderfill{\leaders\hbox to 1em{\hss.\hss}\hfill}


\def\tilde{\widetilde}
\def\bar{\overline}
\def\hat{\widehat}

\def\tg{{\tilde\gamma}}
\def\tu{{\tilde u}}
\def\tbt{{\tilde\beta}}
\def\hu{{\hat u}}

\def\cech{${\rm C}^{\kern-6pt \vbox{\hbox{$\scriptscriptstyle\vee$}\kern2.5pt}}${\rm ech}}
\def\Cech{${\sl C}^{\kern-6pt \vbox{\hbox{$\scriptscriptstyle\vee$}\kern2.5pt}}${\sl ech}}

\def\hlf{\scriptstyle{1/2}}

\def\a{{\alpha}}
\def\aa{{[{\a}]}}

\def\b{{\beta}}
\def\bbb{{[{\b}]}}
\def\d{{\delta}}
\def\g{{\gamma}}
\def\gg{{[{\g}]}}

\def\e{{\epsilon}}

\def\ve{{\varepsilon}}

\def\m{{\mu}}
\def\n{{\nu}}
\def\u{{\Upsilon}}
\def\l{{\lambda}}
\def\s{{\sigma}}
\def\t{{\theta}}

\def\o{{\omega}}


\def\CA{{\cal A}}

\def\CC{{\cal C}}

\def\CE{{\cal E}}
\def\CF{{\cal F}}
\def\CG{{\cal G}}

\def\CJ{{\cal J}}

\def\CL{{\cal L}}

\def\CO{{\cal O}}

\def\CQ{{\cal Q}}

\def\CS{{\cal S}}
\def\CT{{\cal T}}

\def\CV{{\cal V}}
\def\CW{{\cal W}}

\def\CZ{{\cal Z}}


\def\ba{{\bf a}}
\def\bB{{\bf B}}
\def\bb{{\bf b}}
\def\bC{{\bf C}}

\def\bd{{\bf d}}
\def\pd{{\p}^{\prime}}
\def\pdd{{\p}^{\prime\prime}}

\def\bg{{\bf g}}

\def\bk{{\bf k}}

\def\bm{{\bf m}}

\def\bP{{\bf P}}

\def\bR{{\bf R}}

\def\bT{{\bf T}}
\def\bt{{\bf t}}

\def\bV{{\bf V}}
\def\bv{{\bf v}}

\def\bZ{{\bf Z}}

\def\p{\partial}
\def\pb{\bar{\partial}}

\def\dd{{\rm d}}
\def\eb{{e}^{\dagger}}
\def\ib{\bar{i}}
\def\jb{\bar{j}}

\def\wb{\bar{w}}
\def\xb{\bar{x}}

\def\zb{\bar{z}}

\def\uA{{\underline{A}}}
\def\uB{{\underline{B}}}
\def\uC{{\underline{C}}}
\def\uD{{\underline{D}}}

%
%

\hyphenation{anom-aly anom-alies coun-ter-term coun-ter-terms}
\def\inv{^{\raise.15ex\hbox{${\scriptscriptstyle -}$}\kern-.05em 1}}

\def\Dsl{\,\raise.15ex\hbox{/}\mkern-13.5mu D} 
\def\dsl{\raise.15ex\hbox{/}\kern-.57em\partial}

\def\tr{{\rm tr}} 
\font\bigit=cmti10 scaled \magstep1

\def\lspace{\ifx\answ\bigans{}\else\qquad\fi}
\def\lbspace{\ifx\answ\bigans{}\else\hskip-.2in\fi} 
\def\boxeqn#1{\vcenter{\vbox{\hrule\hbox{\vrule\kern3pt\vbox{\kern3pt
      \hbox{${\displaystyle #1}$}\kern3pt}\kern3pt\vrule}\hrule}}}
\def\mbox#1#2{\vcenter{\hrule \hbox{\vrule height#2in
          \kern#1in \vrule} \hrule}}  
%

\def\log{{\rm log}}
\def\cos{{\rm cos}}

\def\darr#1{\raise1.5ex\hbox{$\leftrightarrow$}\mkern-16.5mu #1}

\def\half{{\textstyle{1\over2}}} 
\def\roughly#1{\raise.3ex\hbox{$#1$\kern-.75em\lower1ex\hbox{$\sim$}}}

\def\np#1#2#3{Nucl. Phys. {\bf B#1} (#2) #3}
\def\pl#1#2#3{Phys. Lett. {\bf #1B} (#2) #3}

\def\lmp#1#2#3{Lett. Math. Phys. {\bf #1} (#2) #3}
\def\cmp#1#2#3{Comm. Math. Phys. {\bf #1} (#2) #3}

\def\jhep#1#2#3{JHEP {\bf#1}(#2) #3}

\def\cqg#1#2#3{Class.~Quantum Grav. {\bf #1} (#2) #3}

\def\IB{\relax\hbox{$\inbar\kern-.3em{\rm B}$}}

\def\ID{\relax\hbox{$\inbar\kern-.3em{\rm D}$}}
\def\IE{\relax\hbox{$\inbar\kern-.3em{\rm E}$}}
\def\IF{\relax\hbox{$\inbar\kern-.3em{\rm F}$}}
\def\IG{\relax\hbox{$\inbar\kern-.3em{\rm G}$}}
\def\IGa{\relax\hbox{${\rm I}\kern-.18em\Gamma$}}
\def\IH{\relax{\rm I\kern-.18em H}}
\def\IK{\relax{\rm I\kern-.18em K}}
\def\IL{\relax{\rm I\kern-.18em L}}
\def\IP{\relax{\rm I\kern-.18em P}}
\def\II{\relax{\rm I\kern-.18em I}}

\def\ndt{{\noindent}}

\def\sssec#1{\ndt$\underline{#1}$}

\def\CA{{\cal A}}

\def\CC{{\cal C}}

\def\CE{{\cal E}}
\def\CF{{\cal F}}
\def\CG{{\cal G}}

\def\CJ{{\cal J}}

\def\CL{{\cal L}}

\def\CO{{\cal O}}

\def\CQ{{\cal Q}}

\def\CS{{\cal S}}
\def\CT{{\cal T}}

\def\CV{{\cal V}}
\def\CW{{\cal W}}

\def\CZ{{\cal Z}}

\def\bB{{\bf B}}
\def\bC{{\bf C}}

\def\bP{{\bf P}}

\def\bR{{\bf R}}

\def\bT{{\bf T}}

\def\bV{{\bf V}}

\def\bZ{{\bf Z}}

\def\p{\partial}
\def\pb{\bar{\partial}}

\def\dd{{\rm d}}
\def\ib{\bar{i}}
\def\jb{\bar{j}}

\def\wb{\bar{w}}

\def\zb{\bar{z}}

\def\ch{{\rm ch}}


\def\inbar{\,\vrule height1.5ex width.4pt depth0pt}

\def\boxit#1{\vbox{\hrule\hbox{\vrule\kern8pt
\vbox{\hbox{\kern8pt}\hbox{\vbox{#1}}\hbox{\kern8pt}}
\kern8pt\vrule}\hrule}}
\def\mathboxit#1{\vbox{\hrule\hbox{\vrule\kern8pt\vbox{\kern8pt
\hbox{$\displaystyle #1$}\kern8pt}\kern8pt\vrule}\hrule}}

\def\lime{{\rm Lim}_{\kern -16pt \vbox{\kern6pt\hbox{$\scriptstyle{\e \to 0}$}}}}

\def\naiveq{\qquad =^{\kern-12pt \vbox{\hbox{$\scriptscriptstyle{\rm naive}$}\kern5pt}} \qquad}

\def\lref{\begingroup\obeylines\lr@f}
\def\lr@f#1#2{\gdef#1{\ref#1{#2}}\endgroup\unskip}

\lref\landscape{L.~Sussking, {\it The anthropic landscape of string theory}, hep-th/0302219}

\lref\moore{G.~Moore, {\it Vanishing vacuum energies for nonsupersymmetric strings}, in, Cargese
School on nonpertutbative quantum field theory, 1987}

\lref\natnik{N.~Berkovits, N.~Nekrasov, {\it The character of pure spinors}, hep-th/0503075}
\lref\cartan{
E. Cartan, {\it The Theory of Spinors}, Dover, New York, 1981\semi P. Budinich and
A. Trautman, {\it The Spinorial Chessboard},  Trieste Notes in Physics,
Springer-Verlag, Berlin, 1988\semi
C.~Chevalley, {\it The algebraic theory of spinors}, Columbia University Press, NY 1954}.
\lref\berkovitsbghost{N.~Berkovits, {\it Covariant Multiloop Superstring Amplitudes}, hep-th/0410079\semi
N.~Berkovits, {\it Super-Poincare Covariant Two-Loop Superstring Amplitudes}, hep-th/0503197}
\lref\berkovitsmulti{N.~Berkovits, {\it  Multiloop Amplitudes and Vanishing Theorems using the Pure Spinor Formalism for the Superstring}, hep-th/0406055}
\lref\natser{N.~Berkovits, S.~Cherkis, {\it Higher dimensional twistor transforms using pure spinors},
hep-th/0409243}
\lref\berkovitsnonmin{N.~Berkovits, {\it Pure Spinor Formalism as an N=2 Topological String}, hep-th/0509120}
\lref\berkovitstrieste{N.~Berkovits, {\it ICTP Lectures on covariant quantization of the superstring}, 
hep-th/0209059}
\lref\howe{G.~Moore, P.~Nelson, {\it The aetiology of sigma model anomalies}, \cmp{100}{1985}{83-132}\semi
P.~S.~Howe, G.~Papadopoulos, {\it Anomalies in two-dimensional supersymmetric non-linear $\s$-models}, \cqg{4}{1987}{1749-1766}}
\lref\gravan{L.~Alvarez-Gaume, E.~Witten, {\it Gravitational anomalies}, \np{234}{1983}{269-330}}
\lref\gsw{M.~B.~Green, J.~H.~Schwarz, {\it Covariant description of superstrings}, \pl{136}{1984}{367-370}}
\lref\gsan{M.~B.~Green, J.~H.~Schwarz, {\it Anomaly cancellations in supersymmetric $d=10$ gauge theory and superstring theory}, \pl{149}{1984}{117-122}}
\lref\witohet{E.~Witten, {\it Some properties of $O(32)$ superstrings}, \pl{149}{1984}{351-356}} 
\lref\technique{A.~Schwarz, hep-th/0102182}
\lref\membrane{N. Berkovits, "Toward Covariant Quantization of the Supermembrane", 
JHEP 0209 (2002) 051, hep-th/0201151.}
\lref\vanhove{L. Anguelova, P.A. Grassi and P. Vanhove, {\it
Covariant one-loop amplitudes in D=11}, Nucl. Phys. B702 (2004) 269,
hep-th/0408171\semi P.A. Grassi and P. Vanhove, {\it
Topological M Theory from Pure Spinor Formalism}, hep-th/0411167.}
\lref\golos{A.~Gorodentsev and A.~Losev, around late 2000, unpublished.}
\lref\lmt{A.~Losev, A.~Marshakov, A.~Zeitlin, {\it On first order formalism in string theory}, hep-th/0510065}
\lref\movsch{M.~Movshev and A.~Schwarz,"On maximally supersymmetric
Yang-Mills theories", Nucl. Phys. B681 (2004) 324, hep-th/0311132\semi
M.~Movshev and A.~Schwarz, "Algebraic structure of Yang-Mills
theory", hep-th/0404183.}
\lref\abcd{N.~Nekrasov and S.~Schadchine, {\it ABCD of instantons},
\cmp{252}{2004}{359-391}, hep-th/0404225.}
\lref\FS{L.~D.~Faddeev, S.~L.~Shatashvili, {\it Realization of the Schwinger term in the Gauss law and the possibility of correct quantization of a theory with anomalies}, \pl{167}{1986}{225-228}\semi
L.~D.~Faddeev, S.~L.~Shatashvili, {\it Algebraic and Hamiltonian methods in the theory of non-abelian anomalies}, Theor. Math. Phys. {\bf 60} (1984) 770}
\lref\atiyahsinger{M.~F,~Atiyah, I.~M.~Singer, {\it Dirac operators coupled to vector bundles}, Proc. Natl. Acad. Sciences, {\bf 81} (1984) 2597}
\lref\atiyahbott{M.~Atiyah, R.~Bott, {\it The moment map and equivariant cohomology} , Topology  {\bf  23}, vol. 1(1984) 1-28}
\lref\samsonco{A.~Alekseev, L.~Fadeev, S.~Shatashvili, {\it Quantization of symplectic orbits of compact Lie groups by means of the functional integral}, J.of Geometry and Physics, {\bf 5} (3) (1988) 391-406}
\lref\kliu{K.~Liu, {\it Holomorphic equivariant cohomology}, Math. Annalen, {\bf 303} 1 ( 1995) 125-148}
\lref\pietropeter{P.~A.~Grassi, G.~Policastro , M.~Porrati,  P.~Van~Nieuwenhuizen, {\it Covariant quantization of superstrings without pure spinor constraints}, \jhep{0210}{2002}{054},
hep-th/0112162\semi Y. Aisaka and Y. Kazama, {\it A new first class
algebra, homological perturbation and extension of
pure spinor formalism for superstring}, \jhep{0302}{2003}{017}, hep-th/0212316.}
\lref\opennc{N.~Nekrasov, {\it Lectures on open strings, and noncommutative gauge fields}, hep-th/0203109}
\lref\cdv{
A.~Connes, M.~Dubois-Violette, {\it Yang-Mills algebra}, math.QA/0206205, \lmp{61}{2002}{149-158} }
\lref\feiginfrenkel{B.~Feigin, E.~Frenkel, {\it Affine Kac-Moody Algebras and Semi-Infinite Flag Manifolds}, \cmp{128}{1990}{161-189} }
\lref\ff{B.~Fegin, E.~Frenkel, {\it A family of representations of affine Lie algebras}, Rus. Math. Surveys {\bf 43} (1988), 221-222\semi
{\it Representations of affine Kac-Moody algebras and bosonization}, in {\sl Physics and mathematics of strings}, pp. 271-316, World Scientific, 1990\semi
{\it Representations of affine Kac-Moody algebras, bosonoziation and resolutions}, Lett. Math. Phys.{\bf 19} (1990) 307-317}

\lref\edik{E.~Frenkel, {\it Wakimoto modules, opers and the center at the critical level}, Adv. in Math. {\bf 195} (2005)  197-404, math.QA/0210029}
\lref\jan{J.de Boer, L.~Feher, {\it Wakimoto realizations of current algebras: an explicit construction}, \cmp{189}{1997}{759-793}}
\lref\bln{A.~Losev, L.~Baulieu, N.~Nekrasov, {\it Target space symmetris in topological field theories I}, \jhep{0202}{2002}{021}, hep-th/0106042}
\lref\kapustin{ F.~Malikov, V.~Schekhtman, A.~Vaintrob, "Chiral de Rham complex",  $\qquad\qquad$ math.AG/9803041\semi
F.~Malikov, V.~Schekhtman, "Chiral Poincare duality", math.AG/9905008\semi
F.~Malikov, V.~Schekhtman, "Chiral de Rham complex II", math.AG/9901065\semi
V.~Gorbounov, F.Malikov, V.Schekhtman,  "Deformations of chiral algebras and quantum cohomology of toric varieties", math.AG/0001170\semi
E.~Frenkel, M.~Szczesny, {\it Chiral de Rham complex and orbifolds}, math.ag/0307181\semi
A.~Kapustin, {\it Chiral de Rham complex and the half-twisted sigma model}, hep-th/0504074}
\lref\ztwitten{E~Witten, {\it Two dimensional models with (0,2) supersymmetry: perturbative aspects}, hep-th/0504078}
\lref\alekseev{A.~Alekseev, T.~Strobl, "Current Algebras and Differential Geometry", hep-th/0410183}

\lref\malikov{V.~Gorbounov, F.Malikov, V.Schekhtman, "On chiral differential operators over homogeneous spaces", math.AG/0008154\semi
V.~Gorbounov, F.Malikov, V.Schekhtman,  "Gerbes of chiral differential operators, I-III", math.AG/0005201, math.AG/0003170, math.AG/9906117}
\lref\cov{N. Berkovits, {\it
Super-Poincar\'e Covariant Quantization of the
Superstring}, JHEP 04 (2000) 018, hep-th/0001035.}

\lref\griffitsharris{P.~Griffiths, J.~Harris, {\it Principles of algebraic geometry}, 1978, New York, Wiley \& Sons}

\lref\sharpe{E.~Sharpe, {\it Lectures on D-branes and sheaves}, hep-th/0307245} 
\lref\courant{T.J.~Courant, {\it Dirac manifolds}, Trans. AMS {\bf 319} (1990), 631-661}

\lref\hitchin{N.~Hitchin, {\it Generalized Calabi-Yau manifolds}, math.DG/0209099}
\lref\gualtieri{M.~Gualtieri, {\it Generalized complex geometry}, math.DG/0401221}
\lref\strobl{A.~Kotov, P.~Shaller,  T.~Strobl, {\it Dirac sigma models}, hep-th/0411112}

\lref\fms{D.~Friedan, E.~Martinec, S.~Shenker, {\it Covariant quantization of superstrings}, 
\pl{160}{1985}{55} \semi
{\it Conformal invariance, supersymmetry and string theory}, \np{271}{1986}{93}}
\lref\knizhnik{V.~Knizhnik, {\it Analytic fields on Riemann surfaces}, \pl{180}{1986}{247} \semi
{\it Analytic fields on Riemann surfaces II}, \cmp{112}{1987}{567-590}}
\lref\fln{E.~Frenkel, A.~Losev, N.~Nekrasov, to appear}
\lref\aib{E.~Frenkel, A.~Losev, {\it Mirror symmetry in two steps: A-I-B}, hep-th/0505131}
\lref\swnc{N.~Seiberg, E.~Witten, {\it String theory and noncommutative geometry}, JHEP 09 (1999) 032}
\lref\bcov{M.~Bershadsky, S.~Cecotti, H.~Ooguri, C.~Vafa, {\it Kodaira-Spencer theory of gravity and exact results for quantum string amplitudes}, \cmp{165}{1994}{311-428}, hep-th/9309140}
\lref\bpz{A.~Belavin, A.~Polyakov, A.~Zamolodchikov, {\it Infinite conformal symmetry in two dimensional quantum field theory}, \np{241}{1984}{333-380}}
\lref\polyakov{A.~Polyakov, {\it Quantum geometry of bosonic strings}, \pl{103}{1981}{207-210}\semi
A.~Polyakov, {\it Quantum geometry of fermionic strings}, \pl{103}{1981}{211-213}}
\lref\verlinde{M.~Rocek, E.~Verlinde, {\it Duality, Quotients, and Currents}, hep-th/9110053}
\lref\sammar{E.~Martinec, S.~Shatashvili, {\it Black hole physics and Liouville theory}, \np{368}{1992}{338-358}}
\lref\arkadyalbert{A.~Tseytlin, A.~Schwarz, {\it Dilaton shift under duality and torsion of elliptic complex}, hep-th/9210015}
\lref\wittentdg{E.~Witten, {\it Two dimensional gauge theories revisited}, hep-th/9204083}

\def\ihes{\centerline{\vbox{\hbox{\sl \qquad\qquad\qquad Institut des  Hautes  Etudes  Scientifiques} 
\hbox{Le
 Bois-Marie, 35 route de Chartres,   91440  Bures-sur-Yvette,   France}}}}
\def\itep{{Institute~for~Theoretical~and~
Experimental Physics,~Moscow}}

\Title{
\vbox{
\hbox{IHES-P/05/35}\hbox{ITEP-TH-58/05}\hbox{hep-th/0511008}}}
{}\vskip-15pt
{}\centerline{\titlefont LECTURES ON CURVED BETA-GAMMA SYSTEMS,}  
\bigskip\centerline{\titlefont  PURE SPINORS, AND ANOMALIES }
\bigskip
\vskip 5pt
\centerline{Nikita
A.~Nekrasov\footnote{$^{\dagger}$}{On leave of absence from the {\itep}}}
\vskip 15pt\centerline{{\ihes}
}
\vskip 20pt
\noindent{
The curved beta-gamma system is the chiral sector of a certain 
infinite radius limit of the non-linear sigma model with complex target space. Naively it only depends on the complex structures on the worldsheet and the target space. It may  
suffer from the worldsheet and target space diffeomorphism anomalies which we review. We analyze the curved beta-gamma system on 
the space of  pure spinors, aiming to verify the consistency of Berkovits covariant superstring quantization. We demonstrate that under certain conditions
both anomalies can be cancelled for the pure spinor sigma model, in which case one reproduces the old construction of B.~Feigin and E.~Frenkel.} 
\Date{October 2005}

\vfill\eject
\pageno=1

\centerline{\authorfont TABLE OF CONTENTS}\nobreak
{\bf     \medskip{\baselineskip=12pt\parskip=0pt\input betagamma.tmp \bigbreak\bigskip}}

\newsec{Introduction}

String theory, being a theory of quantum gravity, has intimate relations with the symmetries including diffeomorphisms, both in space-time (target space in the sigma model formulation) and the worldsheet. 
In the conventional string backgrounds, which look approximately as sigma models on small curvature spaces, the general covariance is achieved using the metric in the target space, which describes propagating gravity degrees of freedom, and the metric on the worldsheet, which is fully dynamical in the non-critical backgrounds \polyakov. 

String backgrounds often have interesting limits where most of the string fluctuations decouple, yet the extended nature of the string is still visible. 
One such degeneration is the so-called Seiberg-Witten \swnc\ limit, which is used to expose the noncommutative geometry of the open string field theory in the field theory limit. Another degeneration
is the infinite radius limit of the sigma model with Kahler target space (this condition can be somewhat relaxed), where the $B$-field is adjusted so as to keep the holomorphic maps unsuppressed (this is essentially the ${\bar t} \to \infty$ limit of \bcov, see also the recent work \lmt). In all such cases the limiting sigma model is best described using the first order formalism, which is also useful in the analysis of T-dualities \verlinde\sammar.

The degeneration which we shall discuss in this paper occurs for the sigma models with complex targets. The resulting simplified model (more precisely, its chiral part) is the so-called curved beta-gamma system:
\eqn\cbg{S = {1\over{2{\pi}}} \int_{\Sigma} \ {\b}_{i} {\pb} {\g}^i}
where ${\g}^i$ are the dimension zero fields describing the map of the two dimensional Riemann surface $\Sigma$ to the target space $X$, and ${\b}_i$ are the sections of the pull-back of the holomorphic cotangent bundle ${\CT}_{X}^{*}$ to $X$, tensored with $(1,0)$-forms on $\Sigma$. 
Naively the theory defined by the Lagrangian \cbg\ is the free field theory, which is also conformally invariant. 

However, this statement requires some elaboration, since we should not forget about the global properties of the system \cbg. The coordinate transformations relating different coordinate systems on $X$ may act non-trivially on the operator content of the theory \cbg\ and moreover there are
actually obstructions for gluing the free field theories \cbg\ over all of $X$. Also, the definition
of the path integral measure in the theory \cbg\ is subtle, and it turns out that the conformal invariance may be broken unless the target space $X$ has certain topological properties. 

In this paper we shall quickly remind the physicists these obstructions, which show up as anomalies in the worldsheet and target space diffeomorphism invariance of the theory \cbg. We should stress that mathematicians have worked out these anomalies in \malikov\  (the target space coordinate invariance in the curved beta-gamma systems 
was also studied in \bln\ at the same time, yet the context was slightly different). We had some difficulties in translating these papers so we rederive the results which we need from scratch. Note also that the recent paper \ztwitten\ contains many explanations on the physics of  \malikov\ and also relates it to the phyisics of $(0,2)$-models. 

Our main goal is to verify that the particularly interesting system \cbg\ where
$X$ is the space of pure spinors in ten dimensions, is free from these potential anomalies. It turns out that the result depends on the subtle issue on how one resolves the singularities of the naive space of pure spinors. 

In Berkovits approach \cov,\berkovitstrieste\ to the covariant quantization of the superstring the first-order system
\cbg\ is coupled to the first order fermionic system and the second-order bosonic system describing the physical space-time. In some applications it is convenient to think of the model as the \cbg\ system with
superspace as the target space. In particular, the main ingredient of Berkovits approach is the nilpotent
$Q$-operator, which, in part, can be attributed to the existence of holomorphic symmetries of that superspace. We plan to elaborate on these features in a future work, aiming to clarify the definition of
string amplitudes in Berkovits formalism\berkovitsbghost,\berkovitsmulti. Note that the particular super-target space, namely
${\Pi}TX$, which is free from all anomalies, leads to the well-studied type A topological strings. Their chiral version corresponds to the so-called chiral de Rham complex, introduced and studied in \kapustin\ and recently applied to mirror symmetry in \aib. An interesting superspace $X \times {\Pi}{\CT}{\bar X}$
for particular $X$ was proposed recently by Berkovits \berkovitsnonmin.

The paper is organized as follows. The section $\bf 2$ discusses general systems  \cbg\. We show that
the first Pontryagin class $p_1 (X)$ of the target $X$ is the obstruction for the global definition of the
fields ${\b}_{i}$ and ${\g}^i$ of the model. We also show that the first Chern class $c_1(X)$ of $X$ is
the obstruction to the global definition of the holomorphic stress-energy tensor $T_{zz}$ in the model \cbg. We also analyze in some detail the case where $X$ is the total space of a bundle over some base $B$, with one-dimensional fibers. We show that if the fiber is ${\bC}^{*}$ then the Pontryagin and Chern anomalies can be cancelled if the first Pontryagin class $p_1 (B)$ of the base factorizes as the product of
two Chern classes of some line bundles. This is somewhat analogous to the mechanism of the anomaly cancellation for the $SO(32)$ and $E_{8} \times E_{8}$ heterotic strings \gsw\witohet. 
The section $\bf 3$ is devoted to pure spinors. We remind their definition and the r\^ole they play in Berkovits
formalism. We also discuss the geometry and topology of the space ${\tilde Q}$ of projective pure spinors and that of various cones over it. We compute the first Chern and Pontyagin 
classes of $\tilde Q$ and find that they are non-vanishing\footnote{$\dagger$}{That $p_{1}({\tilde Q}) 
\neq 0$ was stressed by E.~Frenkel and  E.~Witten}, and show that $p_{1}$ is actually proportional to $c_{1}^2$. It follows that by taking the appropriate
${\bC}^{*}$-bundle over ${\tilde Q}$ the anomalies are cancelled. Even though the result sounds almost trivial, it took us some time to get through various obstacles, so we hope the reader will find
some details of our calculations useful. The section $\bf 4$ contains discussion and conclusions.

\newsec{General remarks}
{}\footnote{$^{!}$}{The results of this chapter were developed with the help of A.~Losev and E.~Frenkel, and will be elaborated upon in \fln.}

\subsec{The $\beta$-$\gamma$ system as an infinite radius limit}

Two dimensional sigma models describe maps of a Riemann surface $\Sigma$ into some target
space $X$. The usual approach to the sigma model works under the assumption that 
$X$ is Riemannian manifold as well, and the action of the sigma model is given by the
Dirichlet functional:
\eqn\sigm{S_{0} = \int_{\Sigma} \sqrt{h} h^{ab} G_{\m\n}(X) {\p}_{a} X^{\m} {\p}_{b} X^{\n}}
In addition, one can also couple the sigma model to the antisymmetric tensor on $X$, 
the so-called $B$-field, as well as to other geometric data, such as a functions $T(X)$ (the tachyon),
 the dilaton (Weyl compensator) ${\Phi}(X)$ etc. 
 \eqn\sigmi{S = S_{0} + \int X^{*} B + \int \sqrt{h} T(X) + \int \sqrt{h} R^{(2)}_{h} {\Phi}(X) + \ldots}
 The Lagrangian \sigm\ is conformally invariant. Quantum theory defined with the help 
 of \sigm\ ceases to be conformally invariant, unless the metric $G_{\m\n}$ obeys certain
 equations. In the limit of the large volume, the condition on the target space metric
 to lead to the conformally invariant two dimensional sigma model with the Lagrangian \sigm\ is essentially the Ricci-flatness. 
 In the presence of the other couplings \sigmi\ these conditions get modifications. 
 In what follows we shall be mainly interested in the $B$-field and $\Phi$ (dilaton) couplings in
 \sigmi\ in addition to the basic Lagrangian \sigm. 
  
 Now let us assume that  $X$ is a complex manifold, and that the metric is hermitian.
 In local complex coordinates $x^{i}, {\xb}^{\ib}$ on $X$ it has only the components
 $G_{i\ib}$. We can now rewrite the theory \sigm\ using the first order formalism:
 \eqn\sigmii{\eqalign{& 
 S_{0} \to \int p_{i}{\pb}x^{i} + {\bar p}_{\ib} {\p} {\xb}^{\ib} + G^{i\ib} p_{i} \wedge {\bar p}_{\ib} \cr 
& B \to {\tilde B} = B + G_{i\ib} {\dd} x^i \wedge {\dd} {\xb}^{\ib} \cr
& {\Phi} \to {\tilde\Phi} = {\Phi} + {1\over{8\pi}} {\log} \left( {\det} G_{i\jb} \right) \cr }}
The last line, the dilaton shift, can be understood using the technique of \arkadyalbert.
We can develop a large volume expansion by expanding in $G^{i\ib}$ the correlation functions of the theory
defined by \sigmii, while keeping other couplings, like $\tilde B$ fixed (one can in particular get another view on the Ricci-flatness equations following from the conformal invariance, by doing the conformal perturbation theory around the system \cbg, cf. \lmt). Ignoring for the moment these other
couplings the limit is the sigma model which looks like a holomorphic square of the
{\it curved, or non-linear, beta-gamma system}, i.e. the theory with the action
\eqn\sigmiii{S_{\b\g} = {1\over 2\pi} \int {\b}_i {\pb} {\g}^{i}}
where the fields of the beta-gamma system are related to the fields of the sigma model \sigmi\ via:
\eqn\sigmiv{{\b}_i = p_i \ , \qquad\qquad \ {\g}^i = x^i \ . } 
The identification  \sigmiv\ depends on the choice of the coordinate system on the target space $X$.

When we work locally on $X$, the system \sigmiii\ is a simple free field theory, 
the basic operator product being:
\eqn\opei{{\g}^i (z) {\b}_{j} (w) \sim {\d}^{i}_{j} {{\dd}w \over z - w}}
Out of \opei\ one can construct various local operators, by taking the differential polynomials
in ${\b}_{j}$ and ${\g}^{i}$ and normal ordering. The normal ordering depends on the choice of the
local coordinate $z$ on the worldsheet. In what follows we denote by ${\p} = dz {\p}_{z}$ the holomorphic worldsheet
exterior derivative, by ${\pb}$ the antiholomorphic one, and by $\dd$ the exterior target space holomorphic differential.

\subsec{Useful operator product expansions}

As a preparation, let us discuss the following dimension one operators:
\eqn\jv{\eqalign{& J_{V} =  {\b}_i V^{i}({\g}) (z) \equiv {\lime} \ {\b}_i ( z + {\e} ) V^{i} ( {\g}(z)) + {1\over \e}
{\p}_i V^{i} ( {\g}(z)) \cr
& \qquad \qquad C_{B} = B_{i}({\g}(z)) {\p} {\g}^{i} \cr}}
where $B \in {\Omega}^{1}_{U}$, $V \in {\CT}_{U}$.  
Note that the definition of the current $J_V$ depends on the choice of
local coordinate $z$, more precisely on the differential ${\dd}z$. Two choices of ${\dd}z$ would lead
to two operators $J_{V}$ and $J_{V}^{\prime}$ which differ by an operator of the form $C_{B}$
for some $B$. 
The definition of $C_{B}$ does not depend on the choice of $z$. 

Note also, that the definition of the current $J_V$ is not covariant with respect to the
target space coordinate changes. We shall make this statement more precise
later, right now only note that the subtraction term ${1\over {\e}} {\p}_{i}V^{i}$ 
is not invariant under the coordinate changes, if $V^i$ transform as the components
of the vector field on $X$. The closest  geometric object to the divergence
${\p}_i V^i$ is the divergence defined using some holomorphic top form:
\eqn\holdiv{{1\over \Omega} {\CL}_{V} {\Omega} \sim {\p}_{i} V^{i} + V^{i} {\p}_{i} {\log}{\omega}({\g})}
where
\eqn\omgfrm{ {\Omega}  = {\omega} ({\g}){\dd}{\g}^{1} \wedge \ldots \wedge {\dd}{\g}^{d}}
is the (meromorphic) holomorphic top form on $X$. We shall see later that \holdiv\ indeed shows up in the proper definitions of the currents corresponding to the target space vector fields  $V$.

It is straightforward to calculate:
 \eqn\opeii{\eqalign{J_{V_{\ba}}(z+{\e}) J_{V_{\bb}}(z) & \sim - {\half} {{\Sigma}_{\ba\bb}(z + {\e}) + {\Sigma}_{\ba\bb}(z) \over {\e}^2 } - {J_{[V_{\ba}, V_{\bb}]}(z) \over {\e}}  - {C_{{\Omega}_{\ba\bb}}(z) \over {\e}} \cr
J_{V}(z+ {\e}) C_{B}(z) & \sim - {\iota_{V}B (z) \over {\e}^2} - {C_{{\CL}_{V} B} (z) \over {\e}} \cr }}
where
\eqn\anom{\eqalign{ {\Sigma}_{\ba\bb} = {\tr} {\CV}_{\ba} {\CV}_{\bb}  , & 
\quad  {\Omega}_{\ba\bb} = {\half} {\tr} \left(  {\CV}_{\ba} {\dd} {\CV}_{\bb} - {\CV}_{\bb} {\dd} {\CV}_{\ba} \right)\cr
& {\CV}_{\ba} =  \Vert  {\p}_i V^j_{\ba} \Vert \cr}}

\subsec{Courant bracket}

Let ${\upsilon} = V + {\xi}$ be a section of the ${\CT}_{X}\oplus{\Omega}^{1}_{X}$ bundle. This is the object of
study of the generalized complex structure \hitchin\gualtieri\ (although in our context everything is holomorphic, so we are talking here about the generalized hypercomplex structure), and the generalized Dirac structures \strobl. We can canonically associate 
a dimension one operator to $\upsilon$:
\eqn\dimo{{\CO}_{\upsilon} = J_{V} + C_{\xi}}
From \opeii\ we calculate for ${\varpi}  = W + {\eta}$:
\eqn\opeiii{{\CO}_{{\upsilon} }(z+{\e}) {\CO}_{{\varpi}}(z) \sim {g({\upsilon} , {\varpi}) (z+{\e}) + g({\upsilon} ,{\varpi})(z) \over {\e}^2} -
{{\CO}_{[[{\upsilon} ,{\varpi}]]} + C_{{\Omega}_{VW}} \over {\e}} }
where
\eqn\courb{
[[{\upsilon} ,{\varpi}]] = [V,W] +  {\CL}_{V}{\eta} - {\CL}_{W}{\xi} -{\half} {\dd} \left( {\iota}_{V} {\eta} - {\iota}_{W}{\xi} \right)}
is the so-called Courant \courant\ bracket, 
\eqn\metr{2 g({\upsilon} ,{\varpi}) = {\Sigma}_{VW} + \iota_{V}{\eta} + \iota_{W}{\xi}}
is the (quantum corrected) metric on ${\CT}_{X}\oplus{\Omega}^{1}_{X}$. Because of the ${\Sigma}_{VW}$ correction,
and also because of the term with ${\Omega}_{VW}$ in \opeiii\ the operator product expansion of
the operators ${\CO}_{\upsilon},{\CO}_{{\varpi}}$ does not have an obvious geometric interpretation, since
these corrections do not transform covariantly under the coordinate changes. However, this is in 
accordance with the non-trivial nature of the $\b$-fields, which do not transform naively. Let us note that if instead of the operator product expansion we were only interested in the Poisson
brackets of ${\CO}_{\upsilon}$ and ${\CO}_{{\varpi}}$ viewed as the functionals on the loop space $LX$
then we would get the Courant bracket and the canonical metric on ${\CT}_{X} \oplus {\Omega}^{1}_{X}$. This is a
holomorphic analogue of the observation in \alekseev. 

We now
turn to the exact determination of the transformation properties of the $\b$-fields.

\subsec{The target space coordinate transformations}

We wish to understand the transformation properties of the local operators under the 
coordinate transformations ${\g} \mapsto {\tg}$. Later on we shall use this information in trying to define the $\b - \g$ system globally on the manifold $X$, which has several coordinate charts.  

\subsubsec{Classical theory}

Let $( {\g}^{i}) $ and $( {\tg}^{a} )$, $i,a = 1, \ldots, d$  be the two sets of  local coordinates on some domain $U 
\subset X$. These coordinates are related by the local holomorphic diffeomorphism : $f: {\gamma} \mapsto {\tilde\gamma}$. 
How should we transform the fields ${\b}_i$'s? Classically, ${\b}_i$ transforms as
the $(1,0)$-form on $X$, i.e. ${\b} \mapsto {\tilde\b} = f^{*}{\b}$: 
\eqn\bclas{{\tilde\b}_a = {\b}_i g^i_a ({\g})}
where 
\eqn\jac{g^i_a ({\g}) = \left[ \left( {\p\tilde\g}/{\p\g} \right)^{-1}\right]^{i}_{a} = {\p}{\g}^{i} \slash {\p}{\tg}^{a} ({\g})}
The inverse  Jacobian matrix $g$ in \jac\ will play several important roles in what follows. 
For some 
purposes it is convenient to view $g$ as a collection of vector fields $g_{a}$ on $U$:
\eqn\vecf{g_{a} = g_{a}^{i}{{\p}\over {\p}{\g}^{i}} = {{\p}\over {\p\tg}^{a}} = {\tilde\p}_{a}}
It is also convenient to introduce the one-forms
\eqn\vomg{\eqalign{ {\tilde g}^{a} = {\dd} {\tilde\g}^{a} =& \  {{\p\tilde\g}^{a} \over {\p}{\g}^{i}} {\dd}{\g}^{i} = ( g^{-1} )^{a}_{i} {\dd} {\g}^{i} \equiv {\tilde g}^{a}_{i} {\dd} {\g}^{i} \cr & \cr
& \iota_{g_{a}} {\tilde g}^{b} = {\d}^{b}_{a} \cr}}
Note that
\eqn\cmmt{[g_a , g_b ] = 0, \qquad {\dd}{\tilde g}^a = 0}

\subsubsec{Quantum corrections}

The formula \bclas\ does not quite make sense quantum mechanically, because of the
short distance singularity between ${\b}_i$ and $g^i_{a}$ in \bclas. As in \jv\ we point-split and subtract
the divergent term: 
$$
\sim {\p}_i g^i_a {1\over\e}
$$
to get a well-defined operator. However, this operator may not be the correct one.
It is clear that we should look for the dimension one operators of the form:
\eqn\btt{{\tilde\b}_{a} = {\b}_i g^{i}_{a} + B_{ai} {\p}{\g}^{i}}
where 
the normal ordering is understood. Indeed, it is the form \btt\ which guarantees the correct operator product expansion of ${\tilde\b}$ and $\tilde\g$.

Note that the matrices ${\CG}_a = \Vert {\p}_{i} g^{j}_{a} \Vert$, similar to the matrices ${\CV}_{a}$ introduced in \anom, obey the following Maurer-Cartan
equations:
\eqn\mk{{\CL}_{g_{[a}}{\CG}_{b]} = [{\CG}_{b}, {\CG}_{a}]}
which can also be presented as:
\eqn\mki{{\CG}_{a} = g^{-1} {\CL}_{g_{a}} g} 
We need to find the one-forms $B_{a}$ in \btt. They cannot be set to zero, in general, since
we want to have no singularities in the operator product expansion of
${\tilde\b}_{a}$ with ${\tilde\b}_{b}$. By calculating the operator product
of the fields \btt\ and setting the singular parts to zero we get $d^2$ conditions on $d^2$ unknowns. 
Indeed, the absence of the double pole is an equation, symmetric in $a,b$ , while the absence
of the first order pole is the antisymmetric one. However, as we shall now see, there is a freedom
in choosing $B_a$'s. This freedom is related to the fact that the $\b\g$-system has moduli. Globally, there are also obstructions for choosing $B_a$. This is related to the target space diffeomorphisms anomaly.  

Using \jv\ we calculate the operator product expansions:
\eqn\opei{{\tilde\b}_{a} (w + {\e} ) {\tilde\b}_{b}(w) \sim  -{ {\CS}_{ab}(w) \over {\e}^2} - {{\CA}_{ab} \over {\e}} }
where the symmetric tensor ${\CS}$ and the antisymmetric one-form valued tensor ${\CA}$ are given by:
\eqn\opebt{\eqalign{
{\CS}_{ab}&  = \quad {\Sigma}_{ab} + \iota_{g_{a}} B_{b} + \iota_{g_{b}} B_{a} \cr
& \equiv \qquad {\p}_{i}g_{a}^{j} {\p}_{j}g^{i}_{b} + g_{a}^{i} B_{bi} + g_{b}^{i}B_{ai} \cr
{\CA}_{ab} & = \quad - {\Omega}_{ab} + \iota_{g_{a}} {\dd} B_{b} - \iota_{g_{b}} {\dd}B_{a} + {\half} {\dd} {\m}_{ab} \cr & \qquad 
{\m}_{ab}  = \iota_{g_{a}} B_{b} - \iota_{g_{b}} B_{a} \cr}}
where 
\eqn\omgi{{\Omega}_{ab} = {\half} {\tr} \left(  {\CG}_{a} {\dd} {\CG}_{b} - {\CG}_{b} {\dd} {\CG}_{a} \right) }
We can write:
\eqn\bba{B_{a} = {\half} \left( {\s}_{ab} - {\m}_{ab} \right) {\dd} {\tg}^{b}}
where
\eqn\sab{{\s}_{ab} = {\s}_{ba} = \iota_{g_{a}} B_{b} + \iota_{g_{b}} B_{a} }
Setting ${\CS}_{ab} = 0$ we get 
\eqn\sym{{\s}_{ab} = - {\Sigma}_{ab} = - {\p}_{i} g^{j}_{a} {\p}_{j} g^{i}_{b}  . }
It remains to determine ${\m}_{ab} = - {\m}_{ba}$. 
Since ${\m}_{ab}$ is antisymmetric in $a$, $b$ we can define the following two-form:
\eqn\muf{{\m} = {\m}_{ab} \ {\dd} {\tg}^{a} \wedge {\tg}^b}
which contains all the information about the tensor ${\m}_{ab}$. 
Let us contract ${\CA}_{ab}$ with the vector $g_{c}$ and set the result to zero. We get the
equation:
\eqn\aabc{{\CL}_{g_{c}}{\m}_{ab} +  2 \iota_{g_{c}}
\iota_{g_{a}} {\dd}B_{b} - 2 \iota_{g_{c}}\iota_{g_{b}} {\dd} B_{a} = {\tr} \left( {\CG}_{a}{\CL}_{g_{c}}{\CG}_{b} - {\CG}_{b} {\CL}_{g_{c}} {\CG}_{a} \right)}
which using \mk\ can be transformed to:
\eqn\aabci{\eqalign{ {\CL}_{g_{c}}{\m}_{ba} + {\CL}_{g_{a}}{\m}_{cb} + {\CL}_{g_{b}}{\m}_{ac} = & \
 {\tr} \left( {\CG}_{a}{\CL}_{g_{c}}{\CG}_{b} - {\CG}_{b} {\CL}_{g_{c}} {\CG}_{a} \right) - {\CL}_{g_{a}}{\s}_{bc} + {\CL}_{g_{b}}{\s}_{ac} \cr
 & = {\tr}\left( {\CG}_{a} {\CL}_{g_{[c}}{\CG}_{b]} + {\CG}_{b} {\CL}_{g_{[a}}{\CG}_{c]} + {\CG}_{c} {\CL}_{g_{[a}} {\CG}_{b]} \right) \cr
 & = {\tr} {\CG}_{a} [ {\CG}_{b}, {\CG}_{c}] \  . \cr}}
From the equation \aabci\ we see that the matrix ${\m}_{ab}$ is not determined 
uniquely. Indeed, we can shift:
\eqn\shfts{{\m}_{ab} \mapsto {\m}_{ab} + {\CL}_{g_{a}} f_{b} - {\CL}_{g_{b}} f_{a}}
which is equivalent to 
\eqn\shftsb{B_{a} \mapsto B_{a} - {\half} \iota_{g_{a}} {\dd} f , \qquad f = f_{a} {\dd}{\tg}^{a}}
This undeterminancy is related to the naive symmetry of the action \sigmiii:
\eqn\symsgm{{\b}_{i} \to {\b}_{i} + \left( {\p}_i f_j - {\p}_j f_i \right) {\p} {\g}^j }
Note that the equation \aabci\ simplifies in the ${\tilde\g}$-coordinates:
\eqn\wzm{{\dd} {\m} = - {\tr} \left( {\dd} {\tilde g} {\tilde g}^{-1} \right)^3 = {\tr} \left( g^{-1} {\dd} g \right)^3 }
The freedom \symsgm\ in $\tilde\g$-coordinates reads as: ${\m} \mapsto {\m} + {\dd} f$,
where $f$ is viewed as $(1,0)$-form on $U$. In some special cases the formula \wzm\ was already noted in \jan.

\subsubsec{Coordinate transformations and Wess-Zumino term}  
To summarize, the coordinate transformation ${\g} \mapsto {\tilde\g}$ is accompanied by the
transformation:
\eqn\bttr{\mathboxit{\eqalign{& {\tilde \b}_{a} =  {\b}_{i} g^{i}_{a} ( {\g})   - {\half} \left( {\p}_{j} g^{i}_{a} {\p}_{i} g^{j}_{b} \right) {\p} {\tilde\g}^{b} + {\half} {\m}_{ab} {\p}{\tilde\g}^{b} \cr
& \qquad\qquad g^{i}_{a} ({\g} ) = {{\p}{\g}^{i} \over {\p}{\tg}^{a}} ({\g}) \cr
& \qquad{\dd} \m = {\tr} \left( g^{-1} {\dd} g \right)^{3} ,  \quad {\m} =  {\m}_{ab} {\dd} {\tilde\g}^{a} \wedge {\dd}{\tilde\g}^{b}\cr}}}  
The formulae like \bttr\ can be found in \malikov. Let us note another useful formula:
\eqn\bttrt{\eqalign{{\tilde\b}_{a} = 
& \quad {\b}_{i} g^i_{a} + {\half} {\tr} \left( {\CG}_{a} g {\p} g^{-1} \right) + \iota_{{\tilde\p}_{a}} {\m} \cr
& \quad\qquad {\CG}_{a} = \Vert {\p}_{j} g^{i}_{a} \Vert \cr}}
We can phrase \bttr\bttrt\ in a more invariant way:
\eqn\bttri{{1\over{2{\pi}}} {\tbt} {\pb} {\tg} = {1\over 2{\pi}} {\b} {\pb} {\g} + L_{\rm wzw} \left( g \right)}
where $L_{\rm wzw} (g)$ is the usual (level one) Wess-Zumino-Novikov-Witten Lagrangian:
\eqn\wzwa{L_{\rm wzw} (g) = {1\over 4{\pi}} {\tr} ( g^{-1} {\p} g g^{-1} {\pb} g ) + {1\over 12{\pi}} {\dd}^{-1} 
 {\tr} \left( g^{-1} {\dd} g \right)^{3}}
The deeper meaning of \bttri\  and its generalizations involving the dependence
on the complex structure on the worldsheet will be discussed in \fln.

\subsubsec{Target space symmetry currents}

Suppose $V = V^{i} ({\g}) {\p}_{i}$ is a holomorphic vector field on $X$. In classical geometry it
generates an infinitesimal symmetry of the manifold $X$, the symmetry of its complex structure. Let us see whether this symmetry is preserved in quantum theory. In order for this to be the case we should be able to construct a holomorphic current, which would generate the quantum counterpart of the classical symmetry. Naively, this current should be given by:
\eqn\clcur{J_{V} \naiveq {\b}_{i} V^{i} ({\g})}
where we again use the normal ordering implicitly. However most likely the definition \clcur\ will not be
compatible with the coordinate transformations on $X$. So, we should allow for the correction term 
$C_{B}$, for some $B$:
\eqn\qucur{{\CJ}_{V} = {\b}_i V^i ({\g}) + B_{i}( {\g} ) {\p}{\g}^i}
where $B_{i} {\dd}{\g}^i$ is a locally defined one-form on $U \subset X$, which is clearly a linear functional of $V$. Its behaviour $B \mapsto {\tilde B}$ 
under the coordinate transformations ${\g} \mapsto {\tg}$ can be recovered from \bttr:
\eqn\bbtrr{\eqalign{ {\tilde B}_{a} ({\tg} ) {\dd}{\tg}^{a} - B_{i} ({\g}) {\dd} {\g}^{i} = \ &  \half \left( \iota_{V} {\m} - 2
\left( {\dd} g g^{-1}  \right)^{i}_{j} {\p}_{i}V^{j} +  V^{i} {\tr} \left( g^{-1} {\p}_{i} g g^{-1} {\dd} g  \right) \right)  \cr =  \ 
& {\half} \left( \iota_{V} {\m} - {\tr} \left( {\CV}  {\dd} g g^{-1} \right) +{\tr} \left( {\tilde \CV} {\dd} {\tilde g} {\tilde g}^{-1} \right)  \right) \  , \cr}}
where we have introduced matrices, already familiar from \anom\ :
\eqn\vmatr{
 {\CV} =  \Vert {\p}_{i} V^{j} \Vert,  \qquad {\tilde\CV} = \Vert {\tilde\p}_{a} {\tilde V}^{b} \Vert  \ , }
 where ${\tilde V}^{a} = {\tilde g}^{a}_{i} V^{i}$, ${\tilde g} = g^{-1}$. The matrices ${\CV}, {\tilde
 \CV}$   behave as connections in the $V$ direction:
 \eqn\vcon{{\tilde\CV} = g^{-1} {\CV} g - g^{-1} {\CL}_{V} g \ .}

\subsubsec{Stress-energy tensor}

Now let us discuss the transformation properties of the stress-energy tensor.
In the local coordinate patch $U$ where our theory is represented by the free fields
${\b}_i$ and ${\g}^i$ we have the standard definition of the stress-energy tensor,
following from the naive Lagrangian \sigmiii: 
\eqn\sti{T \naiveq {\b}_i {\p} {\g}^i \equiv  {\lime} \left( {\b}_{i} ( z + {\e} ) {\p}{\g}^i  (z)
  + {d \over {\e}^2} \right)}
  where $d = {\rm dim}_{\bC} X$. 
Now let us see what happens when we perform the target space coordinate transformation
${\g} \mapsto {\tilde\g}$ (cf. \jan\malikov).
\eqn\stiii{\eqalign{{\widetilde T} = &\  {\tilde\b}_{a} {\p}{\tilde\g}^a \equiv
{\lime} \left(   {\tilde\b}_{a} (z+{\e}) {\p}{\tilde\g}^a (z) + {d\over {\e}^2} \right) \cr 
  & = T - {\half} {\tilde g}^{a}_{i}  {\p}^2 g^{i}_{a} - ({\p} g^{i}_{a} ) {\p} {\tilde g}^{a}_{i} +
  B_{ai}{\tilde g}_{j}^{a} {\p}{\g}^{i}{\p}{\g}^{j} \cr
  & = T - {\half} {\p}^{2} {\rm log} \ {\rm det} \Vert g_{a}^{i} \Vert\cr}}
So we see that in order for the stress-energy tensors to be coordinate independent, the coordinate
transformation $ {\g} \mapsto {\tg}$  should better preserve (perhaps up to a constant multuple) a
holomorphic volume form. Indeed, the determinant ${\rm det}   \Vert g_{a}^{i} \Vert$ is the ratio
of the holomorphic volume forms on the coordinate patches ${\g}$ and ${\tilde\g}$. 

The anomalous term in the transformation law for the naive
stress-energy tensor means that the theory actually depends on the choice of the target space coordinates, unless some coupling to the worldsheet
metric curvature
\eqn\tdcrvt{ {1\over{8\pi}} \int R^{(2)} {\log}\ {\omega}({\g})}  
is added. The modification \tdcrvt\ of the action \sigmiii\ modifies the stress-energy tensor to:
\eqn\strssmdf{T = {\b}_{i} {\p} {\g}^{i} - {\half} {\p}^{2} {\log}\ {\omega}({\g})}
where ${\omega}({\g})$ comes from the holomorphic top degree form on $X$. In order for \strssmdf\ to be regular, the argument of the logarithm should not vanish, so the holomorphic top degree form 
$\Omega$ must be non-vanishing and regular on $U$. 

\subsec{The coordinate transformations on the worldsheet}

A priori, the definitions of the stress-energy tensor, and the currents $J_{V}$ depend
on the choice of the local coordinates $z$ on the worldsheet. 
For example, $T$, defined by \sti\ transforms as
  a projective connection under the holomorphic reparameterizations of the $z$ coordinate:
\eqn\zcoor{z \mapsto {\hat z}}
\eqn\stii{T \mapsto {\widehat T} = {1\over \left( {\p}_{z} {\hat z} \right)^{2} }   \left( T  - {d\over 6}
\left\{ {\hat z} ; z \right\} \right)}
where
\eqn\sch{\left\{ {\hat z} ; z \right\}  =  {{\p}_{z}^{3} {\hat z} \over {\p}_{z} {\hat z} } -
 {3\over 2} \left( {{\p}^2_{z} {\hat z} \over {\p}_{z} {\hat z} } \right)^{2}   } 
The formula \stii\ holds for any \bpz\ two dimensional conformal field theory with central charge $c = 2d$.
Technically, the shift in \stii\ comes from the expansion of
\eqn\anomshft{{d\over{\e}^2} - {d ({\p}_{z}{\hat z})^2 \over{({\hat z}(z+{\e})  - {\hat z}(z))^2}}}
However, the  expression \stii\ with the Schwarzian derivative \sch\ does not take seem into account the shift \strssmdf. It would lead to the strange-looking formula
$$
{\hat T} =^{\kern-4pt{\rm ?}} \qquad  {\hat T}^{\rm naive} + {\half} {{\p}^2_{zz} {\hat z} \over ({\p}_{z}{\hat z})^3} 
{\p}_{z} {\log} \ {\o}({\g})
$$
which is inconsistent with \stii. 
The truth is that the modification \strssmdf\ of the stress-energy tensor implies that ${\b}_{i}$'s are
no longer primary fields. They transform under the worldsheet coordinate changes as:
\eqn\bttrws{\eqalign{{\b}_i \mapsto & 
 \ {\hat \b}_i = {\b}_i - {\half} {\p}\left( {\log}\ {\p}_z {\hat z} \right) {\p}_i {\log} \ {\o}({\g}) \cr
 & {\hat\b}_{i {\hat z}} = {1\over {\p}_{z} {\hat z}} {\b}_{i z} - {\half}  {{\p}^2_{zz} {\hat z} \over ({\p}_{z}{\hat z})^2} 
{\p}_{i} {\log} \ {\o}({\g}) \cr}} 
thus making the formula \stii\ true indeed. 

Similarly, the currents $J = J_{V}$ are transforming with the cocycle:
\eqn\currnts{\eqalign{J  \mapsto &  {\hat J} = 
 J - {\half} {\Omega}^{-1} {\CL}_{V} {\Omega} \  {\p} \left(  {\log} {\p}_{z}{\hat z} \right) \cr
 & 
 {\hat J}_{\hat z} = {1\over{{\p}_{z}{\hat z}}} \left( J_{z} - {\half}  {1\over {\o} ({\g})} {\p}_{i} \left( {\o}({\g}) V^{i}({\g}) \right) {\p}_{z} {\log} \ {\p}_{z}{\hat z} 
\right) \cr}}
where if we didn't take into account the anomalous transformation \bttrws\ we would have gotten
${\p}_i V^i$ instead of the covariant divergence 
\eqn\covdiv{
{\Omega}^{-1} {\CL}_{V} {\Omega}= {1\over {\o} ({\g})} {\p}_{i} \left( {\o}({\g}) V^{i}({\g}) \right)
}
Note that the volume ($=\Omega$)-preserving vector fields $V$ correspond to the currents $J_{V}$ which are the primary fields. This is of course
in agreement with the ${\half} c_{1}(X) c_{1}({\Sigma})$ nature of the anomaly \ztwitten\ we are
discussing.

\subsec{Global theory}

We now may pose the problem of formulating the $\b$-$\g$ system on $X$ globally.
We know already that we need to use a holomorphic top form $\Omega \in H^{d,0}(X, {\bC}) = H^{0}(X, K_{X})$. 

In order to do this we may wish to have the following conditions satisfied:
\item{$1.$} 
For each coordinate patch $U_{\a}$ with the coordinates ${\g}_{\aa} = ( {\g}^{i} )$ we have a copy of the standard system \sigmiii\ 
with particular curvature coupling:
\eqn\loclag{L_{\aa} = {1\over 2\pi} \int {\b}_i {\pb} {\g}^{i} + {1\over 4} R^{(2)} {\log} \ {\o} ( {\g} )}
where
\eqn\locomg{{\Omega}\biggr\vert_{U_{\a}} = {\o}({\g}) {\dd}{\g}^1 \wedge \ldots \wedge {\dd} {\g}^{d}}

\item{$2.$}
On the overlaps $U_{\a\b} = U_{\a} \cap U_{\b}$, where the coordinates ${\g}_{\aa} = ( {\g}^{i})$
and ${\g}_{\bbb} = ( {\tg}^{a} )$ are related by the local biholomorphism $f_{\a\b} : {\g}_{\aa} \to {\g}_{\bbb}$,
the   fields of two systems \loclag\ corresponding to $U_{\a}$ and $U_{\b}$ are related by the field
redefinition \bttr

\item{$3.$}
The stress-energy tensors of two systems $L_{\a}$ and $L_{\b}$ transform one into another under
\bttr

\item{$4.$}
The glueings over $U_{\a\b}$'s are correctly defined, in the sense that on every triple overlap
$U_{\a\b\g} = U_{\a} \cap U_{\b} \cap U_{\g}$ the composition of three transformations
$L_{\aa} \to L_{\bbb} \to L_{\gg} \to L_{\aa}$ is the identity.

We shall see in the next subsections that one may encounter anomalies, which obstruct the
existence of the solution, obeying the conditions $2$, $3$ and $4$. If the anomalies are absent, then there may be
several solutions to the condition $2$, they have to do with the moduli of the $\b$-$\g$ sigma model.
Finally, if the manifold $X$ has symmetries, we may also want to have the following property:

\item{$5.$} 
The complex Lie group $G$, acting on $X$, generated by the vector fields $V_{\uA}$, ${\uA} = 1. \ldots , {\rm dim}G$, to be represented by the currents $J_{\uA} \equiv {\CJ}_{V_{\uA}}$, which form the affine Lie algebra ${\hat\bg}$ at
some level $k = k_{X}$, which depends on $X$ (if the group $G$ is not simple, there may be several levels). 

There exist certain obstacles in getting this wish granted as well. We shall not explore this issue in full generality, some aspects of this problem were already discussed in \feiginfrenkel,\edik,\malikov, \jan. In particular, in the case of ${\bC}^{*}$-bundles over the homogeneous spaces $G/H$ the refs. \feiginfrenkel, \edik\ contain the full solution of this problem, see also \ff. We make several comments in the case where $G$ acts on $X$ freely, and discuss some other examples. 

\subsubsec{{\Cech} notions}

A word on notations. As before, ${\dd}$ denotes the holomorphic exterior derivative on $X$. It sends
$(p,q)$ forms to $(p+1, q)$ forms. In this section we shall be also dealing with {\cech} cochains, cocycles and coboundaries (see chapter $0$ of \griffitsharris\ for systematic introduction, \sharpe\ for the introduction for physicists, and \ztwitten\ for the introduction for physicists in the context, maximally close to ours). {\cech} $q$-cochain ${\ba}$ valued in some sheaf ${\CF}$ of abelian groups on $X$ is the assignment of a section ${\ba}_{{\a}_0 {\a}_1 \ldots {\a}_{q}}$ of ${\CF}$ restricted to
$$
U_{{\a}_0 {\a}_1 \ldots {\a}_q} \equiv  U_{{\a}_0} \cap U_{{\a}_1} \cap \ldots \cap U_{{\a}_q} \ ,
$$
so that
$$
{\ba}_{{\a}_0 {\a}_1 \ldots {\a}_{q}} \in {\Gamma} ( U_{{\a}_0 {\a}_1 \ldots {\a}_{q}} , {\CF})
$$
We assume that $
{\ba}_{{\a}_0 {\a}_1 \ldots {\a}_{q}} $ is totally antisymmetric in the indices ${\a}_{0} , {\a}_{1}, \ldots , {\a}_{q}$.

For example, in what follows ${\CF}$ will be often a sheaf of holomorphic vector fields ${\CT}_{X}$ on $X$ or a sheaf of closed holomorphic two-forms ${\CZ}_{X}^{2}$.
There are very few such objects defined globally on $X$,  if any. However, if we only require them to be
well-defined on small domains, such as  $U_{\a}$, $U_{\a\b}$ etc. then they become abundant. 

The space of all such locally defined sections of $\CF$ is denoted by ${\CC}^{q}(X, {\CF})$.

{\cech} differential ${\d}$ maps $q$-cochains to $q+1$-cochains, 
$$
{\d}: \ {\CC}^q \to {\CC}^{q+1}
$$
\eqn\cchd{ \left( {\d}{\ba} \right)_{{\a}_0 {\a}_1 \ldots {\a}_q {\a}_{q+1}} = \sum_{i=0}^{q+1}
(-1)^{i} \ {\ba}_{{\a}_0 \ldots {\a}_{i-1} {\a}_{i+1} \ldots {\a}_{q+1}} \ , }
and obeys ${\d}^2 = 0$. Thus we can define the cohomology groups, $$
H^{q}(X, {\CF})  = {\ker} {\d}\vert_{{\CC}^q} / {\rm im} {\d}\vert_{{\CC}^{q-1}} \ .
$$
 
\subsubsec{Glueing across the patches}

We now proceed with the investigation of the conditions $2$, $4$ on our list. On the intersection 
$U_{\a\b} = U_{\a} \cap U_{\b}$ we have two coordinate systems: ${\g}_{\aa} = ( {\g}^{i})$ and
${\g}_{\bbb} = ( {\tg}^{a} )$. Correspondingly we have a map:
\eqn\jacm{g_{\a\b} : U_{\a\b} \to GL_{d} ({\bC}), \qquad g_{\a\b} = \biggl\Vert {{\p}{\g}^i \over {\p}{\tg}^{a}} \biggr\Vert}
Note,
\eqn\invjac{g_{\a\b}  = g_{\b\a}^{-1} \ .}
The intersections 
$U_{\a\b}$ don't have the complicated topology, so that the three form ${\tr} \left( g_{\b\a} {\dd} g_{\a\b} \right)^{3}$ is exact:
\eqn\wzwi{{\tr} \left( g_{\b\a} {\dd} g_{\a\b} \right)^{3} = {\dd}{\m}_{\a\b}}
The corresponding two-form ${\m}_{\a\b} \in {\Omega}^{2}_{U_{\a\b}}$ enters the relation between the fields ${\b}^{[{\a}]} = ( {\b}_{i} )$ assigned to
$U_{\a}$ and the fields ${\b}^{[{\b}]} = ( {\tilde\b}_{a} )$, assigned to $U_{\b}$:
\eqn\bttrab{{\b}^{[{\b}]} = {\b}^{[{\a}]} g_{\a\b} + {\half} {\tr}\left(  {\CG}_{\a\b} g_{\a\b} {\p} g_{\b\a}\right)
+ {\half} \iota_{{\underline{\p}}}\ {\m}_{\a\b}}
 where
$$
 {\underline{\p}} = {\p}{\g}^{i} {\p}_{i} = {\p}{\tg}^{a} {\tilde\p}_{a} , \qquad
 {\CG}_{\a\b} = \Vert {\p}_{i} (g_{\a\b})^{j}_{a} \Vert  \ {\dd} {\tg}^{a} \ .
$$ 
The equation \wzwi\ does not determine ${\m}_{\a\b}$ uniquely. We must decide on how to choose
the representative modulo exact two-forms. First of all, we may want to insist on the condition
\eqn\inver{{\m}_{\b\a} = - {\m}_{\a\b}}
More precisely, if $f_{\a\b} : U \to U$, $U \approx U_{\a\b}\approx U_{\b\a}$ is the map which sends ${\g}_{\aa}$ to 
${\g}_{\bbb}$, $f_{\a\b} \circ f_{\b\a} = id$, then ${\m}_{\b\a} = - f_{\a\b}^{*} {\m}_{\a\b}$. In what follows we ignore these notational subtleties, by utilizing the coordinate-independent expressions. 

\subsubsec{Moduli of the model}

The relations \wzwi,\inver\ still  allow transformations of the form:
\eqn\bshift{{\m}_{\a\b} \mapsto {\m}_{\a\b} + b_{\a} -b_{\b}} 
where 
$b_{\a}$ is the closed holomorphic $(2,0)$-form, regular on $U_{\a}$, while $b_{\b}$ is the closed $(2,0)$-form, regular 
on $U_{\b}$.

Such a shift can be undone by the
similarity transformation on the fields ${\b}^{\aa}, {\g}_{\aa}$, generated by:
\eqn\symtrn{{\exp} \oint f_{{\a}, i} {\p} {\g}^{i}} and the similarity transformation of the fields
${\b}^{\bbb}, {\g}_{\bbb}$ generated by
\eqn\symtrn{{\exp} \oint f_{{\b}, a} {\p} {\tg}^{a} , }
where $b_{\a} = {\dd} f_{\a}, b_{\b} = {\dd}f_{\b}$ locally (cf. \ztwitten). 
However, \bshift\ does not exhaust all the freedom in solving \wzwi. The most general thing that can happen is the shift
\eqn\bbshift{{\m}_{\a\b} \mapsto {\m}_{\a\b} + b_{\a\b}, \qquad {\dd} b_{\a\b} = 0 }
where $b_{\a\b}$ is a closed two-form, regular on $U_{\a\b}$. The space of such forms, obeying the condition:
\eqn\coci{b_{\a\b} + b_{\b\g} + b_{\g\a} = 0, \qquad {\rm on} \ U_{\a\b\g} = U_{\a} \cap U_{\b} \cap U_{\g}} which follows from certain anomaly cancellation condition, to be discussed in the coming subsection,  modulo the
forms of the form \bshift\ is the first {\cech} cohomology group with coefficients in the sheaf of closed holomorphic two-forms, $H^{1}(X, {\CZ}_{X}^{2})$. Together with the space $H^{1}(X, {\CT}_{X})$ of classical complex structure deformations these parameterize the infinitesimal deformations of the $\b$-$\g$ sigma model:
\eqn\dfrm{\mathboxit{{\bf Deformations} = H^{1} ( X , {\CT}_{X} \oplus {\CZ}_{X}^{2} )}}
Note that this space is very similar to the space of deformations of the generalized (hyper) complex structure, since the sheaf ${\CZ}_{X}^{2}$  is essentially the quotient ${\Omega}^{1}_{X} / {\dd} {\CO}_{X}$.

\subsubsec{Obstructions}

Not every choice of ${\m}_{\a\b}$ obeying \wzwi\inver\ leads to the consistent theory.  Indeed, we have to make sure that given ${\b}^{[{\a}]}$ the fields ${\b}^{[{\b}]}$ defined using the glueing across the patches $U_{\a\b}$ directly, or via the third coordinate chart $U_{\g}$, ${\b}^{[{\a}]} \to {\b}^{[{\g}]} \to {\b}^{[{\b}]}$, coincide. Let us introduce the notations:
\eqn\coordbgtrip{\eqalign{& {\g}_{\aa} = ( {\g}^{i} ) , \ {\g}_{\bbb} = ( {\tg}^{a} ) , \ {\g}_{\gg} = ( {\hat\g}^{A} ) \cr
& {\b}^{\aa} = ( {\b}_{i} ) , \ {\b}^{\bbb} = ( {\tilde\b}_{a} ) , \ {\b}^{\gg} = ( {\hat\b}_{A} ) \cr
& \qquad i, a, A = 1, \ldots, d \cr
& \qquad {\m}_{\a\b} = {\m}_{ab} \ {\dd}{\tg}^{a} \wedge {\dd}{\tg}^{b} \cr
& \qquad {\m}_{\b\g} = {\tilde\m}_{AB} \ {\dd}{\hat\g}^{A} \wedge {\dd}{\hat\g}^{B} \cr
& \qquad  {\m}_{\a\g} = {\hat\m}_{AB} \ {\dd}{\hat\g}^{A} \wedge {\dd}{\hat\g}^{B} \cr
}} 
We should compare the results of two manipulations: one is the
direct change of coordinates:
\eqn\chonethree{{\g}^{i} \mapsto {\hat \g}^{A}}
another is the composition of two coordinate changes: 
\eqn\chott{{\g}^{i} \mapsto {\tg}^{a} \mapsto {\hat \g}^{A}}
The single coordinate changes act as follows:
\eqn\snglcch{\eqalign{& {\tbt}_{a} = {\b}_{i} g^{i}_{a} + {\half}{\tr} \left( {\CG}_{a} g {\p}_{} g^{-1} \right) + {\half}{\m}_{ab} {\p}_{} {\tg}^{b} \cr
& {\hat\b}_{A} = {\tbt}_{a} {\tilde g}^{a}_{A} + {\half}{\tr} \left( {\tilde\CG}_{A} {\tilde g} {\p}_{} {\tilde g}^{-1} \right)  + {\half}{\tilde\m}_{AB} {\p}_{} {\hat\g}^{B} \cr
& {\hat\b}_{A}^{\circ} = {\b}_{i} {\hat g}^{i}_{A} + {\half}{\tr} \left( {\hat\CG}_{A}  {\hat g} {\p}_{} {\hat g}^{-1} \right) + {\half}{\hat\m}_{AB} {\p}_{} {\hat\g}^{B} \cr}}
where:
\eqn\jacs{\matrix{& \qquad g^{i}_{a} = {{\p\g}^{i} \over {\p\tilde\g}^{a}}, \quad & \qquad 
{\tilde g}^{a}_{A} = {{\p\tilde\g}^{a} \over {\p\hat\g}^{A}}, \quad & \qquad {\hat g}^{i}_{A} = {{\p\g}^{i} \over {\p\hat\g}^{A}}\quad \cr
& & & \cr
& \quad {\tilde\CG}_{A} = \Vert {\p}_{j} g^{i}_{a} \Vert , \quad & \quad  {\tilde\CG}_{A} = \Vert {\tilde\p}_{b} {\tilde g}^{a}_{A} \Vert , \quad & \quad 
 {\hat\CG}_{A} = \Vert {\p}_{j} {\hat g}^{i}_{A} \Vert  \cr
 & & & \cr}}
The following identities are useful:
 \eqn\usflnid{\eqalign{
 & \qquad {\hat g}^{i}_{A} = g^{i}_{a} {\tilde g}^{a}_{A} \cr
 & {\hat\CG}_{A} = {\tilde g}^{a}_{A} {\CG}_{a} + g^{-1} {\tilde\CG}_{A} g \cr}} 
We recall that the products like  ${\b}_i  V^i  ({\g})$  in \snglcch\ are understood as the normal ordered
products.  Now we can substitute the first line in \snglcch\ into the second and compare the result with the
third:
\eqn\cmps{\eqalign{{\hat\b}_{A} = & \quad : \left( : {\b}_i g^{i}_{a} : \right) {\tilde g}^{a}_{A} : \ + {\half} {\tr} \left( {\tilde g}^{a}_{A} {\CG}_{a}  g {\p}_{} g^{-1} \right) + {\half} {\tr} \left( {\tilde\CG}_{A} {\tilde g} {\p}_{} 
{\tilde g}^{-1} \right) \cr
& \qquad + {\half} \left( {\m}_{ab} {\tilde g}^{a}_{A} {\p}_{} {\tilde\g}^{b} + {\tilde\m}_{AB} {\p}_{} {\hat\g}^{B}\right)\cr}}  
Now we should remember that in \cmps\ we have the double normal ordering, which has to be
converted into a single one:
\eqn\dblnrm{ \quad : \left( : {\b}_i g^{i}_{a} : \right) {\tilde g}^{a}_{A} : \  = {\b}_{i} {\hat g}^{i}_{A} - \left( {\p}_{} g^{i}_{a} \right) \left( {\p}_{i} {\tilde g}^{a}_{A} \right)}
Now it is straightforward to compare:
\eqn\anomm{\eqalign{ {\hat\b}^{\circ}_{A} - {\hat\b}_{A} = & \quad {\half} {\tr} \left\{ \left( {\tilde g}^{-1} {\hat \p}_{A} {\tilde g} \right) \left( {\p}_{} g g^{-1} \right) - \left( {\tilde g}^{-1}   {\p}_{}  {\tilde g} \right) \left({\hat \p}_{A}{ g} {g}^{-1} \right) \right\} \cr
& + {\half} \left( {\hat \m}_{AB} - {\tilde \m}_{AB} - {\m}_{ab} {\tilde g}^{a}_{A} {\tilde g}_{B} \right) {\p}_{} {\hat\g}^{B} \cr}}
Insisting on the equality ${\hat\b}^{\circ}_{A}  = {\hat\b}_{A}$ is equivalent to the following {\it cocycle} condition on the set of ${\m}_{\a\b}$'s:
\eqn\cocm{\left( {\d}{\m}\right)_{\a\b\g} \equiv {\m}_{\a\b} + {\m}_{\b\g} + {\m}_{\g\a} = {\tr} \left( g_{\a\b} {\dd} g_{\b\g} \wedge {\dd} g_{\g\a} \right)}
where, recall:
\eqn\gab{g_{\a\b} = \Vert g^i_a \Vert , g_{\b\g} = \Vert  {\tilde g}^a_{A} \Vert ,  g_{\a\g} = \Vert {\hat g}^i_{A} \Vert = g_{\g\a}^{-1}}
Note that if we apply the ${\dd}$-operator to both left and right hand sides of \cocm\ then we get the identity, thanks to \wzwi.
Thus, the following {\cech} $2$-cocycle $\psi$,
\eqn\cocmi{{\psi}_{\a\b\g} = {\m}_{\a\b}  + {\m}_{\b\g} + {\m}_{\g\a} - {\tr} \left( g_{\a\b} {\dd} g_{\b\g} \wedge {\dd} g_{\g\a} \right) \  , }
 takes values in closed $2$-forms:
 \eqn\cocmii{{\dd} {\psi}_{\a\b\g} = 0}
 As ${\m}_{\a\b}$'s are defined from \wzwi\ up to an addition of the closed $2$-forms $b_{\a\b}$'s, which are regular on $U_{\a\b}$, our problem is find 
 \eqn\bcob{b = ( b_{\a\b} ), \qquad b_{\a\b} \in {\CZ}^{2}_{U_{\a\b}}, \qquad {\rm s.t.} \quad  {\d} b = {\psi} \ . }
The equations \cocm\bcob\  means that $\psi$
represents a trivial second {\cech} cohomology class:
\eqn\cocmi{0  = \left[ {\m}_{\a\b} + {\m}_{\b\g} + {\m}_{\g\a} - {\tr} \left( g_{\a\b} {\dd} g_{\b\g} \wedge {\dd} g_{\g\a} \right) \right] \in 
H^{2} ( X, {\CZ}_{X}^{2})} with values in closed holomorphic $2$-forms. 
If we drop the contribution of ${\m}$'s in \cocmi\ we would get a condition of vanishing in cohomology
of the sheaf ${\Omega}_{X}^{2}$ of holomorphic $2$-forms, But, in a sense, since  
the failure of $ {\tr} \left( g_{\a\b} {\dd} g_{\b\g} \wedge {\dd} g_{\g\a} \right) $ to be ${\dd}$-closed is 
${\d}$-exact:
\eqn\fbc{\eqalign{& {\dd} \  {\tr} \left( g_{\a\b} {\dd} g_{\b\g} \wedge {\dd} g_{\g\a} \right) = \left( {\d} {\CW} \right)_{\a\b\g}, \cr
& \qquad {\CW}_{\a\b} = {\tr} \left( g_{\b\a}{\dd} g_{\a\b} \right)^{3}\ ,  \cr} }
we don't loose much information (in a more sophisticated language, this reflects the degeneration of certain
spectral sequence at the second term). 
Thus, we need (keeping \fbc\ in mind):
\eqn\cocmi{0  = \left[ {\psi} \right] \in 
H^{2} ( X, {\CZ}_{X}^{2})} 
In general, the $2$-cocycle ${\psi}$, valued in closed holomorphic $2$-forms, may represent a non-trivial cohomology class. In this case one cannot define the ${\b}$-fields consistently over $X$. Mathematically one gets the so-called gerbe of chiral differential operators \malikov.  But physically it means that the model is anomalous, and extra degrees of freedom are needed to define it properly
\FS.

We can phrase the result as follows: the group $H^{2}(X, {\CZ}_{X}^{2})$ parameterizes the obstructions for deforming the model. Together with the classical piece of the complex structure deformations obstructions, they form the group of
\eqn\obst{\mathboxit{{\bf Obstructions} = H^{2} ( X, {\CT}_{X} \oplus {\CZ}_{X}^{2} )}}

\subsubsec{Anomaly and Pontryagin class}

The class $[ \psi ] $ in the cohomology group $H^{2}(X, {\CZ}_{X}^{2})$ is actually the first Pontryagin class of $X$, $p_{1}(X)$, or, in a more holomorphic language, the second Chern class $ch_2 ({\CT}_{X})$ of the holomorphic tangent bundle. Its emergence is quite similar to the emegence of the 
first Pontryagin class of the manifold in the studies of heterotic string compactifications \witohet\ and
in supersymmetric sigma models \howe .
To understand the relation of $[ \psi ]$ to $p_{1}(X)$ let us invoke  the good old descent formalism, well-known in the theory of anomalies, e.g. \FS, \gravan.  

In this section $\bd$ denotes de Rham exteriour derivative acting on smooth differential forms on $X$,
with respect to the complex structure on $X$ it splits as a sum of two nilpotent operators
\eqn\cmsp{{\bd} = {\pd}  + {\pdd}}
where ${\pd}$ maps the ${\Omega}^{p,q}$ forms to ${\Omega}^{p+1, q}$, while ${\pdd}$ maps
${\Omega}^{p,q}$ to ${\Omega}^{p,q+1}$. ${\pd}$ is what we called ${\dd}$ in the rest of the paper.

Take the holomorphic tangent bundle ${\CT}_{X}$ and view it as a complex vector bundle ${\CE}$ over $X$. It can be endowed with hermitian metric, and with some unitary connection $A$. With respect to the 
complex structure on $X$ the connection splits as a sum of $(1,0)$ and $(0,1)$ parts, while the curvature $F$ splits as the sum of three terms:  $(2,0$, $(1,1)$ and $(0,2)$. 
In the coordinate chart $U_{\a}$ over which the bundle ${\CE}$ is trivialized: ${\CE}\vert_{U_{\a}} \approx U_{\a} \times {\bC}^{d}$, the connection is described by the matrix-valued one-form:
\eqn\cont{A_{\a} = A_{\a}^{1,0} + A_{\a}^{0,1} , \qquad A^{1,0}_{\a} = \left(  A_{jk}^{i} {\dd}x^k \right)_{\a} , \quad A^{0,1}_{\a} = \left( A_{\jb k}^{i} {\dd}{\xb}^{\jb} \right)_{\a} = - \left( A^{-1,0}_{\a} \right)^{\dagger} \ , }
and on the overlaps
\eqn\conov{A_{\b} = u_{\b\a} A_{\a} u_{\a\b} + u_{\b\a} {\bd} u_{\a\b}}
where $u_{\a\b} = u_{\b\a}^{-1}  : U_{\a\b} \to U(d)$ are the transition functions for ${\CE}$. 
Now let us demand that the $(0,1)$ part of the connection defines the holomorphic structure on the vector bundle. It means that the holomorphic sections are the ones which are annihilated by ${\pdd} + A^{0,1}$ operator:
\eqn\hs{{\pdd} s^{i} + A_{\jb k}^{i} s^{k} {\dd}{\xb}^{\ib} = 0}  
The consistency of \hs\ demands $F^{0,2}=0$. It implies that on each coordinate chart $U_{\a}$ we can find the complex gauge transformations $e_{\a} = ( e^{i}_{j} )_{\a}$ such that
\eqn\lc{A_{\a}^{0,1} = e_{\a} {\pdd} e_{\a}^{-1} \ , \qquad A_{\a}^{1,0} =  - 
{\eb}_{\a}^{-1}{\pd} {\eb}_{\a}}
where $e_{\a}$ and ${\eb}_{\a}$ are the components of the vierbein of the hermitian metric $h =   {\eb} e$. The overlap relation \conov\ implies:
\eqn\overl{e_{\a} = u_{\a\b} e_{\b} g_{\b\a}} where 
\eqn\overll{{\pdd} g_{\a\b} = 0 , \quad {\pd} g_{\a\b}^{\dagger} = 0 \ ,  \qquad g_{\a\b} (x) \in GL_{d} ({\bC}) \ , x \in U_{\a\b}} are the transition functions \jacm\ of the holomorphic bundle ${\CE} = {\CT}_{X}$. Thus, the complex gauge transformation $e_{\a}$ maps $A$ to the connection ${\Gamma}$, s.t. ${\Gamma}^{0,1}=0$, and
\eqn\gmcn{{\Gamma} = \left( {\Gamma}^{1,0}_{\a} \right) = \Vert \left( {\Gamma}_{jk}^{i} {\dd}x^{j} \right)_{\a}  \Vert = h_{\a} ^{-1} {\pd} h_{\a}}
where $h_{\a} = {\eb}_{\a} e_{\a}$ is the hermitian metric on ${\CE}$. On the intersections $U_{\a\b}$:
\eqn\overlh{h_{\a} = g_{\b\a}^{\dagger} h_{\b} g_{\b\a}}
Now let us calculate the density of the second Chern class of ${\CT}_{X}$ using the metric $h$:
\eqn\dpc{p_{\a} = {1\over 8{\pi}^{2}} {\tr} \left( F_{\a} \wedge F_{\a} \right) = {1\over 8 {\pi}^2} 
{\tr} \left( {\pdd} {\Gamma}_{\a} \wedge {\pdd}{\Gamma}_{\a} \right) = {\bd}\  {\rm CS}_{\a}}
where
\eqn\csa{{\rm CS}_{\a} = {1\over 8 {\pi}^2} {\tr}\left(  {\Gamma}_{\a} \wedge {\pdd} {\Gamma}_{\a} - {1\over 3} {\Gamma}_{\a} \wedge {\Gamma}_{\a} \wedge {\Gamma}_{\a} \right) }
is  $(2,1) \oplus (3,0)$-form, defined on $U_{\a}$. On the overlap $U_{\a\b}$ we have, using \overl\gmcn:
\eqn\overli{{\Gamma}_{\a} = g_{\a\b} {\Gamma}_{\b} g_{\b\a} + g_{\a\b} {\pd} g_{\b\a}}
and, therefore:
\eqn\overlii{( {\d}{\rm CS})_{\a\b}  = {\rm CS}_{\b} - {\rm CS}_{\a} = {\bd} {\rho}_{\a\b}}
with the $(2,0)$-form ${\rho}_{\a\b}$, defined on the double intersections $U_{\a\b}$:
\eqn\overliir{{\rho}_{\a\b} = {\m}_{\a\b} - {1\over 8{\pi}^2} {\tr} \left( g_{\a\b} {\dd} g_{\b\a} \wedge {\Gamma}_{\a} \right)}
Finally, 
\eqn\overliii{\left( {\d}{\rho} \right)_{\a\b\g} = {\rho}_{\a\b} + {\rho}_{\b\g} + {\rho}_{\g\a} = 
({\d}{\m})_{\a\b\g} + {1\over 8{\pi}^2} {\tr} \left( g_{\b\a} {\dd} g_{\a\g} \wedge {\dd} g_{\g\b} \right)}
which is our anomaly two-form ${\psi}_{\a\b\g}$. 
\subsubsec{Automorphisms}

The naive continuation of the sequence \obst\dfrm\ is:
\eqn\infsym{\mathboxit{{\bf Infinitesimal \quad automorphisms} = H^{0} ( X, {\CT}_{X} \oplus {\CZ}^{2}_{X})}} 
Indeed, the group $H^{0} (X, {\CT}_X)$ enumerates the globally defined holomorphic vector fields
 on $X$, which are the symmetries of the manifold $X$ viewed as a complex variety, while
 the group $H^{0}(X, {\CZ}_{X}^{2})$ enumerates globally defined closed $2$-forms ${\tau}  = {\tau}_{ij} {\dd}{\g}^i \wedge {\dd}{\g}^{j}$, which occur in the  theory via the shifts of the Lagrangian
 \eqn\naivlag{\int {\b} {\pb} {\g} \mapsto \int {\b} {\pb} {\g} + \int {\tau}_{ij} {\p} {\g}^{i} {\pb} {\g}^{j}}
Such a shift does not change the equations of motion. It also does not affect the perturbative correlation functions. However, it affects non-perturbative correlation functions, unless $[ {\tau} ] \in H^{2} ( X, {\bZ})$, i.e. it is an integral form, cf. \ztwitten. 

Anyway, we shall now examine whether this is true in more detail. We shall see that the
first, classical geometry piece, ${\CT}_{X}$, may have hard time being realized in the quantum theory.
 We now discuss the question $5.$ on our list.

\subsubsec{Global symmetry currents: abelian case}
 
Let us start with the single vector field. We are given a holomorphic vector field $V$. In the local coordinate patch $U_{\a}$ it is described by the components
$V_{\aa} = ( V^{i} ), V = V^{i} {\p}_{i}$. In passing to the coordinate patch $U_{\b}$ over the intersection
$U_{\a\b}$ we encounter the analogue of the problem \bbtrr. Specifically, 
let us denote by $B_{\a}$ and $B_{\b}$ the  one-forms $B_{i}({\g}) {\dd}{\g}^{i}$ and ${\tilde B}_{a} ({\tg}) {\dd}{\tg}^{a}$, defined on $U_{\a}$ and $U_{\b}$ respectively. We wish to construct the current ${\CJ}_V$ which is a global object on $X$:
\eqn\globc{{\CJ}_{V} = {\b}_i V^i + B_{i} {\p}{\g}^i = {\tilde\b}_a {\tilde V}^a + {\tilde B}_a {\dd} {\tg}^a}
Introduce the familiar by now matrices (cf. \vmatr) ${\CV}_{\a}$ for each coordinate chart $U_{\a}$: 
$$
{\CV}_{\a} = \Vert {\p}_{i} V^{j} \Vert , \qquad {\CV}_{\b} = \Vert {\tilde\p}_{a} {\tilde V}^{b} \Vert
$$
and the following {\cech} $1$-cochain:
\eqn\nucoc{{\n}_{\a\b} \equiv \iota_{V} {\m}_{\a\b} -  \ {\tr} \left( {\CV}_{\a} {\dd}g_{\a\b} g_{\b\a}  \right) + 
  {\tr}\left( {\CV}_{\b} {\dd}g_{\b\a}  g_{\a\b} \right) }
Then \globc\ can be rewritten as:
\eqn\bbtrri{{\n}_{\a\b} = \iota_{V} b_{\a\b} + 
   2 ( {\d}B )_{\a\b} }
   where $b_{\a\b} \in {\CZ}^2_{U_{\a\b}}$. 
The consistency of \bbtrri\ can be checked by applying the operators ${\d}$ and $\iota_{V}$ to ${\n}$:
\eqn\dltnu{\eqalign{ \left( {\d}{\n} \right)_{\a\b\g}\ &  = {\tr}\left( \left(  {\CL}_{V} g_{\a\b}\right)  {\dd} g_{\b\g} g_{\g\a} - {\dd} g_{\a\b} \left( {\CL}_{V} g_{\b\g} \right) g_{\g\a} \right) + \iota_{V} \left( {\m}_{\a\b} + {\m}_{\b\g} + {\m}_{\g\a} \right)  \cr
& \qquad\qquad\qquad = \iota_{V}{\psi}_{\a\b\g}  \ , \cr
 \iota_{V} {\n}_{\a\b}\ & = - {\tr} \left( g_{\b\a} {\CV}_{\a} {\CL}_{V} g_{\a\b}\right) + {\tr} \left( g_{\a\b} {\CV}_{\b} {\CL}_{V} g_{\b\a}\right)  \cr
& \qquad\qquad\qquad = - ( {\d} {\bv} )_{\a\b} \ , \cr
 {\bv}_{\a} \ & = -  {\tr} {\CV}_{\a}^{2} \cr}}
while applying ${\d}$, $\iota_{V}$ to the right hand side of \bbtrri\ gives:
\eqn\bbtrrii{\eqalign{  \iota_{V} \left( \iota_{V} b + 2 {\d} B \right) & \ = 2 {\d} ( \iota_{V} B ) \cr
 {\d} \left( \iota_{V} b + 2 {\d} B \right) & \ = \iota_{V} {\psi} \cr}}
the last equality being true iff the $p_1$ anomaly is absent, cf. \bcob. 

Without assuming the existence of $b, B$, verifying \bbtrri\bcob\ we can still
neatly package \dltnu\ and \bbtrrii\ into the condition of {\it equivariant closedness}:
\eqn\eqcl{( {\d} + t \iota_{V} ) \left( {\bv}t^2 + {\n}t + {\psi} \right) = 0 }
where we realize the {\it holomorphic equivariant} {\cech}-complex of $X$ via {\cech} ${\Omega}^{*}_{X}$-cochains with values in the polynomial functions  of $t$. The grading is defined to be 
\eqn\eqdeg{{\rm equivariant \ degree} = {\rm form \ degree} + 2 t {{\p}\over {\p} t}}
On such cochains  the operator ${\d} + t \iota_{V}$ has degree one, it is nilpotent and its cohomology is what we need.  For example
\eqn\eqpont{{\bv} t^2 + {\nu}t  + {\psi}}
represents a ${\bC}^{*}$-holomorphic equivariant cohomology class of degree four. If, instead of {\cech} we were to use Dolbeault picture, which is more natural in the approach involving
$(0,2)$ supersymmetric models \ztwitten, then the analogue of ${\d}$ would have been the 
${\pb}$ operator acting on ${\Omega}^{p,q}$-forms on $X$, and \eqcl\ would have looked more familiar.
The appropriate cohomology theory, the holomorphic equivariant cohomology, has been developed in \kliu.

Now let us assume $B, b$ exist. Consider the following $0$-cochain ${\bk} $:
\eqn\levv{{\bk}_{\a} = \iota_{V} B_{\a} -  {\half} {\tr} {\CV}_{\a}^{2} \ ,}  
which is a quadratic functional of $V$.  Then \bbtrri\ implies (cf. \opeii):
\eqn\levvi{{\d}{\bk} = 0 \ ,}
so it is actually a $0$-cocycle, and represents certain cohomology class of {\cech} cohomology 
with the coefficients in the sheaf ${\CO}_{X}$ of holomorphic functions. If $X$ is compact, then
it implies that ${\bk}$ is a constant, which is the level of the current algebra. In our examples
it will be always constant, even for non-compact $X$.
Note that the value of ${\bk}$ can be found by analyzing the behavior of $V$ near its zeroes. 
Indeed, let us assume that $V$ generates the action of ${\bC}^{*}$ on $X$. Suppose $p \in X$ is an isolated fixed point of this action, $V(p) = 0$. Then there exist local coordinates $( {\g}^{i} )$ such that near this point 
\eqn\fixnrml{V = \sum_{i} \ m_{i} {\g}^{i} {{\p}\over{{\p}{\g}^{i}}} + {\rm higher \ order\ terms}} 
where $m_{i} \in {\bZ}$. The invariant meaning of the components of ${\vec m} = ( m_{1} , \ldots, m_{d})$ is that these are the  weights of ${\bC}^{*}$-action on the tangent space $T_{p}X$ at $p$. They are defined uniquely up to permutations.
Then 
\levv\ implies:
\eqn\levvi{{\bk}  = - {\half} {\vec m}^2}
It may seem surprising that the square of the weight vector is the same for all fixed points of the ${\bC}^*$-action on $X$, but this is in fact the consequence of the triviality of the equivariant
second Chern class of ${\CT}_{X}$, which is expressed by the equation
\eqn\eqex{\left( {\d} + t \iota_{V} \right) \left(  2 B t  + b \right) = 2{\bk}t^2+{\bv}t^2  + {\n}t + {\psi} } 
combining the formulae \bcob\ and \bbtrri\ into the single condition of the
{\it equivariant exactness}. 
\subsubsec{Global symmetry currents: non-abelian case}

Now let us assume that a complex Lie group $G$ acts on $X$. Let $\bg$ be the Lie algebra of $G$, 
and let ${\phi} = ( {\phi}^{\uA} ) $, $\uA = 1, \ldots, {\rm dim}G$, be the linear coordinates on $\bg$, corresponding to some basis ${\bt}_{\uA}$ in $\bg$. 
The action of $G$ on $X$ is generated by the holomorphic vector fields $V_{\uA}$. We can view them as
the linear function on $\bg$ with values in $H^{0}(X, {\CT}_{X})$, 
\eqn\hom{{\phi} \mapsto 
V( {\phi}) = {\phi}^{\uA} V_{\uA} \ . }
We have the defining relations of $\bg$:
\eqn\cmnt{[V_{\uA} , V_{\uB} ] = f_{\uA\uB}^{\uC} V_{\uC} \ , }
where $f_{\uA\uB}^{\uC}$ are the structure constants of ${\bg}$:
\eqn\strc{[{\bt}_{\uA}, {\bt}_{\uB}] = f_{\uA\uB}^{\uC} {\bt}_{\uC}}
We define the currents, by the local formula in each coordinate chart $U_{\a}$:
\eqn\curns{{\CJ}_{\uA} = {\b}_{i} V^{i}_{\uA} + B_{i \uA} {\p} {\g}^{i} \ , }
where the one-forms $B_{\uA}$ are to be found from the several conditions: i) the  ${\CJ}_{\uA}$'s should form the ${\hat\bg}$ current algebra; \ ii) the ${\CJ}_{\uA}$ should be  independent of 
$\a$, modulo some automorphism of $G$.

Let us first work out the conditions which follow from i). They should hold in each coordinate chart $U_{\a}$. In the next paragraph we omit the index $\a$. 

Requiring the residue at the first order pole of the operator product expansion of ${\CJ}_{\uA}$ and ${\CJ}_{\uB}$ to be $f_{\uA\uB}^{\uC} {\CJ}_{\uC}$ implies:
\eqn\curntalg{{\CL}_{V_{[ \uA}} {B}_{\uB]} - f_{\uA\uB}^{\uC} B_{\uC} - {\half} {\dd} {\bm}_{\uA\uB}  - {\Omega}_{\uA\uB} = 0}
where
\eqn\ommn{\eqalign{& {\Omega}_{\uA\uB} = {\half} {\tr} {\CV}_{[ \uA} {\dd} {\CV}_{{\uB}]}\ , \cr
& {\bm}_{\uA\uB} = \iota_{V_{[\uA}} B_{{\uB}]} \ , \cr}
}
and
\eqn\vmatg{{\CV}_{\uA} =   \Vert {\p}_i V_{\uA}^j \Vert}
The matrices \vmatg\ obey the one-cocycle condition:
\eqn\vmcocl{{\CL}_{V_{[{\uA}}} {\CV}_{{\uB}]} - f_{\uA\uB}^{\uC} {\CV}_{\uC} + [{\CV}_{\uA}, {\CV}_{\uB}] = 0 \ ,} 
where we view ${\CV}_{\uA}$ simply as the matrix-valued functions on $U_{\a}$. 
It is convenient to combine $B_{\uA}'s$ and ${\CV}_{\uA}$'s into the linear functions on $\bg$:
\eqn\bexp{B ( {\phi} ) = \left( \left( B_{\uA} \right)_{\a} {\phi}^{\uA} \right)  , \qquad {\CV}({\phi}) = \left( \left( {\CV}_{\uA} \right)_{\a}  {\phi}^{\uA} \right)} 
The equations \vmcocl\curntalg\ are antisymmetric in $\uA$, $\uB$ indices, and can be written compactly using anticommuting variables $c^{\uA}$, which are of course the BRST ghosts:
\eqn\curtalgi{\eqalign{& {\CL}_{V (c)} {\CV}(c) - {\CV} ( [c,c] ) + [ {\CV}(c) , {\CV}(c) ] = 0 \cr
& 2 {\CL}_{V (c)} B (c) - B ( [c,c] ) - {\dd} \left( \iota_{V(c)} B (c) \right) - {\tr} {\CV} (c) {\dd} {\CV} (c) = 0 \cr}}
Now let us discuss the condition ii) We use the notations \hom\bexp. We have the $0$, $1$, and $2$ cochains, ${\bv}({\phi}, {\phi})$, 
${\n} ({\phi})$, and ${\psi}$, respectively, which are the second, first and zeroth order polynomials in $\phi$:
\eqn\sigmn{{\bv}({\phi}, {\phi})_{\a} = ( {\bv}_{\uA\uB})_{\a} {\phi}^{\uA} {\phi}^{\uB}, \qquad  ({\bv}_{\uA\uB})_{\a} = - {\tr} \left( \left( {\CV}_{\uA} \right)_{\a}\left( {\CV}_{\uB} \right)_{\a} \right)}
\eqn\nuna{{\n} ({\phi} )_{\a\b} = \iota_{V({\phi})} {\m}_{\a\b} - {\tr}\left(  {\CV}({\phi})_{\a} {\dd}g_{\a\b} g_{\b\a} \right) + {\tr} \left( {\CV}({\phi})_{\b} {\dd} g_{\b\a} g_{\a\b} \right) }
The second order pole in the operator product expansion of ${\CJ}_{\uA}$'s:
\eqn\centrt{{\bk}_{\uA\uB} = {\bv}_{\uA\uB} + \iota_{V_{\uA}} B_{\uB} + \iota_{V_{\uB}} B_{\uA} \ , }
is also conveniently packaged into the second order polynomial in $\phi$:
\eqn\centrti{{\bk} ( {\phi} , {\phi} ) = {\bv} ( {\phi} , {\phi} ) + 2 \iota_{V ({\phi})} B ( {\phi} )}
 
 The direct analogue of \eqex\ which follows from the analogue of \bbtrri\ is  the
{\it $G$-equivariant exactness}:
\eqn\eqexi{\left( {\d} + \iota_{V ({\phi})} \right) \left(  2 B ({\phi}) + b \right) = 2{\bk} ( {\phi}, {\phi} ) +{\bv} ( {\phi}, {\phi} )   + {\n} ( {\phi} )  + {\psi} } 
Without assuming the existence of $b, B({\phi})$, verifying \eqexi\ we still
have the {\it $G$-equivariant closedness}:
\eqn\eqclnonab{( {\d} + \iota_{V ({\phi})} ) \left( {\bv} ( {\phi}, {\phi} ) + {\n} ({\phi})  + {\psi} \right) = 0 }
Finally, let us mention that the differential ${\d} + \iota_{V({\phi})}$ is most naturally interpreted in the language of the holomorphic $G$-equivariant cohomology. The generalization of the abelian complex 
calculating the ${\bC}^{*}$-equivariant cohomology is the space of $G$-invariant ${\Omega}^{*}_{X} \otimes   {\rm Fun}({\bg})$ {\cech}-cochains, where $G$ acts on $\bg$ in the adjoint representation, and the grading is defined analogously to \eqdeg\  ( see \atiyahbott,\wittentdg\ for the introduction into the de Rham version of the equivariant cohomology):
\eqn\eqdegnonab{{\rm equivariant \ degree} = {\rm form \ degree} + 2 {\phi} {{\p}\over {\p} {\phi}}}
On such cochains  the operator ${\d} + t \iota_{V}$ has degree one, it is nilpotent and its cohomology is what we need.  For example
\eqn\eqpont{{\bv} ({\phi}, {\phi}) + {\nu}({\phi})  + {\psi}}
represents the holomorphic $G$- equivariant cohomology class of degree four.

\subsubsec{The case of free $G$ action}

Suppose the action of $G$ on $X$ is free. In this case the equivariant cohomology is expected to coincide with the cohomology of the factorspace $X/G$ \atiyahbott.
Let us investigate the solutions to the equations \eqexi. It is convenient to introduce the
connection one-forms ${\Theta}^{\uA} = {\Theta}^{\uA}_{i}{\dd}{\g}^i$ which obey:
\eqn\conof{\eqalign{
{\dd} {\Theta}^{\uC} + {\half} f^{\uC}_{\uA\uB} {\Theta}^{\uA} \wedge {\Theta}^{\uB} & \ = 0 \cr
\iota_{{\bV}_{\uA}} {\Theta}^{\uB} & \ = {\d}^{\uB}_{\uA} \cr
 {\CL}_{{\bV}_{\uA}} {\Theta}^{\uC} + f^{\uC}_{\uA\uB} {\Theta}^{\uB} & \ = 0 \cr}} 

The equations we shall finally get are similar to \aabci\ except that
now we expand the one-forms ${B}_{\uA}$ in ${\Theta}$'s:
\eqn\binth{B_{\uA} = {\half} \left( {\s}_{\uA\uB} - {\m}_{\uA\uB} \right) {\Theta}^{\uB}}
and instead of \aabci\ we get:
\eqn\abbcii{{\sum}_{\kern -30pt \vbox{\kern11pt\hbox{$\scriptstyle{{\rm cyclic} \ {\uA} \to {\uB} \to {\uC}}$}}}
 {\CL}_{V_{\uA}}{\m}_{\uB\uC} + f_{\uA\uB}^{\uD} {\m}_{\uD\uC}   =  - {\tr} {\CV}_{\uA} [ {\CV}_{\uB}, {\CV}_{\uC}]}
 which is the condition that
 the 3-cocycle on the Lie algebra ${\bg}$ with values in the module ${\CO}_{U}$ is the coboundary of 
 ${\m}$ 
 \eqn\mcocc{( {\dd}_{\bg} {\m}) _{ABC} = - {\tr} {\CV}_{A} [ {\CV}_{B}, {\CV}_{C}]}
 From the one-cocycle condition \vmcocl\ 
 follows  the closedness of the three-form
\eqn\thrfrm{{\o}_3 = {\tr}\left(  {\CV}_{A} [ {\CV}_{B}, {\CV}_{C}] \right) \  {\Theta}^{A} \wedge {\Theta}^{B} \wedge {\Theta}^{C}\ .}

\subsubsec{Stress-energy tensor and $c_1$ anomaly}

Now let us address the question $3.$ on our list. 
We can define the naive stress-energy tensors $T_{\aa}^{\rm naive}$ by the formulae \sti\ on each coordinate patch $U_{\a}$. Then, using \stiii\ we immediately conclude that on the overlaps $U_{\a\b}$:
\eqn\nai{T_{\bbb}^{\rm naive} - T_{\aa}^{\rm naive} = - {\half} {\p}^{2} {\log} \ {\det} g_{\a\b}}
The globally defined stress-energy tensor must be defined as:
\eqn\stglob{T = T^{\rm naive}_{\aa} - {\half} {\p}^2 {\log} {\o}_{\aa}}
where ${\o}_{\a}$ are holomorphic functions on $U_{\a}$ such that on the overlaps they are related by
\eqn\overlo{{\o}_{\aa} = {\o}_{\bbb} 
\ {\det} g_{\a\b}}
(up to possible constant factors which we ignore).
In other words, the top holomorphic form
\eqn\tophf{{\Omega} = {\o}_{\aa} {\dd}{\g}_{\aa}^{1} \wedge \ldots \wedge {\dd}{\g}_{\aa}^{d}}
is in fact independent of ${\a}$, is nowhere vanishing and everywhere regular (otherwise the logarithm in \stglob\ would have singularities). 
In other words, it means that $X$ must have a Calabi-Yau structure (in variance with more stringent condition of being a Calabi-Yau manifold, which means in addition that $X$ is Kahler, which we don't need here). 

The obstruction of having the globally defined holomorphic top form is $c_{1}(X)$. Interestingly enough,
this is also an obstruction for the conformal invariance of the $(0,2)$ model, which has fermions and
antiholomorphic coordinates ${\bar\g}^{\ib}$ as the worldsheet fields \ztwitten.
The term \tdcrvt\ is nothing but the Bott-Chern secondary characteristic class, constructed out of the
class ${\half}c_{1}(X) c_{1}({\Sigma})$ on $X \times \Sigma$. This class enters the Riemann-Roch formula for the determinant line bundle associated with the ${\b}{\g}$ system \ztwitten\ and its consequences will be more carefully studied in  \fln.

\subsec{The anomalous example}

The simplest yet very instructive example of the target space which we shall consider first
will be that of the projective space, $X = {\bC\bP}^{d-1}$.

\subsubsec{The Pontryagin anomaly}
 Let
$(x_0 : x_1 : \ldots : x_{d-1} )$ denote the homogeneous coordinates on $X$. 
Let us cover $X$ by $d$ coordinate
patches $U_{\a} \approx {\bC}^{d-1}$, ${\a}=0, \ldots, d-1$. On the coordinate patch $U_{\a}$ the
homogeneous coordinate $x_{\a} \neq 0$ and we can use as the coordinates 
$$
( u_{\a}) = ( x_{i}x_{\a}^{-1}) , 
$$
The coordinate transformation from $U_{0}$ to $U_{1}$, and to $U_2$  are given by:
let $u_{1}, \ldots , u_{d-1}$ be the coordinates on $U_{0}$, ${\tu}_{1}, \ldots, {\tu}_{d-1}$ be the coordinates on $U_{1}$, and
let ${\hu}_{1} , \ldots , {\hu}_{d-1}$ be the coordinates on $U_2$.
Then:
\eqn\cpn{\eqalign{& {\tu}_{1} = {u}_{1}^{-1}, \cr
& {\tu}_{a} =  {u}_{a}{u}_{1}^{-1} , \qquad a = 2, \ldots d-1 \cr}}
The matrix $g = g_{01}$, associated with the change of variables \cpn\ is readily calculated:
\eqn\gcpn{\eqalign{& g_{01} = \pmatrix{ - u_{1}^2 & - u_{a} u_{1} \cr
0 &\  u_{1}\  {\bf 1}_{d-2} \cr} \cr
& \cr
& g^{-1}_{01} = g_{10} = \pmatrix{ - u_{1}^{-2} & - u_{a} u_{1}^{-2}\cr
0 &\   u_{1}^{-1} {\bf 1}_{d-2} \cr} \cr
& \cr}}
The coordinate transformations from $U_{1}$ to $U_{2}$ and from $U_{0}$ to $U_{2}$ are given by:
\eqn\cpni{\eqalign{& {\hu}_{1} = x_{0}x_{2}^{-1} = {\tu}_{1}{\tu}_{2}^{-1} = u_{2}^{-1} \cr 
& {\hu}_{2} = x_{1}x_{2}^{-1} = {\tu}_{2}^{-1} = u_{1} u_{2}^{-1} \cr
& {\hu}_{A} = x_{A}x_{2}^{-1} = {\tu}_{A}{\tu}_{2}^{-1} = u_{A}u_{2}^{-1} \cr
& \qquad\qquad\qquad A = 3, \ \ldots ,\  N \cr}}  
The corresponding matrices $g_{12}$ and $g_{20}$ are given by:
\eqn\gmtr{\eqalign{& g_{12} = \pmatrix{ u_2 u_{1}^{-1} & 0 & 0 \cr
- u_{2} u_{1}^{-2} & - u_{2}^{2} u_{1}^{-2} & - u_{A}u_{2} u_{1}^{-2} \cr
0 & 0 & u_{2}u_{1}^{-1} {\bf 1}_{d-3} \cr} \cr
& \cr
& g_{12}^{-1} = g_{21} = \pmatrix{ u_{1}u_{2}^{-1} & 0 & 0 \cr
- u_{1} u_{2}^{-2} & -u_{1}^{2}u_{2}^{-2} & - u_{A}u_{1}u_{2}^{-2} \cr
0 & 0 & u_{1}u_{2}^{-1} {\bf 1}_{d-3} \cr} \cr & \cr
& g_{02} = \pmatrix{ - u_{1}u_{2} & - u_{2}^{2} & - u_{A} u_{2} \cr
u_{2} & 0 & 0 \cr
0 & 0 & {\g}_{2} {\bf 1}_{d-3} \cr} \cr & \cr  
& g_{02}^{-1} = g_{20} = \pmatrix{ 0 & u_{2}^{-1} & 0 \cr
- u_{2}^{-2} & - u_{1}u_{2}^{-2} & - u_{A}u_{2}^{-2} \cr
0 & 0 & u_{2}^{-1} {\bf 1}_{d-3} \cr} \cr}}
They verify:
\eqn\clascoc{g_{10} g_{21} g_{02} = 1}
The anomaly two-form
\eqn\anomtf{{\psi}_{012} = {\tr} g_{02} {\dd} g_{10} \wedge {\dd} g_{21} = 
d \cdot {\dd} {\log}\ u_{1} \wedge {\dd} {\log} \ u_2 }
It is well-defined on $U_{012} = U_{0} \cap U_{1} \cap U_{2} \approx {\bC}^{*} \times {\bC}^{*} \times {\bC}^{d-3}$. Since it has a non-trivial period over the non-contractible two-cycle ${\Sigma} \approx {\bT}^{2}$
in ${\bC}^{*} \times {\bC}^{*}$ it represents a non-trivial cohomology class. In {\cech} language, it means that it cannot be represented as the sum:
 $$
 b_{01} + b_{12} +  b_{20} 
 $$
 of closed holomorphic two-forms, well-defined on $U_{01}$, $U_{12}$ and $U_{20}$ respectively. 
 Indeed, since these forms would be regular on the domains which have the topology
 ${\bC}^{*} \times {\bC}^{d-2}$, the corresponding integrals over the ${\Sigma}$ of each of these forms
 would vanish, in contrast with what we said about the integral of the form \anomtf.
  
The factor $d$ in \anomtf\ corresponds precisely to the similar factor in the second Chern character of the tangent bundle of ${\bC\bP}^{d-1}$:
\eqn\chrn{{\ch}_{2}\left( {\CT}_{{\bC\bP}^{d-1}}\right) = {d \over 2} \cdot H^2 , \quad H = c_1 ({\CO}(1)),} 
in agreement with the general theory \malikov.
  
\sssec{\rm A \ warning.}\ In the ${\bC\bP}^{d-1}$ example the three-forms ${\tr} ( g_{\b\a} {\dd}g_{\a\b} )^3$ vanish:
\eqn\thfrmvnsh{\eqalign{ & 
g_{10} {\dd} g_{01} =   
\pmatrix{ 2 {\dd}  {\log} u_{1} & {\dd} (u_{1} u_{a} ) u_{1}^{-2}  \cr
0 & {\dd}  {\log} u_1 \ {\bf 1}_{d-2} \cr} \cr
& \cr
& \qquad {\tr} ( g_{10} {\dd} g_{01} )^{3} = 0 \cr}}
yet the anomaly is alive.

\subsubsec{The Chern anomaly}
  
To check the $c_1$ anomaly, the one which affects the stress-energy tensor, 
it is sufficient to study the transition functions which we already 
 listed in \gcpn,\gmtr\ and \thfrmvnsh: 
\eqn\firstchern{\eqalign{& A_{01} = {\tr} \  g_{01}^{-1} {\dd} g_{01} = d \cdot {\dd} {\log} u_1 \cr
& A_{12} = {\tr} \ g_{12}^{-1} {\dd} g_{12} = d 
\cdot {\dd} {\log} (u_2 u_1^{-1} ) \cr
& A_{20} = {\tr} \ g_{20}^{-1} {\dd} g_{20} = - d\cdot  {\dd} {\log} u_{2}  \cr}}
Accordingly, the naive stress-tensors of the local theories are not compatible:
\eqn\strsstens{\eqalign{& T_{[1]}^{\rm naive} = T_{[0]}^{\rm naive} - {d\over 2} {\p}^{2} {\log}( u_{1} )\cr
& T_{[2]}^{\rm naive} = T_{[1]}^{\rm naive} -{d\over 2} {\p}^{2} {\log} ( u_{2}u_{1}^{-1} ) \cr}}

Notice that $d$ in the formulae \firstchern\strsstens\ corresponds to the first Chern class of the tangent bundle of ${\bC\bP}^{d-1}$: 
\eqn\fircpn{ch_{1}({\CT}_{{\bC\bP}^{d-1}}) = d\cdot  H}
In \fln\ a possible improvement of the naive stress tensor will be discussed. 

\subsubsec{The ${\rm PGL}_{d}$ symmetry}
  
Classically, the PGL$_{d}$ symmetry is generated by:
\eqn\pgln{\eqalign{& N_l = v_l , \qquad\qquad\qquad\cr
& N_{l}^{m} = v_{l} u_{m} , \qquad\qquad l,m = 1, \ldots , d-1\cr
& N^{l} = u_{l} \sum_{m} v_{m} u_{m} ,\qquad \cr}} 
The coordinate change \cpn\ acts on the classical currents \pgln\ 
as follows:
\eqn\pglnt{\eqalign{& {\tilde N}_{1} = - N^{1} , \qquad {\tilde N}_{I} = N_{I}^{1} \cr
 {\tilde N}_{1}^{1} =  - \sum_{m} N_{m}^{m} , & \quad {\tilde N}_{1}^{I} = - N^{I} , \quad {\tilde N}_{I}^{1}  = N_{I} , \quad {\tilde N}_{I}^{J} = N_{I}^{J} \cr
& {\tilde N}^{1} = - N_{1} , \quad {\tilde N}^{I} = - N_{1}^{I} \cr
& \qquad I, J  = 2, \ldots , d-1 \cr}}
Now let us see what happens quantum mechanically. 
Already from our failure to glue the $v$-fields globally over $X$ we know we will not be able to
define the PGL$_{d}$ currents \pgln\ globally over $X$ (unless $d=2$, where the Pontryagin anomaly is
irrelevant). But in fact even locally, on the coordinate charts $U_{\a}$ the currents
\pgln\ cannot be promoted to the current algebra. The subalgebra ${\bB}_{d}$, generated by
$N_l$ and $N_{l}^{m}$, makes sense quantum mechanically. In fact, on $U_{0}$,
we have the level $k = -1$ current algebra of ${\widehat {\bf gl}}_{d-1}$, generated by $N_{l}^{m}$.
This algebra is extended by the abelian current algebra generated by 
$N_l$'s. 
The trouble comes when 
we try to adjoin the $N^l$ generators. Indeed, it is not hard to show, by examining the
behavior near $u = 0$, that unless $d=2$ the equations  \curntalg\centrt\ have no nonsingular solutions. 
On the other hand, globally on ${\bC\bP}^{d-1}$, as the transformations \pglnt\ show, one cannot
restrict to the ${\widehat{\bf gl}}_{d-1}$ subalgebra, or its extension by the abelian subalgebra.

\sssec{\rm Another \ warning.} There exist examples where the target space $X$ has vanishing 
first Pontryagin class, admits the realization of the global holomorphic vector fields in the chiral algebra, yet the sigma model is anomalous due to Chern anomaly. The most famous such example is the generalization of the ${\bC\bP}^1$ sigma model. There, one takes $X = G/B$, the space of complete flags for the group $G$. As shown by B.~Feigin and E.~Frenkel in \feiginfrenkel,\edik\
the current algebra ${\hat \bg}$ at the critical level is realized in the curved beta-gamma system on this manifold. Yet, $c_{1}(X) \neq 0$ for all these spaces. 

\subsec{The fibered targets}

In this section we consider the sigma model on the space $X$ which is the total spaces of the fiber bundles:
\eqn\fbrbndl{\eqalign{F \longrightarrow \quad & X \cr & \downarrow \cr & B \cr}}  
where all spaces are complex manifolds, and the maps are holomorphic. We shall only consider
two cases: $F = {\bC}$ and $F = {\bC}^{*}$. In the former case we assume that $X$ is the total space of the line bundle, while in the latter $X$ is the principal ${\bC}^{*}$-bundle. It is not difficult to generalize to
more general cases, but in our applications these are the only two situations we shall need.

We are interested in the
$c_1$ and $p_1$ anomalies of the sigma model on $X$ and their relation to the anomalies of the sigma model on $B$. Of course, the characteristic classes of $X$ and $B$ are easily related. However, since
$X$ is non-compact, it is safer to perform the explicit calculation of the anomaly two-forms and
to check whether they represent non-trivial elements in $H^{2} ( X, {\CZ}_{X}^{2})$ or not, and similarly 
for $c_{1}(X)$. 

\subsubsec{The line bundle}

 Here we assume $F = {\bC}$. 
Suppose $B$ is covered with the coordinate patches $U_{\a}$. Let $u_{\a} = (  u_{\a}^{i} )$ be the coordinates
on $U_{\a}$, $i = 1, \ldots {\rm dim}_{\bC}B$. Then $X$ can be covered
with the coordinate patches ${\tilde U}_{\a} = {\bC} \times U_{\a}$, with the coordinates
$({\g}_{\a} , u_{\a}^{i})$, . If $u_{\b} = g_{\a\b} ( u_{\a} )$
is the transition function relating the coordinates on $U_{\a}$ and on    $U_{\b}$ over the
overlap $U_{\a\b} = U_{\a} \cap U_{\b}$, $g_{\b\a} \circ g_{\a\b} = id$, then the gluing on ${\tilde U}_{\a\b}$ is achieved with the
help of the transition function 
\eqn\ccntr{{\g}_{\b} = {\g}_{\a} {\chi}_{\a\b} ( u_{\a} ), \qquad u_{\b} = g_{\a\b} ( u_{\a} )}
where ${\chi}_{\a\b}$ is a holomorphic  map  from $U_{\a\b}$ to ${\bC}^{*}$. 
On the triple overlaps we should have the cocycle condition:
\eqn\coccnt{\eqalign{&  {\chi}_{\g\a} ( g_{\b\g} \circ g_{\a\b} ( u_{\a} )) {\chi}_{\b\g} ( g_{\a\b} (u_{\a})){\chi}_{\a\b} (u_{\a}) = 1 \cr
& g_{\g\a} \circ g_{\b\g} \circ g_{\a\b} ( u_{\a} ) \equiv g_{\g\a} \left( g_{\b\g} \left( g_{\a\b} \left( u_{\a} \right)\right)\right) = u_{\a} \cr}}
We can now relate the anomalies of the sigma model on $B$ and those of the sigma model
on the total space of the line bundle $X \to B$, defined using the gluing rules
above.

To this end we need an expression for the jacobian of the transformation \coccnt\ and
its inverse (we skip the indices $\a\b$, and also use $u$ for $u_{\a}$ and ${\tu} = g ( u)$ for $u_{\b}$):
\eqn\jac{{\tilde{\bf g}} = \pmatrix{ {\chi} (u) & {\g} {{\p} {\chi} \over {\p}u^{i}} \cr
& \cr
0 & {\tilde g}^{a}_{i}\cr}}
\eqn\invjac{{\bf g} = \pmatrix{ {\chi}^{-1} & - {\g} g^{i}_{a} {\p}_{i} {\log}{\chi} \cr & \cr
0 &  g^{i}_{a} \cr}}
where, as usual:
\eqn\omg{{\tilde g}^{a}_{i} = \left[ {{\p}{\tu} \over {\p} u} \right]^{a}_{i} , \qquad 
g^{a}_{i} = \left[ {{\p}{u} \over {\p} {\tu}} \right]^{i}_{a} } 
We have:
\eqn\connof{{\bf g}^{-1} {\dd} {\bf g} = \pmatrix{ - {\dd}{\log}{\chi} & - {\chi} g^{i}_{a} {\dd} ( {\g} {\p}_{i}{\log}{\chi} ) \cr
 & \cr
0 & g^{-1} {\dd}g \cr}}
It follows, that
the anomaly two-form for  $X$ and $B$ are related by:
\eqn\anomtftb{{\psi}_{\a\b\g}^{X} = {\psi}_{\a\b\g}^{B} + 
{\dd} {\log}{\chi}_{\a\b} \wedge {\dd}{\log} {\chi}_{\a\g}}
In \anomtftb\ we actually mean by ${\psi}^{B}$ the pull-back on $X$ of the corresponding two-form on $U_{\a\b} \subset B$, and the same is understood  below.

\subsubsec{Non-anomalous local Calabi-Yau}

It might happen that  the anomaly two form \anomtftb\ represents the coboundary, i.e. the exact cocycle.
For example, this is the case for $B$ which is the degree $k$ hypersurface in ${\bC\bP}^{2k-2}$ and $X$ -- the total space of the line bundle ${\CO}( 1 - k )$.  Moreover, $X$ in this case is also a non-compact Calabi-Yau manifold, so both Chern and Pontryagin anomalies vanish. 

Let $(x_{0} : x_{1} : \ldots : x_{2k-2} )$ denote the homogeneous coordinates on ${\bC\bP}^{2k-1}$
and let ${\CF} ( x_0 , x_1 , \ldots , x_{2k-2})$ be the homogeneous degree $k$ polynomial defining $B$:
\eqn\homo{\sum_{i=0}^{2k-1} x_{i} {{\p}{\CF} \over  {\p}x_{i}} =  k \ F }
In order for $B$ to be smooth, the equations ${\CF}(x) = 0$, ${{\p}{\CF} \over  {\p}x_{i}} = 0$ must have $x = 0$
as the only solution.   Let $E = \sum_{i} x_{i} {{\p}\over  {\p}x_{i}}$ denote the Euler vector field. The equation 
\homo\ can be written more compactly as:
\eqn\homoe{{\CL}_{E} {\CF} = k \ {\CF}}
Let ${\g}$ denote the coordinate along the fiber of the line bundle ${\CO}(1-k)$ over ${\bC\bP}^{2k-2}$, restricted on $B$. We can think of the total space of the line bundle ${\CO}(1-k)$ over ${\bC\bP}^{2k-2}$
as of the quotient of ${\bC}^{2k} = {\bC}_{\g} \times {\bC}^{2k-1}_{x}$ (with the locus $x =0$ deleted)  by the action of ${\bC}^{*}$:
\eqn\uoneac{( {\g} , x_{0} , x_{1} , \ldots , x_{2k-2}) \mapsto ( t^{1-k} {\g} , t x_{0} , t x_{1} , \ldots  , tx_{2k-2})} 
This action is generated by the vector field
\eqn\geneu{{\bf e} = E +  ( 1- k ) {\g}{{\p} \over {\p}{\g}}}
The following $2k-1$ form:
$$
 \left( {\dd}{\g} \wedge {{\dd}x_{0} \wedge {\dd}x_{1} \wedge \ldots \wedge {\dd}x_{2k-2} \over {\dd}{\CF}} \right)
 $$
 is well-defined on the locus $\{ {\CF} = 0 \} \subset {\bC}^{2k}$, and is ${\bf e}$-invariant, while the $2k-2$ form 
\eqn\holof{{\Omega} = \iota_{{\bf e}} \left( {\dd}{\g} \wedge {{\dd}x_{0} \wedge {\dd}x_{1} \wedge \ldots \wedge {\dd}x_{2k-2} \over {\dd}{\CF}} \right)}
is well-defined on $X$ and is nowhere vanishing. 
The Chern character of the tangent bundle ${\CT}_{X}$ can be formally calculated using the
exact sequences of bundles 
\eqn\excs{\eqalign{& 0 \to {\CO}_{X} \to \left(  {\CO}(1)^{{\oplus} 2k-1} \oplus {\CO}( 1 - k ) \right) \vert_{X} \to T \to 0 \cr
& \qquad 0 \to {\CT}_{X} \to T \to {\CO}(k)\vert_{X} \to 0 \ , \cr}}
thus:
\eqn\chrntx{{\ch} ( {\CT}_{X} ) = 2 k  - 2 + {k ( k-1)(2k-1) \over 3!} H^{3} + \ldots }
where $H = c_{1}({\CO}(1))$, the hyperplane class. For $k=2$ the target $X = {\CT}^{*}{\bC\bP}^{1}$ - the local ${\bf K3}$ manifold, which can be studied using toric methods \fln. 
\subsubsec{Principal ${\bC}^{*}$-bundles} 

Now consider the case $F = {\bC}^{*}$. In this case the analysis is similar to that of the line bundle, except that now we may use
${\varphi} = {\log} {\g}$ as local coordinates on the fibers $F$, and the coordinate change \ccntr\ becomes:
\eqn\ccntrp{{\varphi}_{\b} = {\varphi}_{\a} + {\log}{\chi}_{\a\b}(u)
 \ . }
 Accordingly, the Jacobian \jac\ and its inverse \invjac\ are relaced by the simpler ones:
\eqn\jacp{{\tilde{\bf g}} = \pmatrix{ 1 &  {{\p} {\log}{\chi} \over {\p}u^{i}} \cr
& \cr
0 & {\tilde g}^{a}_{i}\cr}}
\eqn\invjacp{{\bf g} = \pmatrix{1 & -  g^{i}_{a} {\p}_{i} {\log}{\chi} \cr & \cr
0 &  g^{i}_{a} \cr}}
and the anomaly two-forms for $X$ and $B$ coincide.

\subsubsec{Green-Schwarz mechanism for principal bundles}
Suppose \eqn\fctr{p_{1}(B) = - {1\over 4{\pi}^{2}} F_{1} \cap F_2 \ , } 
where $F_{1} , F_{2} \in H^{2} (B, 2{\pi}  i {\bZ})$. Moreover, let us assume that there exist two holomorphic line bundles $L_1$, $L_2$, such that $\left[ {F_j \over 2\pi i } \right] = c_1 (L_j)$, $j = 1,2$. Playing the tic-tac-toe
game we can conclude that up to {\cech} coboundaries 
\eqn\pofac{ {\tr} \left( g_{\a\b} {\dd} g_{\b\g} \wedge {\dd} g_{\g\a} \right) = {\dd} {\log} {\chi}_{1 , \a\b} \wedge {\dd} {\log} {\chi}_{2, \b\g}}
Now consider the total space $X$ of the principal ${\bC}^{*}$-bundle over $B$, such that the associated line
bundle is isomorphic to $L_1$ (everything works also if we replace $L_1$ by $L_2$). 
Using \connof\ we get:
\eqn\conbf{\eqalign{ {\psi}^{X}_{\a\b\g} & = {\dd}{\log}{\chi}_{1,\a\b} \wedge {\dd} {\log}{\chi}_{2, \b\g} + 
{\dd} {\log}{\chi}_{1,\a\b} \wedge {\dd} {\log} {\chi}_{1,\b\g} \cr
 \qquad\qquad & =(  {\d}  {\log}{\g} \wedge {\log}(  {\chi}_{1} {\chi}_{2} ))_{\a\b\g} \cr
 \qquad\qquad {\g}_{\b} & = {\g}_{\b}\  {\chi}_{1,\a\b} (u) \cr}}
 We can state the result in a more gauge-theoretic language. The 
starting point \fctr\ means that the first Pontryagin class of $B$ can be expressed 
 as the product of the curvatures of two $U(1)$ gauge fields. On the total space
 of any of the corresponding $U(1)$-bundles
the pull-back of the curvature is exact, being ${\dd}$ of the corresponding connection one-form:
\eqn\crvtcn{p_{i}^{*}F_{i} = {\dd} {\t}_{i}, \ i = 1,2 , \qquad p_{i} : L_{i} \to B}
Thus
\eqn\fctri{- 4{\pi}^{2} p_{1}(B) + F_{1} \wedge F_{1} = {\dd} \left( {\t}_{1} \wedge ( F_{1} + F_{2} ) \right) }
Note the similarity of the mechanism of the anomaly cancellation to that of Green-Schwarz in ten dimenions  \gsan

\subsubsec{${\bC}^{*}$-bundles over ${\bC\bP}^{d-1}$ -- anomalies cancelled and symmetry restored}

We now consider the ${\bC}^{*}$-cone over the projective space ${\bC\bP}^{d-1}$. The sigma model
will be described by the fields $u^{l}$ and ${\varphi}$ representing the local coordinates on the 
projective space and on the ${\bC}^{*}$ fiber respectively.
Topologically, the cones over ${\bC\bP}^{d-1}$ are classified by an integer $s$, the first Chern class
of the associated line bundle. The total space of the ${\bC}^{*}$ bundle can be covered by $d$
coordinate patches. Each of them looks like ${\bC}^{*} \times {\bC}^{d-1}$. Let us describe the typical coordinate transformation relating these patches:
\eqn\coordcone{\eqalign{ & {\tilde u}_{1} = u_{1}^{-1} \cr
& {\tu}_{a} = u_{a}  u_{1}^{-1} , \qquad  \ a = 2, \ldots , d-1\cr
& {\tilde\varphi} = {\varphi} + s \ {\log} u_{1} \cr}}
The corresponding momenta are $v_{l}$ and $p$.

As we learned in the previous subsection, the anomaly two-form for $X$ is that one for ${\bC\bP}^{d-1}$, i.e.
\eqn\anomcp{{\psi}_{012}^{X} = d \cdot {\dd} {\log}u_{1} \wedge {\dd} {\log} u_{2}}
on the coordinate patch ${\tilde U}_{012} = {\bC}^{*} \times U_{012}$ where $u_{1} \neq 0 $ and $u_{2} \neq 0$. Now, however, in variance with the situation for $X = {\bC\bP}^{d-1}$, the form \anomcp\ represents a coboundary. Indeed, recall that
on ${\tilde U}_{01}$ we have ${\varphi}_{1} = {\varphi}_{0} + s \ {\log}u_{1}$, on ${\tilde U}_{12}$: 
${\varphi}_{2} = {\varphi}_{1} + s \ {\log} ( u_{2} u _{1}^{-1} )$ and on ${\tilde U}_{02}$: 
${\varphi}_{0} = {\varphi}_{2} + s \ {\log} u_{2}^{-1}$. Then:
 \eqn\anomcn{{\psi}^{X}_{012} = -  {d\over s} \cdot \left( {\dd} {\varphi}_{1} \wedge {\dd} {\log} u_{1} +
 {\dd} {\varphi}_{2} \wedge {\dd} {\log} ( u_{2} u_{1}^{-1} ) + {\dd} {\varphi}_{0} \wedge {\dd} {\log} u_{2}^{-1} \
 \right)}
We can read off \anomcn\ the corresponding ${\m}_{\a\b}$ forms, and get the transformation properties of the $v_{l}, p$ fields. Not surprisingly, the fields $v_l$ transform into the currents which form the ${\widehat{\bf sl}_{d}}$ algebra, which include $p, {\varphi}$ fields. In addition we get another ${\bC}^{*}$-symmetry
which rotates the fiber. 
\eqn\pgltns{\eqalign{& N_{l} = v_l \cr
& N_{l}^{m} = v_{l} u_{m}  -  {\d}_{lm} \left( {s\over d} p - {1 \over 2 s} {\p}{\varphi} \right) \cr
& N^{l} = {\p}u_{l} + u_{l} \left( s p - {d\over 2 s} {\p}{\varphi} \right)  - \sum_{m} v_{m} u_{m}u_{l}\cr
& N = - \sum_{m} N_{m}^{m} \cr
& J = p +  {d \over 2 s^{2}}  {\p}{\varphi} \cr}}
The currents
\eqn\pltnmo{{\tilde N}_{l}^{m} = N_{l}^{m} + {1\over d-1} {\d}_{l}^{m} N }
form a closed subalgebra, isomorphic to ${\widehat {\bf sl}_{d-1}}$, of level $-1$, while
the currents \pgltns\ form the algebra $\widehat{{\bf sl}_{d}}$ of level $-1$ and ${\hat {\bf u(1)}}$ of level
$-d/s^2$.
The resulting current algebra realization \pgltns\ can be mapped to the construction in \feiginfrenkel\ of the chiral algebras associated with more
general cosets $G/H$ and line bundles over them.

The total space $X$ of the principal ${\bC}^{*}$-bundle over ${\bC\bP}^{d-1}$ has a holomorphic top
degree form
\eqn\tpfrm{{\Omega} = {\exp} \left( {d\over s}{\varphi}\right)  {\dd}{\varphi} \wedge {\dd} u_{1} \wedge \ldots {\dd} u_{d-1}}
Accordingly, the stress-energy tensor
\eqn\strsscpd{T = v_{l} {\p} u^{l} + p {\p}{\varphi}  + {d \over 2 s} {\p}^{2} {\varphi}}
has the correction term with ${\p}^2 {\varphi}$.
Note that $T$ has the Sugawara form:
\eqn\sugawaracp{T = - {1\over 2 ( d-1) } 
\left( N_{l} N^{l} + N^{l} N_{l} + N_{l}^{m} N_{m}^{l} + N N \right) + {s^2 \over 2d} J J + {s\over 2} {\p} J}
and that the $U(1)$ charge corresponding to the $J$ current has an anomaly
$$
{d\over s} (g -1) = q (g-1)
$$
on genus $g$ Riemann surface, as can be deduced, among other things, from the operator product expansion:
\eqn\tj{J(y) T(z) \sim {d/s \over ( z- y)^3 } - {1\over ( z - y)^2 } J(z) }
or, more invariantly, from \currnts.

\sssec{\rm A \ simple \ topological \ argument.} Note that in our example the absence of Pontryagin and Chern anomalies
is easy to understand. The total space $X$ is homotopy equivalent to the lens space
${\bf S}^{2d-1}/{\bZ}_{|s|}$. Its rational cohomology is trivial in even degrees, hence there is no room for
$c_1$ or $p_1$. 

\newsec{The pure spinor sigma model}

\hfill\vbox{\hbox{\cyr "Gore ot uma"}
\hbox{\qquad \smallcyr A.Griboedov}
\hbox{"Wit Works Woe"}
\hbox{\qquad $\textstyle A.Griboedov$}}

We now come to the main application of the general theory above, which was in fact our motivation for
the whole endeavour. 

\subsec{Motivation: the covariant superstring quantization}

Superstrings are the basis of our belief in the consistency of string theory, the theory unifying
quantum gravity and all other interactions. The perturbative string is defined as a  two dimensional (super)conformal field theory (e.g. a sigma model) coupled to the two dimensional (super)gravity. The sum over topologies
of the two dimensional manifolds -- worldsheets -- is interpreted as the string loop expansion. 
The fermionic counterpart of the sum over topologies, the sum over spin structures in the Neveu-Schwarz-Ramond (NSR) formulation of the superstring, leads to the GSO projection and the space-time supersymmetry. In the Green-Schwarz approach \gsw, where the sigma model taking values in the supermanifold is coupled to the ordinary two dimensional gravity, the target-space supersymmetry is manifest, but the sigma model is very hard to quantize due to its non-linear nature. In the NSR
approach the worldsheet sigma model is represented via free fields (for Minkowski background, and in the conformal gauge), but the space-time supersymmetry is not manifest. Also, the NSR approach
becomes infinitely complicated for non-trivial Ramond-Ramond backgrounds, where spin fields
for the worldsheet fermions must be exponentiated.

All these difficulties led to the long search for a better formulation of the perturbative theory. Five years ago such a formulation has been proposed by N.~Berkovits, who suggested to use the twistor-like
description of the GS sigma model. In his formulation the worldsheet sigma model
had manifest target space supersymmetry yet it is realized using essentially free fields.
The story is not finished yet, since the fully covariant formulation is not yet known, but for most of practical purposes Berkovits' program is fully operational.

The goal of this section is to raise some concerns about the last statement, and then to eliminate them,
at least when certain assumptions are made. 

In Berkovits' approach, the superstring on flat ten dimensional Minkowski background is described by the following sigma model:
\eqn\btdsgm{\int {\half} {\p} x^{m} {\pb} x^m + p_{\a} {\pb} {\t}^{\a} + {w}_{\a} {\pb} {\l}^{\a}}
where we only write the right-movers (holomorphic sector) for the first order fields. 

The fields $x^m$ are the standard free bosons describing ${\bR}^{10}$, $m=1, \ldots , 10$, the
fields $p_{\a}, {\t}^{\a}$ form the fermionic system of fields of spins $1$ and $0$ respectively,
they transform as the sixteen component Weyl spinors in target space (of opposite chirality). 
Note that in euclidean signature ${\t}^{\a}$ is a complex fermion, but we don't have its complex conjugate. Finally, the most interesting part of the worldsheet theory is the curved $\b\g$ system, represented by the $w_{\a} {\pb} {\l}^{\a}$ term in \btdsgm. The field ${\l}^{\a}$, of spin $0$, takes values in the space
$X$ of the so-called {\it pure spinors} \cartan\ for $SO(10)$. 
These are simply bosonic variables $ {\l} = ( {\l}^{\a} )$, ${\a} = 1, \ldots , 16$,  which obey the following equations:
\eqn\prsp{X = \{ {\l} \ \vert \ {\l} {\g}^{m} {\l} \equiv {\l}^{\a} {\l}^{\b} {\g}^{m}_{\a\b}  =\ 0 , \qquad m = 1, \ldots, 10 \} \ }
The space of solutions to \prsp\ is the cone over the space ${\tilde Q}_{10}$ of projective pure spinors, which is the space of solutions to \prsp\ with the trivial solution ${\l}=0$ deleted, considered up to the ${\bC}^{*}$ rescaling. It is the classical result (which we remind in the next section) that 
\eqn\cos{{\tilde Q}_{10} = SO(10)/U(5)}
In IIA string the left-moving sector would involve the similar fields of the opposite chirality, ${\tilde p}_{\dot \a} , {\tilde\t}^{\dot \a}, {\tilde w}_{\dot \a}, {\l}^{\dot \a}$, in IIB string the chirality of the left movers is the same, and in the heterotic string the left movers are represented in the standard way. 

The action \btdsgm\ is written in the conformal gauge. The prescription for calculation of 
string amplitudes \berkovitsbghost,\berkovitsmulti\ involves a proper definition of the physical states and the $b$-ghost. This is done
using the remarkable nilpotent BRST-like operator $\CQ$:
\eqn\qop{{\CQ} = \oint {\l}^{\a} d_{\a}}
where 
\eqn\dcon{d_{\a} = p_{\a} + ({\g}_{m} {\t})_{\a} {\p}x^{m} + {\textstyle{1\over 2}} ({\g}_{m}{\t})_{\a} ( {\t}{\g}^{m}{\p}{\t})}
The stress-energy tensor
\eqn\stt{T = {\p} x^m {\p} x^m + p_{\a} {\p} {\t}^{\a} + w_{\a} {\p} {\l}^{\a}}
(this formula needs clarification for the $w-{\l}$ part, and we shall make it very explicit in the coming sections) is $\CQ$-exact:
\eqn\sttb{T = \{ {\CQ} , G \} }
where the spin $2$ fermionic field $G$ is not defined globally on $X$. It is defined in patch by patch,
and difference of two expressions on the overlap of two coordinate patches is $\CQ$-exact \berkovitsbghost, \berkovitsnonmin.

\subsec{Pure spinors: a reminder}

In this section the letter $d$ denotes the complex dimension of the Euclidean vector space. The relevant
target spaces will have complex dimensions like ${\half} d(d-1) +1$. We hope this will not lead to any confusion. 

\subsubsec{Cartan and Chevalley definitions, complex structures etc.}

The $SO(2d)$ pure spinor \cartan\ $\l^\a$ is constrained
to satisfy 
\eqn\crtps{\l^\a (\s^{m_1 ..  m_j})_{\a\b} \lambda^\b =0 , \qquad  {\rm for} \qquad 0\leq j< d \ ,}
where $m=1$ to $2D$, $\a=1$ to $2^{d-1}$, and $\s^{m_1 ...m_j}_{\a\b}$ is the
antisymmetrized
product of $j$ Pauli matrices. This implies that
$\l^\a \l^\b$ can be written as
\eqn\pures{\l^\a \l^\b =
{{1}\over{n! ~2^d}}\s_{m_1 ...m_d}^{\a\b} ~(\l^\g \s^{m_1 ... m_d}_{\g\d}
\l^\d)} where
$\l\s^{m_1 ... m_d} \l$ defines an $d$-dimensional complex plane ${\bC}^{d} \subset {\bR}^{2d} \otimes {\bC}$.
This
$d$-dimensional complex plane is preserved  by a $U(d)$
subgroup of $SO(2d)$
rotations. Also, multiplying ${\l}$ by a non-zero complex number does not change this plane. So if we consider the space of ${\l}$'s obeying \crtps\ up to rescalings, the space of {\it projective pure spinors} ${\tilde Q}_{2d}$ in $D=2d$ Euclidean dimensions, then:
\eqn\projps{{\tilde Q}_{2d} = SO(2d)/U(d)}
The real dimension of this space is $d (d-1)$. The space $Q_{2d} \subset S_{2d}$ of pure spinors is a cone over ${\tilde Q}_{2d}$. 
The space $X_{2d}$, which is $Q_{2d}$ with the point ${\l} = 0$ deleted, can be thought of the moduli space of Calabi-Yau complex structures on ${\bR}^{2d}$, i.e. the space of pairs 
$$
( {\rm identification} \  {\bC}^{d} \approx {\bR}^{2d} , {\Omega} \in {\Lambda}^{d} {\bC}^{d} )
$$
This is an important space in the context of B type topological strings. 

\subsubsec{A little bit of geometry and topology}

For $d < 5$ the spaces $Q_{2d}$ are simple. For $d < 4$ they coincide with $S_{2d} = {\bC}^{2^{d-1}}$, for $d = 4$
$Q_{8}$ is a quadric hypersurface in $S_{8} = {\bC}^{8}$, a lightcone. For $d \geq 5$ it is not a complete
intersection, the number of defining equations \cartan\ being strictly greater then the codimension
of $Q_{2d}$ in $S_{2d}$.

The representation \projps\ shows that ${\tilde Q}_{2d}$ is actually a (co)adjoint orbit of
$SO(2d)$, and, in particular, is a compact K\"ahler manifold. We can parametrize ${\tilde Q}_{2d}$ by matrices of the form:
\eqn\matr{J = g^{-1} J_{0} g}
where
\eqn\jnought{J_0 = \pmatrix{ 0 & {\bf 1}_{d} \cr - {\bf 1}_{d} & 0}}
and $g \in SO(2d)$, $gg^{t} = {\bf 1}_{2d}$. Indeed, the matrices $g \in SO(2d)$ which commute
with $J_0$ belong precisely to $U(d)$. The group $SO(2d)$ acts on ${\tilde Q}_{2d}$ in a Hamiltonian fashion. The transformation: ${\d}g = g {\Phi}$, ${\Phi}^t = - {\Phi}$, is generated by the Hamiltonian
\eqn\genham{H_{\Phi} = {\tr} \left( g^{-1} J_{0} g\  {\Phi} \right)}
For generic ${\Phi}$ the corresponding $H_{\Phi}$ is a Morse function, i.e. it has non-degenerate
critical points.  There are precisely $2^{d-1}$ such points, and they are in one-to-one correspondence with the elements $w$ of the coset ${\CW}_{D_{d}}/{\CW}_{A_{d-1}} = {\bZ}_{2}^{d-1}$ of the Weyl group of $SO(2d)$
by that of $SU(d)$. Namely, the element $w = \left( \pm 1, \ldots , \pm 1 \right)$ (the total number of $\pm 1$'s is $d$ and their product is equal to $+1$), corresponds to the point \matr\ of the form
\eqn\crtpnt{J_{w} = \pmatrix{ 0 & w \cr - w & 0 }}
which is a critical point of $H_{\Phi}$ for
\eqn\diagph{{\Phi}  = \pmatrix{ 0 & {\phi}_{d} \cr - {\phi}_d & 0}}
with ${\phi}_{d} = \left( {\phi}_{1} , \ldots , {\phi}_{d} \right)$ (every $\Phi$ can be brought to this form by
the action of $SO(2d)$, the "eigen-values" ${\phi}_i$ are uniquely defined up to the action of ${\CW}_{D_{d}}$, i.e. up to permutations and the even number of sign flips).

The points $J_{w}$ are also in one-to-one correspondence with the components of an unconstrained chiral spinor in $2d$ dimensions. This is not a coincidence. In fact, by "quantizing" ${\tilde Q}_{2d}$
in the sense of geometric quantization, using the smallest possible multiple of the Kirillov-Kostant form as the symplectic form,  one gets precisely $S_{2d}$. The maximal torus of $SO(2d)$, $U(1)^{d}$, acts on $S$ and the eigenvectors, the weight subspaces, are precisely the components of the spinor. 
On the other hand, this action is obtained by quantizing $H_{\Phi}$. The critical points of $H_{\Phi}$, 
are the fixed points of $\Phi$ action on ${\tilde Q}_{2d}$. One can relate these by using the coadjoint orbit quantization \samsonco.

Now let us discuss the parameterization of ${\tilde Q}_{2d}$. Using \matr\ we can parametrize
the vicinity of each point $J_{w}$ by taking the appropriate components of $g$. Let us consider the neighborhood of $J_0$ for simplicity. Then, in a first approximation:
\eqn\gfap{g = {\bf 1}_{2d} + \pmatrix{ {\rm Re}u & {\rm Im}u \cr {\rm Im}u & - {\rm Re}u } + \ldots}
where $u$ is a complex antisymmetric $d\times d$ matrix:
\eqn\uma{u = \Vert u_{ab} \Vert_{a,b =1 , \ldots d}, \qquad u_{ab} = - u_{ba}}
which parametrizes the quotient ${\bf so}(2d)/{\bf u}(d)$ of Lie algebras. 
Moreover, the expansion of $H_{\Phi}$ near $J_0$ looks as follows:
\eqn\hfex{H_{\Phi} =   - 2\sum_a {\phi}_a + 2 \sum_{a < b} ( {\phi}_a  + {\phi}_b ) | u_{ab} |^2 + \ldots }
It is not difficult to show that near $J_{w}$ the expansion looks similar, with the only change 
${\phi}_a \mapsto w_{a} {\phi}_{a}$, where $w_{a}  = \pm 1$, $\prod_a w_{a} = 1$.
The significance of this result is twofold. First of all, it allows to set up the Morse complex for ${\tilde Q}_{2d}$. Indeed, Morse theory states that the cohomology of ${\tilde Q}_{2d}$ can be computed using the complex, whose generators are in one-to-one correspondence with the critical points $w$ of
any Morse function, $H_{\Phi}$ in particular, the degree of the generator being the index of the corresponding critical point, i.e.
the number of negative eigenvalues of ${\p}^2 H_{\Phi}$. In our case all degrees are even, and the
differential (which acts between the critical points whose indices differ by one) is trivial, so the cohomology is read off the critical points immediately. By ordering ${\phi}_1 > {\phi}_2 > \ldots > {\phi}_{d}$ we ensure that the point $J_0$ which corresponds to $w = ( +1 +1 \ldots +1)$ is the absolute minimum of $H_{\Phi}$, i.e. it corresponds to the degree zero cohomology. 
The point $w_2 = ( +1+1 \ldots +1 -1-1)$ is the only critical point of index $2$, and the point
$w_4 = ( +1+1 \ldots -1 +1 -1)$ is the only critical point of index $4$. Thus:
\eqn\lowcoh{H^{2i } ( {\tilde Q}_{2d} ) =  {\bZ}, \qquad i = 0,1,2}
This leads to the following important consequence, namely, whatever $p_1 ({\tilde Q}_{2d}) $ is, it is proportional to $c_{1}^{2} ( {\tilde Q}_{2d})$.

The coefficient of proportionality, which is not needed for our general argument, but might be useful in applications, can be calculated most simply using equivariant cohomology. The localized expression
for $p_1$ is:
\eqn\locpone{p_{1} = {\half} \sum_{a < b} ( {\phi}_a + {\phi}_b )^2  \sim {\half} ( \sum_a {\phi}_a )^2 }
while that for $c_1$:
\eqn\loccone{c_{1} = \sum_{a < b} ( {\phi}_a + {\phi}_b ) = ( d - 1) \sum_{a} {\phi}_a }
The logic behind the formula \locpone\ is that the ${\CW}_{D_{d}}$-invariant polynomials in ${\phi}_a$'s
represent trivial classes in the cohomology of ${\tilde Q}_{2d}$, while all the cohomology is generated by the characteristic classes of the $U(d)$ bundles associated with the principal bundle 
$$
SO(2d)   \to {\tilde Q}_{2d}
$$
These characteristic classes are the symmetric polynomials in $d$ variables ${\phi}_a$. 

For completeness, for $d=5$ the full cohomology of ${\tilde Q}_{10}$ is given by:
\eqn\cohtend{H^{2i} (X)  = H^{20-2i}(X) = {\bZ},\   i = 0,1,2 \qquad H^{2i} (X)  = H^{20 - 2i} (X) = {\bZ}^{2}, \qquad i = 3,4,5}

\subsubsec{The character of pure spinors}

One can also calculate the $c_1$ and $p_1$ classes of ${\tilde Q}_{2d}$ by using the character
of the algebra of polynomial functions on the space of 
pure spinors in $2d$ dimensions. It  was analyzed in \natnik, for example, by using the fixed point techniques of H.~Weyl: 
\eqn\chari{\eqalign{{\chi}_{2d} (t, g) = & \sum_{w} {1\over{1- t \prod_{a} e^{-{w_a{\phi}_{a} \over 2}}}} \prod_{a < b}
{1\over{1-  e^{w_{a}{\phi}_{a} + w_{b}{\phi}_{b}}}} \cr
& \cr
& w = ( w_{1} , w_{2} , \ldots , w_{d} ), \qquad
 w_{a} = \pm 1 , \prod_{a} w_{a} = 1 \cr}}
where
$$
{\chi}_{2d} ( t, g ) = {\tr}_{{\rm Fun}(X_{2d})} \left( t^{K} \ g \right)
$$
$g = {\exp} {\Phi}$ is the element of $SO(2d)$, and $K$ is the generator of ${\bC}^{*}$ which acts on $X_{2d}$ by rescaling of ${\l}$.

For $d=5$, in the limit ${\phi}_a \to 0$ the character reduces to 
\eqn\tend{{\chi}_{10} (t) = {( 1+t)(1+4 t +t^2) \over ( 1- t)^{11}}}
It can be expressed through the characteristic classes of the tangent bundle
${\CT}_{{\tilde Q}_{10}}$:
\eqn\tendi{{\chi}_{\bf 10}(t) = \int_{{\tilde Q}_{10}} {1\over{1 - t e^{c_{1}(L)}}} \ {\rm Td}({\CT}_{{\tilde Q}_{10}})}
where $L$ is the line bundle, associated with the principal ${\bC}^{*}$-bundle $X_{10} \to {\tilde Q}_{10}$.
Expanding both equations \tendi\ and \tend\ near $t = 1$ and equating the coefficients at the first
singular terms, we get:
\eqn\tendii{\eqalign{& {1\over{10!}} \int_{
{\tilde Q}_{10}} c_{1}(L)^{10} = 12 \cr
& ch_{1}({\CT}_{{\tilde Q}_{10}}) = - 8 c_{1}(L) \cr
& p_{1} ({\tilde Q}_{10}) = ch_{2} ({\CT}_{{\tilde Q}_{10}}) = 2 c_{1}(L)^{2} \cr}}
where in the last line we have used the fact that the cohomology of ${\tilde Q}_{10}$ is one-dimensional
in degrees $2$ and $4$. In order to relate \tendii\ and \locpone,\loccone, note that
$$
c_{1}(L) = - {\half} \sum_{a} {\phi}_{a} 
$$

\subsubsec{Coordinates on ${\tilde Q}_{2d}$}

The space ${\tilde Q}_{2d}$ can be covered with $2^{d-1}$ coordinate charts, $U_{w}$,
where $w$ are the $d$-tuples of $\pm 1$'s with the total number of minus signs being even. 
In each of these charts, the coordinates are $u_{ab}$, $1 \leq  a < b \leq d$, and 
$U_{w} \approx {\Lambda}^{2} {\bC}^{d}$. The space $X_{2d}$ is covered by the corresponding charts
${\tilde U}_{w} = {\bC}^{*} \times U_{w}$. The pure spinor ${\l}$ can be written in terms of the local coordinates ${\g} = {\exp} {\varphi}$ and $u_{ab}$ as:   
\eqn\decone{(\l^{d\over 2} = \g,\quad \l^{{d-4}\over 2}_{[ab]}
= \g u_{[ab]}, \quad
\l^{{{d-8}\over 2}}_{[abcd]} = -{1\over 8} \g u_{[ab} u_{cd]}, \quad
\l^{{{d-12}\over 2}}_{[abcdef]} = -{1\over {48}} \g u_{[ab} u_{cd} u_{ef]},
\quad  ... )}
where the superscript on $\l$ is the $U(1)$ charge,
$\g$ is an $SU(d)$ scalar with $U(1)$ charge $d\over 2$,
and $u_{ab}$ is an $SU(d)$ antisymmetric two-form with $U(1)$ charge $-2$.

\subsubsec{The warmup example: six dimensional pure spinors}

In a sense, the simplest pure spinor space is that one in six dimensions. The purity constraint
is vacuous, so one naively is dealing with the space $Q_{6} = S_{6}$ of all spinors, i.e. ${\bC}^4$.
Let ${\l}_1 , {\l}_2 , {\l}_3 , {\l}_4$ denote the coordinates on $S_{6} \approx {\bC}^4$. 
 The space of projective pure spinors is ${\bC\bP}^{3}$ so in order to prepare ourselves for more
complicated problems, 
we might want to treat $S$ as a cone over the space of projective pure spinors, i.e.
replace it by the total space ${\hat Q}_{6}$ of an appropriate line bundle ${\CO}(-1)$ over ${\bC\bP}^3$. 

This total space is known as a blowup of an origin.  This procedure removes the origin in ${\bC}^4$
and replaces it by a copy of ${\bC\bP}^3$. Here is an explicit coordinatization of the resulting space.
It can be covered by four coordinate patches $U_{\a}$, ${\a} = +++, +--, -+-, --+$.
The coordinate patch $U_{\a}$ corresponds to the region ${\l}_{\a} \neq 0$ on the original space
${\bC}^4$.  The latter region is isomorphic to ${\bC}^{*} \times {\bC}^3$. On the blown up space
$Q$ this region is partly compactified to ${\bC}^4$. Let ${\g}^{({\a})} , u_1^{({\a})} , u_2^{({\a})} , u_3^{({\a})}$ be the coordinates
on $U_{\a}$.   The coordinate ${\g}^{({\a})}$ is equal to ${\l}_{\a}$ while
the other three coordinates $u_{i}^{({\a})}$ are the ratios ${\l}_{\b}/{\l}_{\a}$, ${\b}\neq {\a}$.
The difference between $Q$ and ${\bC}^4$ is that these ratios are well-defined on $Q$.

The coordinate transformations, gluing $U_{\a}$ and $U_{\b}$ are easy to figure out.
Let us consider, for example, the transformations from $U_{+++}$ to $U_{--+}$:
\eqn\uonetwo{\eqalign{& {\g}^{(2)} = {\g}^{(1)} u_{1}^{(1)} \cr
& u_{1}^{(2)} = 1/u_{1}^{(1)} \cr
& u_{2,3}^{(2)} = u_{2,3}^{(1)}/u_{1}^{(1)} \cr}}
Note that the holomorphic top form 
$$ 
{\dd} {\g} \wedge {\dd} u_{1} \wedge {\dd} u_{2} \wedge {\dd} u_{3}
$$
is not preserved by these transformations. Instead, the form 
\eqn\hlfrm{{\Omega} = {\g}^{3} {\dd} {\rho} \wedge  {\dd} u_{1} \wedge {\dd} u_{2} \wedge {\dd} u_{3}
= p^{*} {\dd} {\l}_{+++}  \wedge {\dd} {\l}_{+--} \wedge {\dd} {\l}_{-+-} \wedge {\dd} {\l}_{--+} , }
where
\eqn\prjct{p: {\hat Q}_{6} \to S_{6}}
is the projection, 
is preserved. This already indicates that there is some difference between the 
sigma models on $S_{6}$ and $Q_{6}$. While the former is well-defined (up to the usual subtleties
with the integration over the non-compact zero modes which we shall address momentarily),
the latter suffers from Chern anomaly, and from Pontryagin anomaly as well.

However, if we remove the zero section, i.e. do not allow $\g$ to vanish, then both anomalies go away.
In fact, we get a particular case of a ${\bC}^{*}$ bundle over ${\bC\bP}^{d-1}$, for $d=4$, which we
already discussed.

\subsubsec{Blowup versus surgery}

Already this example indicates that blowing up the apex of the cone $Q_{2d}$ will not lead
to a consistent sigma model. Let us sketch the general situation. 

The holomorphic top degree form on ${\hat Q}_{2d}$, the total space of the appropriate 
line bundle over ${\tilde Q}_{2d}$ is given by (cf. \natser\natnik):
\eqn\holtop{{\Omega} = {\g}^{2d-3} {\dd} {\g} \wedge \bigwedge_{a< b} {\dd} u_{ab}}
On ${\hat Q}_{2d}$ the radial variable ${\g}$ is allowed to vanish, hence ${\Omega}$ will has a vanishing locus there, hence the improved stress-energy tensor
\eqn\imrp{T = {\half} v^{ab} {\p} u_{ab} + p {\p} {\log}{\g} + (d-1) {\p}^2 {\log}{\g}}
will have a singularity. It means that the sigma model on ${\hat Q}_{2d}$ suffers from Chern anomaly.

It is not difficult to show using our general theory of fibered targets that the Pontryagin anomaly is present there as well. 

Finally, the sigma model on ${\hat Q}_{2d}$ has instantons -- non-trivial holomorphic maps which land
at ${\g} = 0$. Their interpretation in the pure spinor approach to superstring quantization is bizzare, to say the least.

However, by removing the locus ${\g} = 0$, i.e. by deleting the point ${\l} = 0$ on $Q_{2d}$
we obtain the space $X_{2d}$ which is the total space of the ${\bC}^{*}$-bundle over ${\tilde Q}_{2d}$
with non-trivial first Chern class. This allows us to kill the $c_1$ and $p_1$ anomalies
(even  at the level of integral cohomology, actually). By the same token the worldsheet instantons also disappear.

\sssec{\rm A \ little\ bit \ of \ topology.} The cohomology of $X_{2d}$ can be calculated using Leray spectral sequence \griffitsharris, whose second term is $E_{2}^{p,q} = H^{p} ( {\tilde Q}_{2d} , H^{q} ({\bf S}^{1}))$ since the 
fiber is homotopy equivalent to the circle ${\bf S}^{1}$. The differential $d_2$ sends
$E_{2}^{p,q}$ to $E_{2}^{p+2, q-1}$. The analysis which leads to \conbf\ can be interpreted by saying that this differential is non-trivial at the term $E_{2}^{2,1} \to E_{2}^{4,0}$, and also at
$E_{2}^{4,1} \to E_{2}^{6,0}$. In particular, we can show that
for $d=5$:
\eqn\cohtot{H^{i} (X_{10}) = {\bZ}, \quad i = 0, 6, 15, 21}
and trivial otherwise. For general $d$ we can show that $H^{i} (X_{2d}) = 0$ for $i=1, \ldots , 5$.  

\subsec{The pure spinors in ten dimensions}

We shall now illustrate these statements by the physically most interesting example, $d=5$.

\subsubsec{Coordinates on the space of pure spinors}

 Choose some identification 
 \eqn\cmplxs{{\bR}^{10} \approx W = {\bC}^{5} \ .} 
 The space of $SO(10)$ spinors can be decomposed as:
 \eqn\spnrs{\eqalign{S = & \ S_{+} \oplus S_{-} , \qquad S_{+} \approx S_{-}^{*} \cr
  S_{+} = & \ L^{-\hlf}  \otimes {\Lambda}^{\rm even} W =  L^{-\hlf} \oplus L^{-\hlf} \otimes {\Lambda}^{2}W \oplus L^{\hlf} \otimes W^{*}\cr
  S_{-} = & \ L^{-\hlf} \otimes {\Lambda}^{\rm odd} W = L^{-\hlf} \otimes W \oplus L^{\hlf} \otimes {\Lambda}^{2}W^{*} \oplus L^{\hlf} \ , \cr}}
  where $L =  {\Lambda}^{5} W$ is the one-dimensional representation of the double cover of $U(5)$, which is
  the subgroup of $Spin(10)$, the double cover of $SO(10)$, preserving the identification \cmplxs. 
 The space of pure spinors in ten dimensions is the quadric in $S_{+}$:
 \eqn\pspsp{{\l} {\g}^{m} {\l} = 0 , \qquad {m} = 1, \ldots , 10}
 
 Then \pspsp\ can be rewritten as:
 \eqn\pspsi{\eqalign{ & {\ve}^{abcde} \left( {\l} {\l}_{abcd} + {\l}_{[ ab} {\l}_{cd]} \right) =0 , \qquad e = 1, \ldots , 5\cr
 & {\l}_{abcd} {\l}_{ef} {\ve}^{abcdf} = 0 , \qquad e = 1, \ldots , 5\cr}}
 where we decomposed the sixteen component spinor as
 \eqn\spinr{{\l} = \left( {\l} , {\l}_{ab} , {\l}_{abcd} \right)}
 according to the $U(5)$ decomposition \spnrs, i.e. ${\l} \in L^{-\hlf}$, ${\l}_{ab} \in L^{-\hlf}\otimes {\Lambda}^{2}W$, ${\l}_{abcd} \in L^{-\hlf} \otimes {\Lambda}^{4} W \approx L^{\hlf} \otimes W^{*}$.
 If ${\l} \neq 0$ then the second equation in \pspsi\ follows from the first, so the space of solutions to
 \pspsp\ is eleven (complex) dimensional, and not six dimensional, as naively one could have expected. 
 
 It is also convenient to use the "five-signs" notations, where the components of the $S_{+}$ spinor
 are labelled by the sequences of five plus or minus signs, with the restriction that the number of
 minus signs is even:
 \eqn\sns{\eqalign{& {\l} = {\l}_{+++++}\ ,   {\l}_{1234} = {\l}_{----+} \ ,
  \ldots \ ,
  {\l}_{2345} = {\l}_{+----} \cr
 &  {\l}_{12} = {\l}_{--+++}\ , 
  {\l}_{13} = {\l}_{-+-++}\ ,
  \ldots \ ,  {\l}_{45} = {\l}_{+++--}\cr
 }}
 We  now discuss the coordinatization of the pure spinor space. We have sixteen coordinate patches, 
 which are characterized by the non-vanishing of one of the sixteen components of the spinor ${\l}_{\a}$. 
 In the first one, we shall call it $U_{+++++}$ the components of the pure spinor are
 parameterized via:
 \eqn\ppppp{\eqalign{& {\l}_{+++++} = {\l} \cr
 & {\l}_{+++--} = {\l} u_{45} \cr
 & \ldots \cr
 & {\l}_{--+++} = {\l} u_{12} \cr
 & {\l}_{+----} = {\l} \left( u_{23} u_{45} - u_{24} u_{35} + u_{25} u_{34} \right) \cr
 & \ldots \cr
 & {\l}_{----+} = {\l} \left( u_{12} u_{34} - u_{13} u_{24} + u_{14} u_{23} \right) \cr}}
 \bigskip
 
 \subsubsec{Coordinate transformations on the pure spinor space}
 
 The space of pure spinors (with the apex of the cone blown up or removed) can be covered by sixteen coordinate
 patches, which are in one-to-one correspondence with the critical points of the Hamiltonian \hfex. 
 Let us discuss the coordinate transformation which occurs on the overlap 
 $U_{+++++}$ and $U_{+++--}$. It is straightforward to calculate:
 \def\tu{{\tilde u}}
 \def\tp{{\tilde p}}
 \def\tv{{\tilde v}}
 \def\tf{{\tilde\varphi}}
 \def\tN{{\tilde N}}
 \eqn\chvar{\eqalign{{\tu}_{ij} = u_{ij} + \left( u_{i5}u_{j4} - u_{i4}u_{j5} \right)/u_{45} &\qquad  i,j = 1,2,3 \cr
 {\tu}_{i5} = u_{i4}/u_{45} ,  &\qquad  {\tu}_{i4} = u_{i5}/u_{45}  \cr
 {\tf} = {\varphi} + {\log} u_{45},   &\qquad  {\tu}_{45} = 1/u_{45} \ \cr }}
Let us now discuss the coordinate transformation which occurs on the overlap 
 $U_{+++++}$ and $U_{+----}$. 
 Let us instroduce the notation:
 \eqn\chardelta{{\chi} = {1\over u_{25}u_{34} - u_{24}u_{35} + u_{23}u_{45}}}
 Then,  straightforward calculation shows, for $i,j,k,l  = 2,3,4,5$:
 \eqn\chvarui{\eqalign{
 & {\tu}_{1i} = {\half} \chi\  {\ve}_{1ijkl} \ u_{1j}\ u_{kl}  \cr
 & {\tu}_{ij} =  {\half} \chi\   {\ve}_{1ijkl} \ u_{kl}  \cr}}
 Explicitly: 
 \eqn\explchvarui{\eqalign{
 {\tu}_{12} = & \chi \left( u_{15}u_{34} - u_{14}u_{35} + u_{13}u_{45} \right) \cr
{\tu}_{13} = & \chi \left( u_{15}u_{24} - u_{14}u_{25} + u_{12}u_{45} \right) \cr
{\tu}_{14} = & \chi \left( u_{15} u_{23} - u_{13} u_{25} + u_{12}u_{35} \right) \cr
& \quad  {\tf} = {\varphi} - {\log}{\chi}  \cr}}
and, finally, 
$${\tu}_{23} =   \chi u_{45} ,  \quad {\tu}_{45} =  \chi u_{23}, \quad
 {\tu}_{25} =  \chi u_{34} ,  \quad {\tu}_{34} =\chi u_{25}, \quad {\tu}_{35} =    - \chi u_{24} ,   \quad  {\tu}_{24} =  - \chi u_{35} . $$

\subsec{Anomaly two-form and anomaly cancellation}

We can easily calculate the anomaly two-form, using \chvar\chvarui:
\eqn\anomtfps{{\psi}_{+++++, +++--,+----} = -4 {\dd} {\log} u_{45} \wedge {\dd} {\log} {\chi}}
Similarly, 
\eqn\anomtfpss{{\psi}_{+++++, +++--, ++-+-} = -4 {\dd} {\log} u_{45} \wedge {\dd} {\log} u_{35}}
On $X_{10}$, as we indicated above, this is a coboundary, e.g.
\eqn\anomtfpci{{\psi}_{+++++, +++--, ++-+-} =
4{\dd} {\tilde\varphi} \wedge {\dd} {\log} {\tu}_{45} -4 {\dd} {\varphi} \wedge {\dd} {\log} u_{45}}
In order to write it as coboundary we must be able to use the expressions
like ${\dd}{\varphi}$ which are well-defined on ${\bC}^*$ but will not be so well-defined on 
${\bC}$, i.e. they do not extend to the zero section of $L$, the line bundle above. 
This is of course a particular case of a general phenomenon we discussed in the section devoted to
$\b\g$-systems taking values in the general cones.
One can illustrate this general result with the explicit calculation.
We study, for simplicity, the triple intersection of the coordinate
patches $U_{+++++}$, $U_{+++--}$, and $U_{++-+-}$.  Let us label them with the
indices $1$, $2$ and $3$. We shall denote the corresponding fields
$p, {\varphi}, u_{ab}, v^{ab}$ via
\eqn\pfuv{p[{\a}] , {\varphi}[{\a}] , u_{ab} [{\a}] , v^{ab}[{\a}], \qquad\qquad {\a} = 1,2,3}
We have:
\eqn\onetwo{\eqalign{& p[2]= p[1] - 2{\p} {\log} u_{45}[1] \cr
& v^{ij} [2] = v^{ij} [1] , \qquad i,j = 1,2,3 \cr
& v^{i4}[2] = \sum_{a} v^{ia}[1] u_{4a}[1], \qquad i = 1,2,3 \cr
& v^{i5}[2] =\sum_{a} v^{ia}[1] u_{a5}[1], \qquad i = 1,2,3 \cr
& v^{45}[2] = 3 {\p} u_{45}[1] + ( p[1] - 2{\p}{\varphi}[1] ) u_{45}[1] - \sum_{a,b} v^{ab}[1] u_{a4}[1] u_{b5}[1] \cr}}
\eqn\twothree{\eqalign{& p[3]= p[2] - 2{\p} {\log} u_{34}[2] \cr
& v^{ij} [3] = v^{ij} [2] , \qquad i,j = 1,2,5 \cr
& v^{i3}[3] = \sum_{a} v^{ia}[2] u_{3a}[2], \qquad i = 1,2,5 \cr
& v^{i4}[3] =\sum_{a} v^{ia}[2] u_{a4}[2], \qquad i = 1,2,5 \cr
& v^{34}[3] = 3 {\p} u_{34}[2] + ( p[2] - 2{\p}{\varphi}[2] ) u_{34}[2] - \sum_{a,b} v^{ab}[2] u_{a3}[2] u_{b4}[2] \cr}}
\eqn\onethree{\eqalign{
& p[3]= p[1] - 2{\p} {\log} u_{35}[1] \cr
& v^{ij} [3] = {\ve}_{j} v^{ij} [1] , \qquad i , j = 1,2, 4, \quad  i < j  \cr
& v^{i3}[3] = {\ve}_{i} \sum_{a} v^{ia}[1] u_{3a}[1], \qquad i = 1,2, 4 \cr
& v^{i5}[3] = {\ve}_{i} \sum_{a} v^{ia}[1] u_{a5}[1], \qquad i = 1,2,4 \cr
& v^{35}[3] = 3 {\p} u_{35}[1] + ( p[1] - 2{\p}{\varphi}[1] ) u_{35}[1] - \sum_{a,b} v^{ab}[1] u_{a3}[1] u_{b5}[1] \cr}}
where all the products are understood with the normal ordering, and the ${\ve}$-symbol is
$$
{\ve}_{1} = {\ve}_{2} = 1, \qquad {\ve}_{4} = -1
$$
Now if we substitute \onetwo\ into \twothree, with the normal ordering understood,
we would get \onethree. This means that the Pontryagin anomaly is cancelled.

\subsubsec{The $SO(10)$ and ${\bC}^{*}$ current algebras and Virasoro algebra}

The $SO(10)$ and ${\bC}^{*}$ currents are defined as follows ( $ a,b = 1, \ldots , 5$ ):
\eqn\soten{\eqalign{ N^{ab} = \ & v^{ab} \cr
 N_{a}^{b} = \ & u_{ac}v^{bc} + {\d}_{a}^{b} \left( {\p}{\varphi} - {\half} p \right) \cr
 N_{ab} = \ & - 3 {\p} u_{ab} + u_{ab} \left( 2 {\p}{\varphi} - p \right) +  v^{cd} u_{ac} u_{bd}   \cr
 J = \ & p + 2 {\p}{\varphi} \cr}}
We list here for completeness the operator product expansion of these currents:
\eqn\opeca{\eqalign{ N^{ab}(z) N_{cd}(0) \qquad\sim \ & - {3 {\d}^{[a}_{c}{\d}^{b]}_{d} \over z^2}
 - {{\d}^{[a}_{c} N^{b]}_{d} + {\d}^{[b}_{d} N^{a]}_{c} \over z} \cr
  N_{a}^{b}(z) N_{c}^{d} (0) \qquad\sim \ & - {3 {\d}_{a}^{d}{\d}_{c}^{b} \over z^{2}}  - {N_{a}^{d} {\d}^{b}_{c} - N_{c}^{b} {\d}^{a}_{d} \over z} \cr
  N_{ab} (z) N_{c}^{d}(0) \qquad\sim \ &\qquad\qquad\quad  {N_{ac} {\d}^{d}_{b} - N_{bc} {\d}^{d}_{a} \over z} \cr
 N^{ab} (z) N_{c}^{d} (0) \qquad\sim\ &\qquad \qquad\quad {N^{bd} {\d}_{c}^{d} - N^{ad} {\d}_{c}^{b} \over z} \cr
  J (z) J (0) \qquad\sim\ & - {4 \over z^{2}} \qquad \qquad  \cr
  N (z) J (0 ) \qquad\sim \ & \qquad\qquad \rm regular \qquad \qquad  \cr
  }}
 Finally, the stress-energy tensor is given by:
\eqn\strt{T = \sum_{a < b} v^{ab} {\p}u_{ab} + p {\p}{\varphi} + 4 {\p}^{2} {\varphi}}
The ${\p}^{2}{\varphi}$ term in \strt\ is due to the ${\varphi}$-dependence of the holomorphic nowhere vanishing
$SO(10)$-invariant holomorphic top form ${\Omega}$:
\eqn\omgps{{\Omega} = e^{ 8 {\varphi}} {\dd} {\varphi} \wedge \bigwedge_{a < b} {\dd} u_{ab} }

\sssec{\rm Remark \ on \ orbifolds.} Note that by performing the ${\bZ}_{8}$ orbifold of $X_{10}$
\eqn\lcy{{\l} \sim e^{2\pi i k \over 8} {\l} , \qquad  k = 0, 1, \ldots, 7}
(and the ${\bZ}_{2d-2}$ orbifold of $X_{2d}$) and by gluing the zero section ${\g} = e^{{\varphi}} = 0$ we would get a local Calabi-Yau manifold. However, the $p_1$ anomaly immediately appears. 

If we don't add the zero section, then we face another problem:  Berkovits ${\CQ}$ operator \qop\ is not invariant under the ${\bZ}_{8}$ symmetry, however it can be made invariant under its ${\bZ}_{4}$ subgroup
by making ${\theta}$ and $X$ transform appropriately. The ${\bZ}_{4}$ orbifold would lead to a type 0 string on
a space with ${\bR}^{10}/{\bZ}_{2}$ singularity (see \ref\vijay{V.~Balasubramanian, S.F.~Hassan, E.~Keski-Vakkuri, A.~Naqvi, {\it Space-time orbifold: A toy model for a cosmological singularity}, hep-th/0202187}\ref\typez{N.~Nekrasov, S.~Shatashvili, {\it On non-supersymmetric CFT in four dimensions}, hep-th/9902110 \semi
I.~Klebanov, N.~Nekrasov, S.~Shatashvili, {\it An orbifold of type 0B strings and non-supersymmetric gauge theories}, hep-th/9909109} for related work). It is
interesting to investigate this model, in particular its twisted sector, further. Even smaller subgroup
${\bZ}_{2}$, which flips the signs of $\l$ and $\t$ simultaneously, leads to type 0 strings in ${\bR}^{10}$\ref\natpriv{N.~Berkovits, private communication}.

\subsubsec{Field transformations and currents}

 Following our general formulae \bttr, had we set ${\m} = 0$ we would have gotten:
 \eqn\chvarb{\eqalign{& {\tp} = p \cr
 & {\tv}^{ij} = v^{ij} , \qquad i,j = 1,2,3 \cr
 & {\tv}^{i4} = - \sum_{a} v^{ia} u_{a4} \cr 
 & {\tv}^{i5} = \sum_{a} v^{ia} u_{a5} \cr
 & {\tv}^{45} = 5 {\p} u_{45} + p u_{45} - \sum_{a,b} v^{ab} u_{a4} u_{b5} \cr}}  
 These formulae are not quite satisfactory, since the current $J$ does not transform to itself,
 and the currents $N^{ab}$ are not quite mapped to the $SO(10)$ currents. Luckily, this is
 a problem a smart choice of the two-form ${\m}$ can fix. Indeed,
 the two-form:
 \eqn\mmf{{\m} = {\m}_{+++++, +++--} = 4 {\dd} {\tf} \wedge {\dd} {\log} {\tu}_{45}\ ,}
 which is the fellow entering \anomtfpci, 
 changes \chvarb\ to:
 \eqn\chvarbi{\eqalign{& {\tp} = p  - 2 {\p} {\log} u_{45} \cr
 & {\tv}^{ij} = v^{ij} , \qquad i,j = 1,2,3 \cr
 & {\tv}^{i4} = \sum_{a} v^{ia} u_{4a} = N^{i}_{4}  \cr 
 & {\tv}^{i5} = \sum_{a} v^{ia} u_{a5} = - N^{i}_{5} \cr
 & {\tv}^{45} = 3 {\p} u_{45} + ( p - 2{\p}{\varphi})  u_{45} - \sum_{a,b} v^{ab} u_{a4} u_{b5}  = - N_{45} \cr}}  
Also, by explicit calculation we can verify that  ${\tilde J} =  \, J$,  and  for $ i,j = 1,2, 3$, $m = 4,5$:
\eqn\crpppmm{\eqalign{{\tN}_{i}^{j} = N_{i}^{j} , & \qquad {\tN}_{m}^{m} = - N_{m}^{m}, \cr
{\tN}_{5}^{i} = - N^{i5} , & \qquad {\tN}_{i}^{5} = N_{i5} \cr
{\tN}_{i}^{4} = - N_{i4}, & \qquad {\tN}_{4}^{i} = N^{i4} \cr
{\tN}_{4}^{5} = N_{5}^{4}, & \qquad {\tN}_{5}^{4} = N_{4}^{5} \cr
{\tN}_{ij} = N_{ij}, &\qquad  {\tN}^{ij} = N^{ij} \cr
{\tN}_{i4} = - N_{i}^{4}, & \qquad {\tN}_{i5} = N_{i}^{5} \cr
{\tN}^{i4} = N_{4}^{i} , & \qquad {\tN}^{i5} = - N_{5}^{i} \cr
{\tN}^{45} = - N_{45}, & \qquad {\tN}_{45} = - N^{45} \cr}} 
which is the action on ${\bf so}(10)$ of a particular element of the group ${\bZ}_{2}^{4} = {\CW}_{\bf so(10)}/{\CW}_{\bf su(5)}$.
Now, we go on to the transformation of the $p$ and $v$ fields, corresponding to the
change of the coordinates \chvarui. Based on our experience
with the previous example, we find that we have to use the closed two-form ${\m}$:
\eqn\mform{{\m} = {\m}_{+++++, +----} = - 4 {\dd} {\tf} \wedge {\dd} {\log} {\chi}}
which leads to the following transformation law:
\eqn\pvtr{\eqalign{ & {\tp} = p + 2 {\p} {\log} {\chi} \cr
&   {\tv}_{12} = N_{2}^{1}, \qquad {\tv}_{13} = - N_{3}^{1} \cr
& {\tv}_{23} = - N_{23}, \qquad {\tv}_{14} =  N_{4}^{1}\cr
& {\tv}_{24} = N_{24}, \qquad {\tv}_{34} = -N_{34} \cr
& {\tv}_{15} = - N_{5}^{1}, \qquad {\tv}_{25} = - N_{25} \cr
& {\tv}_{35} = N_{35}, \qquad {\tv}_{45}  = - N_{45} \cr}}

\subsubsec{Current algebras on pure spinors in $D = 2d$ dimensions}

It is clear from our discussion of ${\bC}^{*}$-bundles that the ten dimensional pure spinors are not special as far as the consistency of the curved beta-gamma system is concerned (they are very special for the construction of manifestly covariant superstring action, of course). In this section we remind the
formulae for the $SO(2d)$ and $U(1)$ currents for the sigma model on the space $X_{2d}$ of pure spinors
in $D=2d$ dimensions (again, with the point ${\l}  = 0$ deleted). These currents were written in \natnik\ using Friedan-Martinec-Shenker \fms\ bosonization, and could also be constructed with the help
of Feigin-Frenkel \feiginfrenkel,\edik\ approach,  but in our approach they are most straightforwardly obtained using the coordinate transformations \bttr\ for appropriate ${\m}$-forms.

Thus, the formulas for the currents in $D=2d$ are given by
\eqn\firstp{
J = -p - 2{\p}{\varphi},}
$$
N^{ab} = v^{ab},$$
$$N^b_a =  - u_{ac}v^{bc} + {\d}_{a}^{b} \left( {\p}{\varphi} - {\half} p \right) ,$$
$$N_{ab} =   (d-2) {\p} u_{ab} + u_{ac}u_{bd} v^{cd} + u_{ab} \left( 2{\p}{\varphi} - p \right),$$
$$T =  \half v^{ab} {\p} u_{ab} + p {\p} {\varphi} + ( d - 1 ) {\p}^{2} {\varphi} 
$$ 
where $T$ is the stress tensor and the $p, v^{ab} = - v^{ba}, {\varphi}, u_{ab}  = - u_{ba}$, $1 \leq a, b \leq d$ fields have 
the operator product expansions \opei:
\eqn\opeone{p(y) {\varphi} (z)\sim -{1\over y-z},\quad
v^{ab}(y) u_{cd} (z) \sim -{\d_c^{[a} \d_d^{b]} {\dd y} \over y-z}.}
The operator product expansions of the currents  \firstp\ can
be computed to be
\eqn\OPE{ N_{mn}(y) \l^\a(z) \sim \half {1\over y-z} (\gamma_{mn}\l)^\a, \quad
J(y) \l^\a(z) \sim {1\over y-z} \l^\a,}
$$N^{kl}(y) N^{mn}(z) \sim
{2-d\over  (y-z)^{2}}
(\eta^{n[k} \eta^{l]m}) +
{1\over y-z}(\eta^{m[l} N^{k]n} -
\eta^{n[l} N^{k]m} )
,$$
$$ J(y) J(z) \sim -{4 \over (y-z)^{2}}, \quad J(y) N^{mn}(z) \sim 0, $$
$$N_{mn}(y) T(z) \sim {1\over (y-z)^{2}} N_{mn}(z) ,\quad
J(y) T(z) \sim  {2-2d\over (y-z)^{3}} + {1\over (y-z)^{2}} J(z),$$
$$T(y) T(z) \sim  {\half}{{d(d-1)+2}\over{(y-z)^{4}}} + {2\over{(y-z)^{2}}} T(z) +
{1\over{y-z}} \p T.$$
So the central charge of Virasoro algebra, generated by $T$,  is $c = d(d-1)+2$, the ghost-number anomaly is
$q = 2-2d$, the level of ${\hat {\bf so(2d)}}$ is $k_{{\bf so(2d)}} = 2-d$, and the ghost-number central
charge is $k_{\bf u(1)} = -4$.

The geometrical meaning of these results is the following \natnik:
\eqn\geomm{\eqalign{& c = {\rm dim}_{\bC} (X_{2d}) \cr
& q = c_{1} ( {\tilde Q}_{2d}) \cr}}
while the geometry behind $k_{\bf so(2d)}$ and $k_{\bf u(1)}$ is explained in \feiginfrenkel,\edik (for example, $k_{\bf so(2d)} = h_{{\bf su(d)}} - h_{\bf so(2d)}$). 
One can also verify the consistency of these charges by considering the
Sugawara presentation of the stress tensor
\eqn\sugawara{T = 
{1\over 2(k_{\bf so(2d)}+h_{\bf so(2d)})} \left( N_{ab}N^{ab} + N^{ab} N_{ab} - 2 N^{a}_{b} N^{b}_{a} \right) +
{1\over 8} J J + {d-1 \over 4} {\p} J }
where $k_{\bf so(2d)} = 2 - d$ is the $\hat{\bf so(2d)}$ current algebra level,  $h = 2d - 2$ is the dual Coxeter number
for $\bf so(2d)$, and the coefficient of $\p J$ has been chosen to give
the ghost-number anomaly $2-2d$.
Setting $k_{\bf so(2d)}=2-d$, one finds that the $\hat{\bf so(2d)}$ currents contribute
$(2d-1)(2-d)$ to the Virasoro central charge while the $\hat{\bf u(1)}$  current
contributes $1+ 3(d-1)^2$. So the
total conformal central charge is $c = d(d-1)+2$ as expected from geometry.

\newsec{Conclusions and the outlook}

 Let us summarize.
We have discussed the general curved beta-gamma systems, and reviewed the constraints on their
conformal invariance and coordinate invariance. We found that the conformal invariance is obstructed
if there isn't a holomorphic nowhere vanishing top degree form on the target space. The topological counterpart of this obstruction is $c_{1}(X)$, the first Chern class of the target space. We found that
the coordinate invariance is obstructed by the first Pontryagin class $p_{1}(X)$ of the target space. 

We then applied the techniques developed for general targets to the case of $X$ being the space of
pure spinors in Euclidean space ${\bR}^{2d}$, with $d=5$ case being the most interesting for the physical applications. The space $Q$ of pure spinors is a surface in vector space given by some quadratic equations. As such, it has a singularity at the origin. One needs to deal with this singularity in order to define the sigma model on this space. One option is to remove the singular point, and work with the space $X = Q - \{ 0 \}$.  Another option is to blow up the singularity, replacing $Q$ by the total space  $\hat Q$ of the appropriate line bundle  over the smooth space
of projective pure spinors. 

We showed that the first option removes all the anomalies and also removes the possible worldsheet instantons. Also, having negative powers of ${\g} = e^{\varphi}$ is important in construction of the $G$-field, the $\CQ$ partner of the stress-energy tensor. We feel, however, that this brutal removal of the singular point has to be better motivated. In particular the resulting non-compactness of the target space needs to be better treated (at present there
are some unclear issues with the definitions of string measure when $X$ is used). Moreover,
if, for some reason, ${\hat Q}$ is preferred over $X$ then  the superstring on ${\bR}^{10}$ would cease to be
consistent beyond tree and one-loop level, thereby killing at once the landscape \landscape\ problem. This is of course one of the unrealized, so far,  hopes to solve some pressing predictive issues of string theory by capitalizing on its unusual, from the conventional quantum field theory point of view, perturbation theory \moore.

We believe there are some lessons to be learned from our exersize.

\vskip 5cm

\ndt{\bf Acknowledgments.} I have benefited from  discussions with D.~Gross, D.~Kazhdan, A.~Losev, Y.~Oz, S.~Shatashvili, R.~Shklar, P.~Vanhove and especially from patient explanations of N.~Berkovits and E.~Frenkel. Research was
partly supported by European RTN under the contract 005104  "ForcesUniverse", by the grants {\cyr RFFI} 03-02-17554
and {\cyr NSh}-1999.2003.2. The hospitality of  Tel-Aviv University, Hebrew University of Jerusalem, Hangzhou Center for Mathematical Sciences, Harvard University and KITP UC Santa Barbara  during various stages of preparation of the
manuscript is gratefully acknowledged.  I also thank the organizers of the ZMP opening colloquium at Hamburg University for providing an opportunity to present there some of the results of this work. 

\footatend\vfill\supereject\immediate\closeout\rfile\writestoppt
\baselineskip=14pt\centerline{{\bf References}}\bigskip{\frenchspacing%
\parindent=20pt\escapechar=` \input refs.tmp\vfill\eject}\nonfrenchspacing
\bye